\documentclass{aa}
\usepackage{graphicx}
\usepackage{txfonts}
\usepackage{subcaption}

\begin{document}

\title{Monitoring dusty sources in the vicinity of Sgr A*}

\author{F. Pei{\ss}ker\inst{\ref{inst1}}
   \and S.E. Hosseini\inst{\ref{inst1},\ref{inst2}}
   \and M. Zaja\v{c}ek\inst{\ref{inst1},\ref{inst2},\ref{inst5}}
\and A. Eckart\inst{\ref{inst1},\ref{inst2}}
\and R. Saalfeld\inst{\ref{inst1}}
   \and M. Valencia-S.\inst{\ref{inst1}}
   \and M. Parsa\inst{\ref{inst1},\ref{inst2}}
   \and V. Karas\inst{\ref{inst6}}
   }
\institute{{I.Physikalisches Institut der Universit\"at zu K\"oln, Z\"ulpicher Str. 77, 50937 K\"oln, Germany\label{inst1} \\
\email{peissker@ph1.uni-koeln.de}}
\and Max-Plank-Institut f\"ur Radioastronomie, Auf dem H\"ugel 69, 53121 Bonn, Germany\label{inst2}
\and Center for Theoretical Physics, Polish Academy of Sciences, Al. Lotnikow 32/46, 02-668 Warsaw, Poland\label{inst5}
\and Astronomical Institute, Czech Academy of Sciences, Bo\v{c}n\'{i} II 1401, CZ-14100 Prague, Czech Republic\label{inst6}
}
\date{Received ?? / Accepted ??}

\abstract
{We trace several dusty infrared sources on their orbits around Sgr A* with SINFONI and NACO mounted at the VLT/Chile. These sources show near-infrared excess and Doppler-shifted line emission. We investigate these sources in order to clarify their nature and compare their relationship to other observed NIR objects close to Sgr A*.}
{By using SINFONI, we are able to determine the spectroscopic properties of the investigated dusty infrared sources. Furthermore, we extract spatial and velocity information of these objects. We are able to identify X7, X7.1, X8, G1, DSO/G2, D2, D23, D3, D3.1, D5, and D9 in the Doppler-shifted line maps of the SINFONI H+K data. From our K- and L$'$-band NACO data, we derive the related magnitudes of the brightest sources located west of Sgr A*.}
{For determining the line of sight velocity information and to investigate single emission lines, we use the near-infrared integral field spectrograph SINFONI data-sets between 2005 and 2015. For the kinematic analysis, we use NACO data-sets between 2002 and 2018. This study is done in the H, $K_s$, and L$'$ band. From the 3D SINFONI data-cubes, we extract line-maps in order to derive positional information of the sources. In the NACO images, we identify the dusty counterpart of the objects. If possible, we determine Keplerian orbits and apply a photometric analysis.}
{The spectrum of the investigated objects show a Doppler-shifted Br$\gamma$ and HeI line emission. For some objects west of Sgr A*, we find additionally [Fe III] line emission that can be clearly distinguished from the background. A one-component blackbody model fits the extracted near-infrared flux for the majority of the investigated objects, with the characteristic dust temperature of $500\,{\rm K}$. The photometric derived \textit{H}- and \textit{K$_S$}-band magnitudes are between mag$_H > 22.5$ and mag$_K = 18.1^{+0.3}_{-0.8}$ for the dusty sources. For the \textit{H}-band magnitudes we can provide an upper limit. For the bright dusty sources D2, D23, and D3, the Keplerian orbits are elliptical with a semi-major axis of  a$_{D2} = (749 \pm 13)$ mas, a$_{D23} = (879 \pm 13) $, and a$_{D3} = (880 \pm 13 )$ mas. For the DSO/G2, a single-temperature and a two-component blackbody model is fitted to the H-, K-, L$'$-, and M-band data, while the two-component model that consists of a star and an envelope fits its SED better than an originally proposed single-temperature dusty cloud.}%22.98$
{The spectroscopic analysis indicates, that the investigated objects could be dust embedded pre-main-sequence stars. The  Doppler-shifted [Fe III] line can be spectroscopically identified in several sources that are located between 17:45:40.05 and 17:45:42.00 in DEC. However, the sources with a DEC less than 17:45:40.05 show no [Fe III] emission. Therefore, these two groups show different spectroscopic features that could be explained by the interaction with a non-spherical outflow that originates at the position of Sgr A*. Followed by this, the hot bubble around Sgr A* consists out of isolated sources with [Fe III] line emission that can partially account for the previously detected [Fe III] distribution on larger scales.}{}


\maketitle

\section{Introduction}
\label{Introduction}

At a distance of approximately 8 kpc at the center of the Milky Way, a supermassive black hole (SMBH) named Sagittarius A* (Sgr A*) with a mass of $\sim 4 \times 10^6 M_\odot$ can be found \citep{Eckart1996a,Schoedel2002,Genzel2010,2017FoPh...47..553E,Gravity2019}. From a scientific point of view, it provides a unique opportunity to understand the dynamical mechanisms of an accreting supermassive black hole. Almost 40 years ago, \cite{Wollman1977} observed radial velocities of several hundred $km\, s^{-1}$ of ionized gas that suggested the presence of a supermassive black hole. This finding was then underlined by the observation of massive young stars that orbit the SMBH with high velocities (\citealp[see, e.g.,][]{Eckart1996a, Ghez1998}). For a long time, it was believed that star formation in the vicinity of Sgr A* is not possible because of the harsh environment. Stellar winds, the dense concentration of stars, a possible magnetic field, shock waves, the gas densities, and strong radiation are considerable arguments for the lack of \textit{in situ} star formation.

In contrast, \cite{Morris1996} suggested that the constant inflow of gas might trigger star formation that could play a role when accretion processes are investigated. Followed by this, the detection of young stars in the Galactic Center, mostly O/B and Wolf Rayet WR stars, indicate a recent star-forming epoch \citep{Eisenhauer2005, Habibi2017}. \cite{Ghez2003} formulated the 'paradox of youth' where the authors reflect the contrary situation. 

\cite{Gillessen2012} reported the discovery of a gas cloud named G2 (\citealp[see e.g.][for a review]{Burkert2012, Meyer2012, Schartmann2012, Scoville2013}) that is moving towards Sgr A*. It is following the gas cloud G1 (\cite{Clenet2005}, \cite{Pfuhl2015} \cite{Witzel2017}) and it was supposed to be dissolved during periapse. However, G1, as well as G2, are detected during and after the periapse in 2014 as compact sources that leads to different interpretations \citep{EckartAA2013, Witzel2014, Zajacek2014, Valencia-S.2015}. Based on these findings, there is a high probability that G2 is a dust enshrouded star with a possible bow-shock \citep{Shahzamanian2016, Zajacek2017}. The authors use a radiative transfer model, that match the detected L$'$- and K-band magnitudes. From this they conclude, that the detected Doppler-shifted emission source could be indeed a young stellar object (YSO) with a dust envelope. This would be in line with the mentioned recent star formation epoch. To emphasize the dusty and stellar nature of the object, \cite{EckartAA2013} named it Dusty S-cluster Object (DSO). To avoid confusion, we will refer to this object in the following as DSO/G2. In this work, we present our H- and K-band detection of the DSO/G2. With the known L$'$- and M-band flux from \cite{Gillessen2012} and \cite{EckartAA2013} we fit a two-component (star+dust envelope) model to the data.
\begin{figure*}[htbp!]
\centering
\includegraphics[width=1.\textwidth]{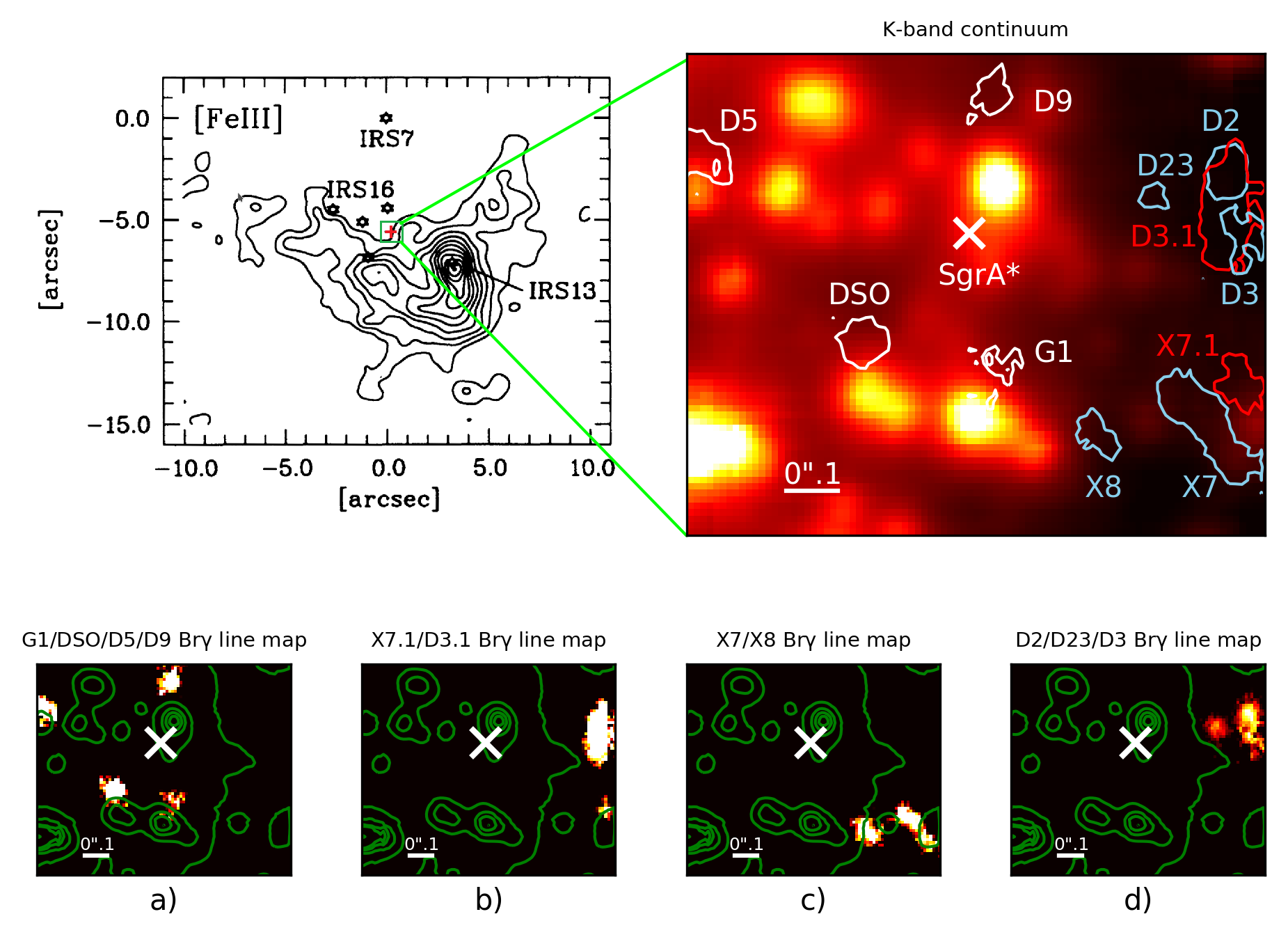}
\caption{[FeIII] distribution, K-band continuum, and Doppler-shifted line maps of the GC. North is up, east is to the left in every image. The line maps show the dusty sources in 2008 and 2015 (lower 4 plots). For a better comparability we show the line maps on top of the 2015 K-band continuum maps in all 4 panels a) to d). Due to the limited FOV of the SINFONI cubes not all sources are visible in all years. Therefore in a) we show the 2008 line map to highlight 4 sources. In b) to d), we show 2015 line maps to highlight the remaining sources. They are based on the related Doppler shifted Br$\gamma$ line extracted from the SINFONI data cubes. In the upper left corner, a [FeIII] line map adapted from \cite{Lutz1993} is presented. It shows the V-shaped [FeIII] distribution that results in the detection of a bow shock like feature. Also, IRS7, IRS13, and IRS16 are indicated. It is 20".0 x 15".0 overview where we mark the S-cluster with a green box. Next to the [FeIII] map, a K-band continuum image from the data cube of 2015 is presented. The size of the image matches the green square in the [FeIII] distribution figure next to it and provides a 1".0 x 1".0 zoom into that region. In this SINFONI K-band continuum image, the contour lines of the, in this work investigated, dusty sources are at $55\%$ of their intensity peak and are based on the Doppler shifted Br$\gamma$ line maps. The white contour lines represent the sources that show no Doppler-shifted [FeIII] emission lines. The blue and red contour lines can be associated with objects, that do show Doppler-shifted [FeIII] emission lines. The red and blue color also indicates the Doppler-shift direction. In the lower four line maps of the dusty sources, the green contour lines are based on the upper right presented K-band continuum image at $10\%$, $20\%$, $30\%$, $40\%$, $50\%$, and $100\%$ of the peak intensity. The related line map velocities, widths, errors, and emission source sizes are listed in Tab. \ref{tab:fig1data}.}
\label{finding_chart}
\end{figure*}
\begin{table*}[hbt!]
\tabcolsep=0.1cm
\centering
\begin{tabular}{|c|cccc|cc|cc|ccc|}
\hline
\hline
 Br$\gamma$ line map properties & G1 & DSO & D5 & D9 & X7.1 & D3.1 & X7 & X8 & D2 & D23 & D3 \\
\hline          
Velocity in km/s & -1195 & +1218 & +115 & -152 & +327 & +46 & -650 & -226 & -392 & -360 & -710\\
\hline
FWHM in km/s & 346 & 207 & 318 & 110 & 152 & 124 & 138 & 166 & 152 & 180 & 180\\
\hline 
Uncertainty in km/s & 45 & 11 & 42 & 11 & 17 & 17 & 11 & 17 & 58 & 29 & 17\\
\hline
Measured size in PSF & 1.1$\pm$ 0.3 & 1.0 $\pm$ 0.05 & - & 0.7 $\pm$ 0.2 & - & - & - & 1.5 $\pm$ 0.1 & 0.8 $\pm$ 0.05 & 0.9 $\pm$ 0.08& 0.8 $\pm$ 0.21\\
\hline
\end{tabular}
\caption{Br$\gamma$ line map properties related to Fig. \ref{finding_chart}. Due to the limited FOV of the SINFONI cubes, not all sources are visible in all years. Therefore for the first 4 sources, the data correspond to 2008, whereas the remainder of the table data corresponds to 2015. The uncertainties representing the standard deviation in km/s. The measured object size is given in units of the PSF extracted from S2. The given uncertainties are an indication for the data-quality and the not avoidable noise. Some sources are located at the edge of the FOV and could be part of future observations.}
\label{tab:fig1data}
\end{table*}
Additionally, \cite{EckartAA2013} and \cite{Meyer2014} reported several other dusty sources that can be found in the direct vicinity ($\sim\, 1"\,\approx\,0.04pc$) of Sgr A*. In order to investigate the nature of the dusty sources we emphasize in this paper the analysis of the D-complex that consists out of D2 and D3 as well as the newly discovered ones D23 and D3.1. This complex of dusty sources can be found about one arcseconds to the west of Sgr A*. This region contains several bright sources whereas the most prominent ones are D2 and D3 \citep{EckartAA2013}. At the same time, they are sufficiently far away from other, bright sources. This allows a detailed investigation of their NIR excess and opens the door to investigate the relation to their surrounding. We are combining the results of this investigation with already published data \citep[see][for more information]{Lutz1993, muzic2008, muzic2010, Yusef-Zadeh2012, EckartAA2013, Meyer2014, Zajacek2017} on other dusty sources to come to conclusions about their nature.

The sources of the D-complex investigated in this study show similar positional and spectroscopically properties as the X7 source \citep{muzic2010} and X8 \citep{Peissker2019}. Concerning their brightness, the dusty sources are comparable to G1 \citep{Pfuhl2015, Witzel2017} and the DSO/G2 \citep{EckartAA2013, Witzel2014, Valencia-S.2015}. The spectroscopic analysis of these sources indicates that several dusty sources could be interpreted as Young Stellar Objects (YSOs)/pre-main sequence stars surrounded by a dusty envelope that is generally non-spherical \citep{Zajacek2014, Zajacek2017}. The mystery of their origin could be answered via the numerical hydrodynamic simulations of the interaction of molecular clouds that originate from distances of several parsecs. \cite{Naoz2018} showed that stellar binaries add a velocity component to the star population that is comparable to the observed properties of the GC disk. Additionally, \cite{alig2013} modeled the interaction of a molecular cloud that interacted with a gas disk. The authors discuss that the stellar disks resulting from this interaction could be responsible for the observed young stars and their orbital parameters. However, \cite{Jalali2014} showed that the gravitational influence of Sgr A* could affect the star formation rate inside approaching clouds with an initial mass of several $100\,M_\odot$. We investigate if other known and unknown sources could possibly contribute to this scenario or if the sources are rather dissolving gas clouds that are formed through other processes \citep{Calderon2018}. Additionally, the spectral analysis shows that next to the HeI and Br$\gamma$ emission lines there is also a Doppler-shifted ${}^3 G\,-\,{}^3 H$ [Fe III] multiplet. Based on this finding, we can divide the investigated sources in the field of view (FOV) of SINFONI into two groups: those with [Fe III] emission and the rest without these iron lines. We investigate whether this finding influences the view on the Sgr A* bubble that is reported by \cite{Lutz1993} and \cite{Yusef-Zadeh2012}. \cite{Lutz1993} show the [Fe III] distribution at the position of the mini-cavity (see Fig. \ref{finding_chart}, upper left image). The authors not only report a bubble around Sgr A* but also a V-shaped bow-shock feature caused by a wind that most probably originates at the position of Sgr A*. \cite{Yusef-Zadeh2012} show several emission lines observed with the Very Large Array (VLA). The presented data also indicate the presence of a bubble around Sgr A*. They additionally report a $2"$ hole in the investigated bubble that is located to the northern side of Sgr A* and resembles the shape of the mini-cavity. The authors discuss a possible wind and wind-wind interaction from Sgr A* (see also \citealp{muzic2010}) but also the presence of young massive stars \cite{muzic2007}. \cite{Yusef-Zadeh2012} conclude that wind from Sgr A* explains the observed features. We will follow this setting and present observational evidence that underlines the presence of a high ionizing wind that is expected to originate at the position of Sgr A*. Additionally, the D-complex reveals several dusty sources that emphasize the idea, that YSOs form in groups in the vicinity of Sgr A*. This process is realized by infalling molecular clumps that disperse after forming a few solar masses main sequence stars \citep{Jalali2014}. We will focus our analysis on the brightest sources in this complex because of the high chance of confusion. 
\\
We are using adaptive optics (AO) for the NACO imaging in the K$_S$- and L$'$-band and spectroscopic data from SINFONI with the H+K grating. Both instruments are mounted by the time of the observations at the Very Large Telescope (VLT).
\begin{figure*}[htbp!]
\centering
\includegraphics[width=1.\textwidth]{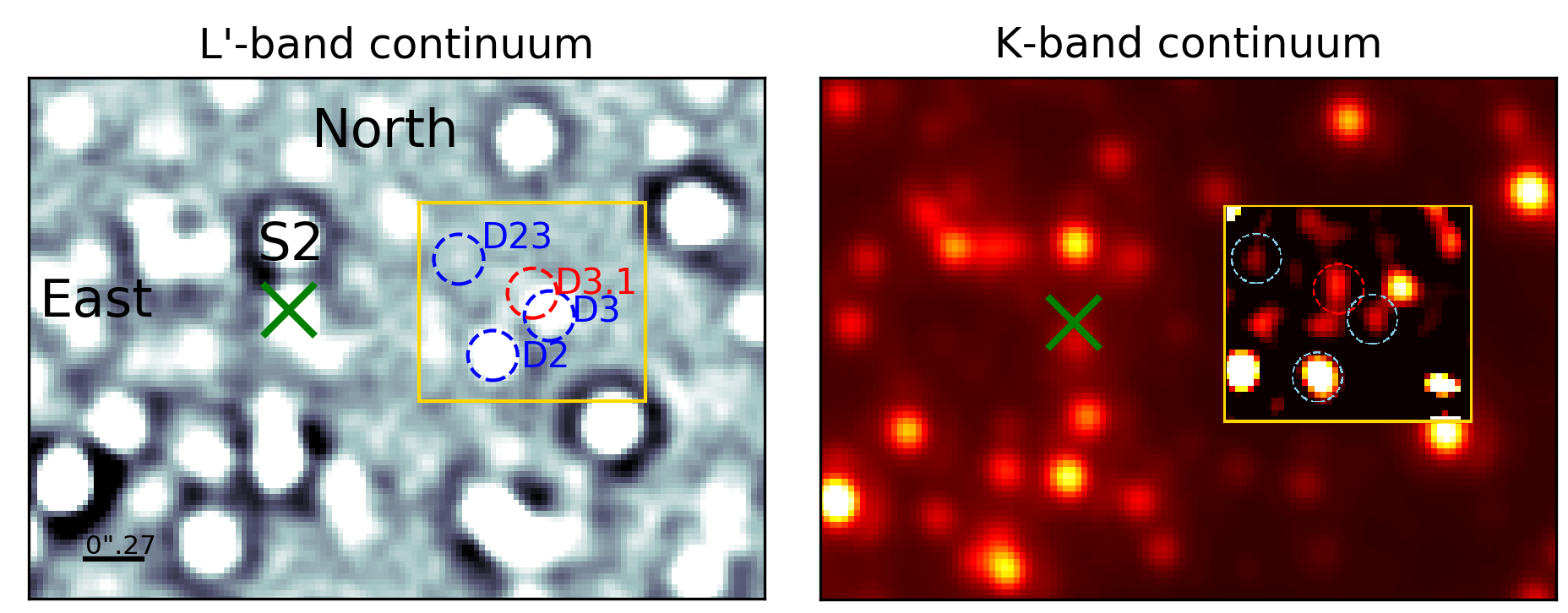}
\caption{L- and K-band continuum of the GC in 2006. The approximate position of Sgr A* is indicated by a green x. Above the SMBH, the bright B2V S-star S2 can be found. The D-complex is indicated by a golden rectangle box. At the position of D2 in 2006, the S-star S43 located. However, the L-band flux of S43 is low compared to D2 (flux D2/S43 $\sim$ 2/1). The detected emission in this NACO image is therefore dominated by D2 in 2006. Because of the chosen contrast, D3 and D3.1 seem to be confused. The right image shows the same FOV in the K-band continuum. Even though this image is from 2004, the D-sources are visible at roughly the same position as in the 2006 L$'$-band image (left). Since their magnitudes are close to the detection limit, we are using high- and low-pass filter to identify a K-band counterpart of the dusty sources. The inlet in the right image shows a 3 px smooth-subtracted detection of these objects.}
\label{finding_chart_naco}
\end{figure*}
Section \ref{Observations} explains the observations with a short overview of the on-source integration time. We also list the NACO and SINFONI data that is used for the analysis. Followed by that, Sec. \ref{Analysis} provides more information about the analysis tools like the high- and low-pass filtering. We also explain how to derive positions and velocities from the data. The results are presented in Sec. \ref{Results} and discussed in Sec. \ref{sec:discussion}. Our conclusions are presented in Sec. \ref{Conclusion}. The same section will discuss the impact on the existing dialogue about the origin and nature of the dusty sources that can be found in the S-cluster. It will also connect the emission of the objects with the [Fe III] distribution that is located mostly at the mini-cavity. 

\section{Observations}
\label{Observations}
This section will give a brief overview of the observational conditions, the used instrumentation, the
structure of the data and the employed data reduction algorithms. The observations have been carried out between 2002 and 2018 with the VLT as described in the next sections in detail. Usually, the weather conditions are fair with mostly clear skies. The L- and K-band seeing is in general good with values ranging between 0".2 and 1".2 measured directly on the Natural Guide Star (NGS) that is used for the SINFONI and NACO observations. For all observations, AO is used with a NGS in order to avoid the cone effect that could influence the data whilst using a Laser Guide Star (LGS).  
\subsection{SINFONI}
SINFONI is a near-infrared integral field spectrograph (IFS) that works between $1.1\,\mu m\,-\,2.45\,\mu m$ \citep[see][for a detailed description]{Eisenhauer2003, Bonnet2004}. The instrument uses an optical wavelength sensor. The structure of the resulting 3d data cubes is organized in a way, that 2 dimensions are associated with the spatial x-y plane. Additional spectral information is provided by the third axis of the cube. The Galactic center data is obtained with the H+K grating and enabled AO. We are using the smallest plate scale with a Field of View (FOV) of $0".8\,\times \,0".8$ and a spatial pixel scale of 25 mas/pixel. The used AO star is located at $15".54$ north and $8".85$ east of Sgr A*. The optical magnitude of this star is around mag $=\,14.6$. The DIT of single exposures is set to 400s in order to individually react to fast-changing weather conditions. Observations, that are carried out by other groups use in contrast a DIT setting of 600s. The subtracted sky is located $5'36"$ north and $12'45"$ west of the central SMBH. Of course, the exposure time setting of the sky fits the object observations and is set to the H+K domain. Before the data reduction, we apply a cosmic ray correction \citep{Pych2004} and a flat field correction to the data. After the standard data reduction \citep{Modigliani2007}, we stacked single data-cubes to increase the Signal/Noise (S/N) ratio \citep{Valencia-S.2015, Peissker2019}. For that, we use a reference frame that is created from single exposures. From this, we extract the positions of several stars (S1, S2, S4, and S62). These coordinates are then used to shift the single exposures in a $100\,\times\,100$ pixel array. The single exposures are calibrated to avoid effects that are caused by the overlapping single data cubes. We select just the best cubes for the stacking by picking data with a Point Spread Function (PSF) that provides FWHM values $<\,6.0-6.5$ pixel.

\subsection{NACO}
NAOS+CONICA (NACO) is a near-infrared imager with a spectrograph mode \citep[][]{Lenzen2003, Rousset2003}. We used the imager mode of NACO to observe the GC in the L$'$- and in the K$_s$-band. IRS7 with an K-band magnitude of $7.7$ and $5".5$ north of Sgr A* is a suitable AO star (see Fig. \ref{finding_chart}). The used number of exposures and the total exposure time of the data in K$_s$-band is listed in Tab. \ref{data} with the observation date and project ID. In the auto-jitter template, the jitter box is set to $4".0$ with a DIT of 10 sec and an NDIT setting of 3. In total, a single Observation Block (OB) contains 42 offset positions. The K-band observations are carried out with the S13 camera. In the AO configuration file, we are setting the dichroic mirror to N20C80 which results in 20 $\%$ light for NAOS. The remaining 80 $\%$ are directed to the detector (CONICA). For this setting, we only use high-quality data with infrared seeing values of $<\,1".0$. The spatial range of the L$'$-band data is set to 27 mas/pixel with a FOV of $28"\times 28"$. Because of the robustness of the L$'$ observations when it comes to atmospheric influences, we are able to select a wider range of data compared to the K-band domain.
Table \ref{data} is an overview of the used NACO data and lists the observation date, the associated program ID and the number of exposures with the resulting exposure time. The majority of the investigated NACO data is the same as discussed in \cite{muzic2010}, \cite{Dodds-Eden2011}, \cite{Witzel2012}, \cite{Sabha2012}, \cite{Shahzamanian2016}, and \cite{Parsa2017}.

\section{Analysis}
\label{Analysis}
In this section, we will describe the used analyzing techniques and tools. We briefly introduce the flux calibration procedure and present the emission line extraction.

\subsection{Line maps}
The Doppler-shifted line maps are extracted from the final SINFONI data-cubes that consist out of several individual observations (see Tab. \ref{tab:data_sinfo1}, \ref{tab:data_sinfo2}, and  \ref{tab:data_sinfo1}). By subtracting the continuum, the Doppler-shifted emission lines of the dusty sources can be studied. From the isolated lines, we are able to create line maps of the related source. These maps can also be associated with the line of sight (LOS) velocity. This procedure can reveal the shape of the Doppler-shifted sources but also eliminates the influence of bright stars that are confusing the spatial analysis of these objects.

\subsection{Low- and high-pass filter}

When it comes to the PSF of the data (NACO and SINFONI), we are using the low- and high-pass filter depending on the scientific goal \citep[][]{Lucy1974, muzic2010}. For the analysis, this is significant because the GC is a crowded FOV especially close to Sgr A* \citep{Sabha2012}. 
A low-pass filter can be used to blur an image, i.e. smear out non-linear pixel information. Unfortunately, this filter also smooths edges of objects that result in a decreased positional accuracy. However, noise can also be associated with low frequencies. Hence, a high-pass filter can be applied in order to isolate signals that are close to the detection limit. It can be used mainly as an edge-detector to determine the positions of stars and decrease the chance of confusion for nearby objects.

We select a 3px Gaussian when we apply the low-pass filter to the SINFONI and NACO data as discussed in \cite{muzic2010}. Followed by this, a 3-pixel Gaussian-shaped FWHM is again used to smooth the data. The size of the used Gaussian should be adjusted to the overall data-quality. The main advantage of the resulting image is reflected in an improved noise level with slightly blurred edges of the stars compared to the input file.

Applying the high-pass filter with the Lucy-Richardson algorithm to the NACO data is discussed in \cite{Witzel2012}. A natural PSF is used for the de/convolution. Starfinder is used to extract the PSF and to apply the high-pass filter. In contrast, this process needs minor adjustments using SINFONI data. First, we are extracting H- and K$_S$-band images from the collapsed data-cube in a spectral range of $1.45\mu m\,-\,1.9\mu m$ and $2.0\mu m\,-\,2.2\mu m$, respectively. Since noise can influence the resulting H- and K-band image, a local background subtraction must be applied. The validity of the subtracted amount is cross-checked with known stellar positions. Also, the number of iterations can be decreased to a level, where the possibility of creating artificial sources can be neglected. The major influence on the data is the PSF. Isolated stars that qualify for the extraction of a PSF are rare in the SINFONI FOV of the GC. Most stars in the GC have a close-by companion. Therefore, the wings of these PSF show strong signs of interactions with other nearby stars. Hence, we are using an artificial PSF for the de/convolution process to minimize the influence of overlapping wings in the SINFONI FOV of the GC. It shows, that a rotated Gaussian-shaped PSF between $40^{\circ}\,-\,50^{\circ}$ with an FWHM of $4.0\,-\,4.5$ pixel in the x-direction is suitable for the process. We use varying axis ratios between 1.0-1.2 with respect to the x-direction. This delivers the most accurate results when it comes to reference positions of objects in the FOV.

\subsection{Orbital fit}

Fitting an orbit to the data requires positions and the line-of-sight velocities. Since multiple Doppler-shifted emission lines are detected for the dusty sources, we focus on the Br$\gamma$ peak because of the S/N ratio compared to HeI or [FeIII]. %The blue-shifted Br$\gamma$ line map of D2 is extracted at $2.162\,\mu m\,\pm\,0.001$, for D3 at $2.164\,\mu m\,\pm\,0.001$. %\subsubsection{Positions, line-of-sight velocity, Orbital fit}
To avoid source-confusion because of the crowded FOV, we identified the investigated sources in the Doppler-shifted Br$\gamma$ line maps extracted from the SINFONI data-cubes. If possible, we verify the spatial information from the SINFONI line maps with L$'$ and $K_S$ NACO continuum images. %From that, we are able to determine the position of D2 and D3.
The position of Sgr A* can be determined either by searching for flares in the single SINFONI data-cubes or by using the position of the NIR counterparts of the SiO Maser in the $27\,mas\,\times\,27\,mas$ NACO images. From there, the position of Sgr A* with respect to the SiO Maser in the radio-domain can be determined. The procedure is described in detail in \cite{Parsa2017} where the authors applied all necessary corrections discussed in \cite{Pfuhl2015}. By using the position of S1, S2, and S4, the position in the SINFONI data-cubes can be extrapolated. By applying a Gaussian fit to the dusty sources in the SINFONI line maps and the NACO L$'$ and $K_S$ continuum images, the distance to the SMBH is measured.

To derive the velocity, the Doppler-shifted Br$\gamma$ and HeI spectral emission lines are fitted with a Gaussian. The fit of the blue and red-shifted lines is done in two steps. First, the continuum spectrum is fitted and subtracted. After that, the various emission lines of the sources and the ambient line emission like, e.g., Br$\gamma$ at 2.1661$\mu m$ or HeI at 2.0586 are fitted with a Gaussian function. These ambient lines are then subtracted from the spectrum and the emission lines of the sources were fitted again without the ambient emission.
From this fit, we derive the central wavelength and the Full Width at Half Maximum (FWHM). However, we compare also the line peak values that are in a satisfying agreement with the Gaussian fit. Whenever necessary, we are using this approach in order to derive the LOS velocity of the dusty sources. 

With these spatial and velocity information, we are applying a Keplerian fit to the data. The method is described in \cite{Parsa2017} in detail and will be shortly outlined in the following.

We are using a Keplerian approach in order to model the orbits of the dusty sources. With this, we are assuming that the motion of the objects is dominated by a stellar counterpart that is shielded by a dusty envelope. However, we use six initial parameters for the orbital fit: semi-major axis, eccentricity, inclination, periapsis, longitude, and the time t$_{\rm closest}$ of the periapse passage. The likelihood function minimizes the residuals for the orbital parameters.
Here, the likelihood function is defined as the weighted sum of the squared residuals. It delivers a direct feedback about the quality of the fit. This iterative process is then repeated with adjusted initial parameters.

For all measured LOS velocities discussed in this work, a barycentric and heliocentric correction is automatically applied by the SINFONI pipeline.

\subsection{Flux calibration}

For the flux determination, the S-star S2 is chosen as a calibrator since it is the brightest and the most isolated star in the S-cluster compared to other candidates. The spectrum of S2 is taken from the SINFONI data cubes with a circular aperture with radius $r = 4$ pixels. From this spectrum, we subtract a background taken at a nearby region that shows no contamination of stars. From this background aperture, the averaged flux is multiplied by the number of pixels in the S2 aperture and subtracted from the spectrum taken from S2. We determine the number of counts at 2.2 $\mu$m  from the resulting background subtracted spectrum. Afterward, we transform the known S2 K$_S$-band magnitude \citep{Schoedel2010} to the flux density of $9.12\times 10^{-12}$ erg\,s$^{-1}$\,cm$^{-2}\,\mu$m$^{-1}$. In this way, the flux density for one pixel can be derived. We multiply this number to the whole data cube, which thereby is flux calibrated. This is done for the SINFONI data cubes between 2005 and 2016.

\subsection{Emission lines of the dusty sources}

From the flux calibrated SINFONI data cubes, we create a source spectrum with a 75 mas circular aperture. We are using an iris background subtraction where we define a "ring" around the circular aperture with different inner and outer radii. Comparing this to background apertures placed in a 3-5 pixel distance to the source aperture shows, that the circular-annular method (iris subtraction) can be applied more consistently throughout the data analysis because of the crowded environment of Sgr A*. The generated averaged background is then subtracted from the formerly obtained spectrum. We focus on the spectral range between $2.0\mu m$ and $2.4\mu m$ where we identify several Doppler-shifted isolated Br$\gamma$ and HeI emission lines. 
To distinguish spectral features of the sources from features of the background, we subtract the background from the obtained spectrum. The resulting emission peaks are then fitted with a Gaussian. From the integrated fit, the flux can be estimated.

\subsection{Photometry}
\label{Photometry}
Here, we will describe the steps that are performed to determine the K- and H-band (upper limit) magnitudes of D2 and D23 from the SINFONI data. We are using appropriate calibrator stars that are listed in Tab. \ref{tab:calib_stars} and adapted from \cite{Schoedel2010}.
\begin{table}[hbt!]
    \centering
    \begin{tabular}{cccc}
         \hline
         \hline
         name & number & \textit{H} & \textit{K$_S$} \\
         \hline
         S2 & 3 & 16.00& 14.13\\
         S7 & 17 & 17.04 & 15.14\\
         S10 & 10 & 16.23& 14.12\\
         \hline
    \end{tabular}
    \caption{H- and K$_S$-band magnitudes of the three calibrator stars. These values are adapted from \cite{Schoedel2010}.}
    \label{tab:calib_stars}
\end{table}
These three calibrator S-stars S2, S7, and S10 are present in every data cube of the chosen SINFONI FOV between 2005 and 2016. The used S-stars are located in an area that is not too crowded to avoid spurious flux from other stars. 
For the calibrator stars and the sources, the counts in an r = 3-pixel aperture are extracted from the median of the SINFONI data-cube. The counts of the detector correspond to the flux of the sources plus the background emission. This background emission has to be subtracted from the continuum SINFONI data cubes for an accurate magnitude calculation.
Three sets of backgrounds are created and for each set, the mean value of five apertures with a consistent radius is derived. In this way, the magnitude of D2 and D23 is calculated three times with different backgrounds and, within the uncertainties of less than 13 mas, differently centered apertures on the sources themselves. Afterward, the mean value of these results is derived. Moreover, the uncertainty of the magnitudes with their standard deviation can be extracted with this procedure.
The fluxes of the sources are computed with the magnitudes from \cite{Schoedel2010} and the zero fluxes from \cite{Tokunaga2007} using
\begin{equation}
F\,=\,F_0 \,\times\,10^{-0.4\,m}    
\end{equation}
where $F_0$ donates the zero flux. The background-subtracted counts of the calibrator stars and their fluxes are used to estimate the calibration ratio. Furthermore, the calibration ratio is used to convert the background-subtracted counts of D2 and D23 to obtain their flux. We correct the fluxes for extinction. The extinction coefficients for both wavelength bands (H- and K$_S$-band) are adapted from \cite{Fritz2011}. It is 4.21 for the H-band and 2.42 for the K$_S$-band. 

\subsection{Determination of uncertainties}

The knowledge of detector related uncertainties should be considered for a critical evaluation of data points. In the following, we want to frame some of the errors that influence the analysis.

\begin{itemize}

\item \textup{SINFONI}\newline
As shown in the SINFONI user manual\footnote{www.eso.org}, the shape of a spectral PSF strongly depends on the position of the photons on the detector. Mainly because the selected pre-optics illuminate the gratings inconsistently. Additionally, the over- and under-subtraction of the sky correction \citep{Davies2007} plays a major role when it comes the identification of compact sources (like e.g. the DSO/G2, D23, and G1) that are close to the detection limit. It should also be mentioned, that a possible velocity gradient should be analyzed with high caution. Due to image motion, exposures of several hours can lead to rates of more than $\sim 20$ mas/hr \citep[see][for more information]{Eisenhauer2003}. This apparent movement of pixels can lead to the false interpretation of a velocity gradient.

In \cite{Peissker2019}, we already discuss the spectral uncertainty and determine a velocity uncertainty of $0.00048\mu m$ or $67$km/s. Depending on the S/N ratio, it is reasonable to use a positional line map uncertainty of $\pm 20$ mas since this value scales linearly with the total on-source exposure time.
\end{itemize}

\begin{itemize}

\item \textup{NACO}\newline
The spatial uncertainty for the NACO data is already discussed in \cite{Gillessen2009}, \cite{Plewa2015}, and \cite{Parsa2017}. These authors conclude in their analysis, that using a Gaussian fit the error of stellar positions is better than a tenth of a pixel. Since the spatial pixel scale is 13 mas, this would result in less than 1.3 mas. Because of the crowded environment around Sgr A*, it is reasonable to assume that the error range for K-band positions of the dusty sources should be at least $\pm 6.5$mas. If the confusion level of the investigated objects is increased because of the close distance to other sources like for example stellar emission or dust, we assume an error range of $\pm 13.0$mas. Since most of the dusty objects show an increased L$'$ magnitude compared to the K$_S$ counterpart, this error range can be reduced even further.
\end{itemize}

Another origin of uncertainties, that affect data from both instruments, is the position of Sgr A*. In the NACO data, the positional error for the SMBH in the GC ranges from 1 to 2 mas. Using faint flares for positioning can lead to values up to 6 mas. Also, the presence of S17 influences the exact position of Sgr A*. Regarding the SINFONI observations, the data suffer from similar problems. However, since we are combining single exposures of 400s-600s, this process results in data cubes with an integration time of several hours where we detect in almost every year at least one prominent flare.

\section{Results}
\label{Results}
In this section, we present the results of our analysis. As mentioned before, we emphasize the detection of D2 and D3 since they are the brightest and most isolated members of the here investigated sources. The related spectrum taken from the SINFONI data cubes is presented. Additional flux and magnitude information are also provided as well as an orbital fit for D2 and D3. Additionally, we present the H- and K-band detection of the DSO/G2 source in 2010 and 2013.
\subsection{Sources west of Sgr A*}
%In the following, we divide the sources into several groups: D2, D23, D3, and D3.1 belong to the D-complex. Even though X7, X7.1, and X8 are also located west of Sgr A*, we group these sources separately. East of Sgr A*, the objects G1, DSO/G2, D5, and D9 can be found.
Here, we will present known and unknown sources that are located west of Sgr A*. 
\subsubsection{D2}
Reported and discussed in \cite{EckartAA2013} and \cite{Meyer2014}, we find the dusty source in the blue-shifted Br$\gamma$ line map (Fig. \ref{finding_chart}). We choose that year because the FOV covers D2 and several other sources for comparison in the SINFONI data. As a comparison, Fig. \ref{finding_chart_naco} shows the same source in the L$'$-band in 2006. The error of the positions is between $0.5\,-\,1.0$ pixel in the NACO and SINFONI data-sets that cover the L$'$- and H+K-band. A positional error, the data quality, and the fit variations are summarized in the resulting uncertainty. 
The Doppler-shifted Br$\gamma$ and HeI velocities, that are detected in the SINFONI data are summarized in Table \ref{tab:LOS}. 
\begin{figure*}[htbp!]
\centering
\includegraphics[width=1.\textwidth]{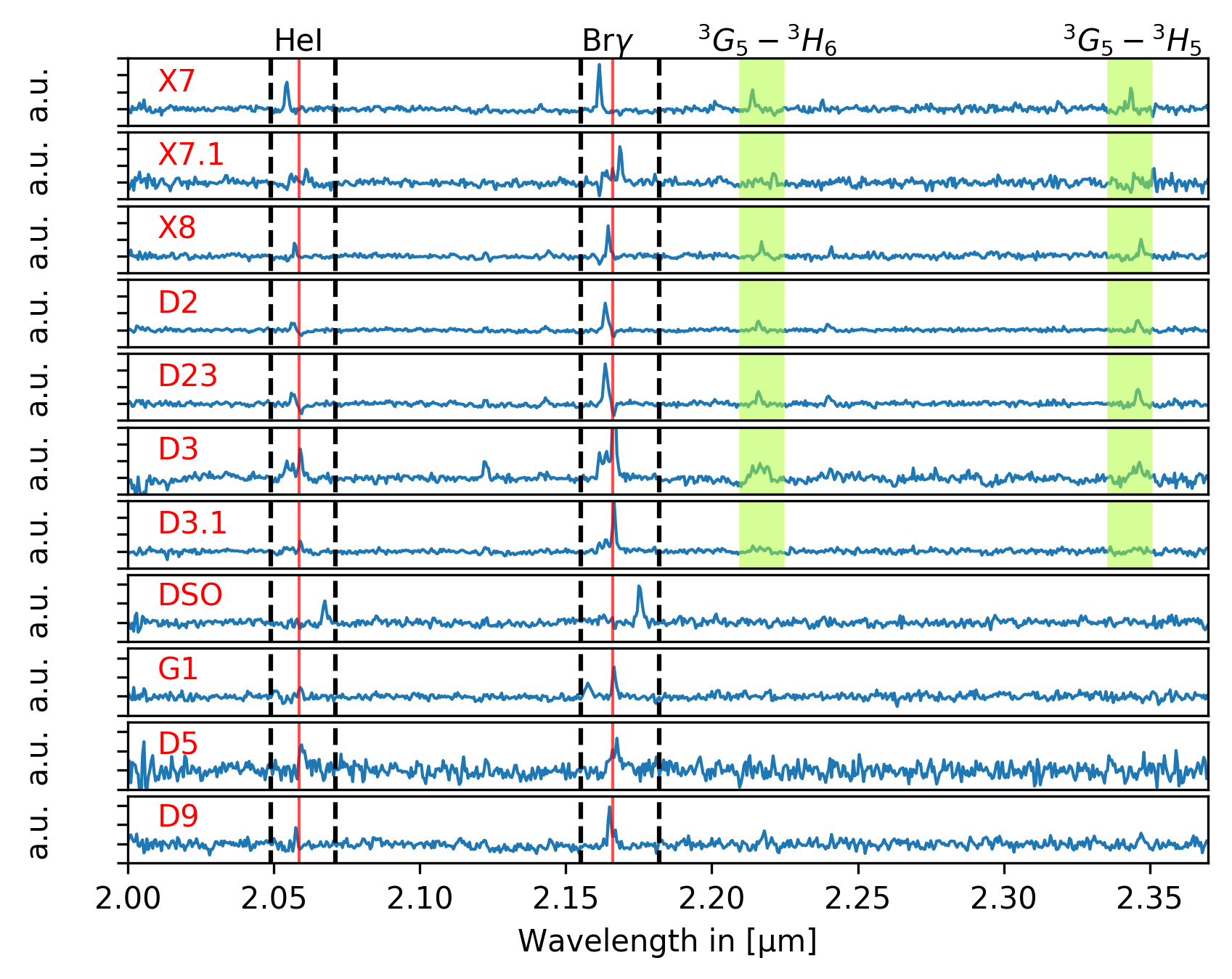}
\caption{Spectrum of the dusty sources in the GC in 2008 and 2015. The K-band spectrum between $2.0\mu m$ and $2.4\mu m$ is continuum subtracted. The red solid line marks the HeI and Br$\gamma$ rest wavelength at $2.0586\mu m$ and $2.1661\mu m$ respectively. The dashed black lines mark the range of the detected blue and red-shifted emission lines. The green bars represent the range of the detected Doppler-shifted [Fe III] lines. The ${^3}G_5\,-\,{^3}H_6$ and ${^3}G_5\,-\,{^3}H_5$ line are the most prominent of the [Fe III] multiplet. For reasons of clarity and comprehensibility, the weaker ${^3}G_3\,-\,{^3}H_4$ and ${^3}G_4\,-\,{^3}H_4$ line are not marked. We only include spectra of sources that show a line map counterpart in the SINFONI data cubes. Because of the proper motion of the investigated objects and the limited FOV, the observation of the sources is not possible throughout the entire set of SINFONI data cubes between 2005 and 2016. The intensity of the spectrum is in arbitrary units (a.u.).}
\label{spectrum}
\end{figure*}
In Fig. \ref{spectrum}, we show the spectrum of D2 in 2015. We find a blue-shifted HeI and Br$\gamma$ peak at $2.0566\mu m$ and $2.1638\mu m$. Also, [FeIII] emission lines are detected at $2.1433\mu m$, $2.2160\mu m$, $2.2403\mu m$, and $2.3459\mu m$ (see Fig. \ref{tab:felines}). The related velocities can be found in Tab. \ref{tab:LOS} and Tab. \ref{tab:Fe_rv}. Also, in Tab. \ref{tab:k_band_pos} and Tab. \ref{tab:L-band-pos} we list the K- and L-band positions of the NACO data-sets between 2002 and 2018. Before 2008, the detection of D2 in the K-band NACO data is confused with S43 (see Fig. \ref{finding_chart_naco}). In contrast, a clear detection of D2 in the L-band is possible for every data-set between 2002 and 2018.

\subsubsection{D23}
We find this newly discovered source not only in the L$'$-band NACO images (Fig. \ref{finding_chart_naco}). The source can also be detected in the blue-shifted Br$\gamma$ line maps. D23 shows several Doppler-shifted K-band emission lines like HeI, Br$\gamma$, and the [FeIII] multiplet (Tab. \ref{tab:LOS} and \ref{tab:felines}).
\begin{table*}[hbt!]
    \centering
    \begin{tabular}{cccc}
         \hline
         \hline
         source & spectral line & central wavelength [$\mu$m] & flux [10$^{-16}$ erg/s/cm$^{2}$]\\
         \hline
         &&&\\
         D2 & [Fe III]$\lambda$2.145 $\mu$m & 2.14338 $\pm$ 0.00017 & 4.039 $\pm$0.009\\
         &[Fe III]$\lambda$2.218 $\mu$m & 2.21605 $\pm$ 0.00016 & 7.951 $\pm$ 0.014\\
         &[Fe III]$\lambda$2.243 $\mu$m & 2.2403 $\pm$ 0.0003 & 6.392 $\pm$ 0.016\\
         &[Fe III]$\lambda$2.348 $\mu$m & 2.3459 $\pm$ 0.0002 &  11.25 $\pm$ 0.03\\
         &&&\\
         D23& [Fe III]$\lambda$2.145 $\mu$m & 2.14323 $\pm$ $7\times10^{-5}$ & 2.898 $\pm$ 0.008\\
         &[Fe III]$\lambda$2.218 $\mu$m & 2.21615 $\pm$ $6\times10^{-5}$ & 8.186 $\pm$ 0.014\\
         &[Fe III]$\lambda$2.243 $\mu$m & 2.24006 $\pm$ 0.00013 & 4.661$\pm$ 0.012\\
         &[Fe III]$\lambda$2.348 $\mu$m & 2.34580 $\pm$  $8\times10^{-5}$&  7.171 $\pm$ 0.014\\
         &&&\\
          D3& [Fe III]$\lambda$2.145 $\mu$m & - & - \\
         &[Fe III]$\lambda$2.218 $\mu$m & 2.21380 $\pm$ 0.00017 & 6.285 $\pm$ 0.017\\
         &[Fe III]$\lambda$2.243 $\mu$m & - & - \\
         &[Fe III]$\lambda$2.348 $\mu$m & 2.34450 $\pm$ 0.00058 & 8.893 $\pm$ 0.023\\
         &&&\\
         \hline
    \end{tabular}
    \caption{Doppler shifted [Fe III] emission lines of D2, D23, and D3.}
    \label{tab:felines}
\end{table*}
\begin{table*}[htbp!]
    \centering
    \begin{tabular}{ccccccccc}
        \hline
        \hline
        & \multicolumn{2}{c}{D2} & \multicolumn{2}{c}{D23} & \multicolumn{2}{c}{D3} & \multicolumn{2}{c}{D3.1}\\
        year & $v_{HeI}$ [km/s] & $v_{Br\gamma}$ [km/s] & $v_{HeI}$ [km/s] & $v_{Br\gamma}$ [km/s] & $v_{HeI}$ [km/s] & $v_{Br\gamma}$ [km/s] & $v_{HeI}$ [km/s] & $v_{Br\gamma}$ [km/s] \\
        \hline
        2005 & -430 & -490 & -250  & -239 &  -  &  -  &  -  &  -  \\
        2006 & -564 & -531 & -216  & -217 &  -  &  -  &  -  &  -  \\
        2007 & -523 & -479 & -226  & -215 &  -  &  -  &  -  &  -  \\
        2008 & -395 & -490 & -259  & -253 &  -318  &  -318  &  145  &  116  \\
        2009 & -458 & -440 & -     & -    &  -  &  -  &  -  &  -  \\
        2010 & -377 & -410 & -243  & -257 &  -  &  -  &  131  &  138  \\
        2011 & -401 & -387 & -281  & -266 &  -  &  -  &  89  &  198  \\
        2012 & -333 & -367 & -262  & -266 &  -  &  -  &  160  &  152  \\
        2013 & -353 & -349 & -292  & -284 &  -  &  -  &  136  &  145  \\
        2014 & -346 & -368 & -335  & -332 &  -  &  -  &  -  &  -  \\
        2015 & -294 & -319 & -324  & -325 &  -660  &  -710  &  58  &  46  \\
        2016 & -356 & -309 & -     &-     &  -  &  -  &  -  &  -  \\
        \hline
    \end{tabular}
    \caption{Line-of-sight velocities of D2, D23, D3, and D3.1. For D2, D23, and D3.1 we can extract these values directly from the spectroscopic analysis of the SINFONI data cubes. The spectroscopic information for D3 in 2008 are extracted from \cite{Meyer2014}. We determine an uncertainty of 0.0005 $\mu$m. This results in a velocity of $\sim36$ km/s for the HeI and Br$\gamma$ emission line.}
    \label{tab:LOS}
\end{table*}
In the years between 2014 and 2018, it can be easily confused with D3 (see Tab. \ref{tab:k_band_pos}). Unlike D2 and D3, the projected trajectory of the object is directed towards west. As listed in Tab. \ref{tab:fig1data}, the measured Doppler-shifted Br$\gamma$ line map size is comparable to D2 and D3.

\subsubsection{D3}

Like D2, this source is part of the analysis in \cite{EckartAA2013} and \cite{Meyer2014}. D3 shows a blue-shifted Br$\gamma$ velocity of several hundred km/s (see Tab. \ref{tab:LOS}). These two Doppler-shifted Br$\gamma$ emission lines are shown in Fig. \ref{spectrum}. Like D2 and D23, the blue-shifted [FeIII] lines ${}^3 G_{5}\,-\,{}^3 H_{6} $ and ${}^3 G_{5}\,-\,{}^3 H_{5} $ can be detected (Tab. \ref{tab:felines}). However, the spectroscopic information of D3 is limited to a few years because of the chosen location of the SINFONI FOV. Because of the reduced S/N ratio at the border of the SINFONI data-cube, a detection of the [FeIII] ${}^3 G_{3}\,-\,{}^3 H_{4} $ and ${}^3 G_{4}\,-\,{}^3 H_{4} $ lines is not possible without confusion. 

\subsubsection{D3.1}
Additionally, we find a source close to D3 that we name for simplicity D3.1. The analysis of the L$'$-band NACO data shows, that D3.1 is moving in the same proper direction as the neighboring source D3. The spectroscopic information extracted from the SINFONI data cube of D3.1 shows a red-shifted Br$\gamma$ line with a velocity of around $50\,km s^{-1}$ close to the Br$\gamma$ rest wavelength in 2015. There are detectable traces of the [FeIII] multiplet but we can not rule out the possibility of confusing the source spectrum with other close by objects as D2, D23, and D3. However, the clear red-shifted Br$\gamma$ line map detection shows indications of an elongated source close to the mentioned objects (see Fig. \ref{finding_chart}). The HeI and Br$\gamma$ LOS velocity is listed in Tab. \ref{tab:LOS}. Because of the close distance of D2, D23, and D3 between 2014 and 2018, a detection in the NACO L$'$ and K-band images is not free of confusion.

\subsubsection{X7}

The blue-shifted X7 \citep[see][for more information]{muzic2007,muzic2010} and X8 \citep{Peissker2019} source in the GC are spatially close and can be characterized as (slightly) elongated sources that point towards Sgr A*. However, X7 is prominent in the L-band and can be identified in every available NACO data-set. Like X8, the position angle (North-East) of X7 is around $\sim\,45\deg$ and pointing towards Sgr A*. The object contributes to the idea of the existence of an ionizing wind coming from Sgr A* \citep[consider][]{Lutz1993, muzic2010, Yusef-Zadeh2012}. Also we find blue-shifted HeI and Br$\gamma$ lines with a velocity of around $600\,km\,s^{-1}$. The prominent [FeIII] multiplet lines ${}^3 G_{5}\,-\,{}^3 H_{6} $ and ${}^3 G_{5}\,-\,{}^3 H_{5}$ are shown in Fig. \ref{spectrum} with a similar blue-shifted velocity of around $500\,km\,s^{-1}$.

\subsubsection{X7.1}

We additionally identify another source north of X7 with a projected distance of around 50.0 mas - 100.0 mas that we name X7.1. Because of the S27 NACO camera pixel scale, an confusion-free L$'$-band identification can not be guaranteed. This red-shifted source is also close to the projected position of X7. It is furthermore located at the border of the SINFONI data cube where we expect a non-linear behavior of the detector. However, an individual identification of this source is presented in Fig. \ref{finding_chart}. The related spectrum can be found in Fig. \ref{spectrum}. It shows red-shifted HeI and Br$\gamma$ as well as [FeIII] ${}^3 G_{5}\,-\,{}^3 H_{6} $ and ${}^3 G_{5}\,-\,{}^3 H_{5}$ lines. Since the object is close to the FOV border, measuring the size is not satisfying. The LOS velocity of the ${}^3 G_{5}\,-\,{}^3 H_{6} $ and Br$\gamma$ line is around $380\,km\,s^{-1}$ (see Tab. \ref{tab:fig1data}, \ref{tab:fig16ata}).

\subsubsection{X8}

As discussed in \cite{Peissker2019}, X8 is a slightly elongated Doppler-shifted source that shows signs of bipolar morphology. Compared with X7, both sources share similar projected positions. We derive also, as mentioned before, a comparable position angle (North-East) for X8 of $\sim\,45\deg$. The analyzed K-band spectrum shows Doppler-shifted HeI and Br$\gamma$ lines. Like for X7, the [FeIII] multiplet is detected at $2.1441\,\mu m$ (${}^3 G_{3}\,-\,{}^3 H_{4} $), $2.2169\,\mu m$ (${}^3 G_{5}\,-\,{}^3 H_{6} $), $2.2410\,\mu m$ (${}^3 G_{4}\,-\,{}^3 H_{4} $), and $2.3469\,\mu m$ (${}^3 G_{5}\,-\,{}^3 H_{5} $) in 2015. The detected velocity for the ${}^3 G_{5}\,-\,{}^3 H_{6} $ [FeIII] line is around $120\,km\,s^{-1}$ whilst the blue-shifted Br$\gamma$ line is at $220\,km\,s^{-1}$ (see Tab. \ref{tab:fig1data}, \ref{tab:fig16ata}).

\subsection{Sources east of Sgr A*}
In this subsection, we show the results of the analysis of sources that are located east of Sgr A*.

\subsubsection{G1}
The spectroscopic analysis of G1 delivers a LOS velocity of around $1200\,km\,s^{-1}$ at a blue-shifted Br$\gamma$ emission line of $2.1575\mu m$ in 2008 which is consistent with \cite{Pfuhl2015} and \cite{Witzel2017}. In their work, the authors trace G1 in the K- and L-band. We find the source in the blue-shifted SINFONI line maps in a distance of around 170 mas to Sgr A* in 2008. The compactness is indicated by the measured size of the object (Fig. \ref{tab:fig1data}). Doppler-shifted [FeIII] lines are not detected.

\subsubsection{DSO/G2}
\begin{figure*}[htbp!]
\centering
\includegraphics[width=1.\textwidth]{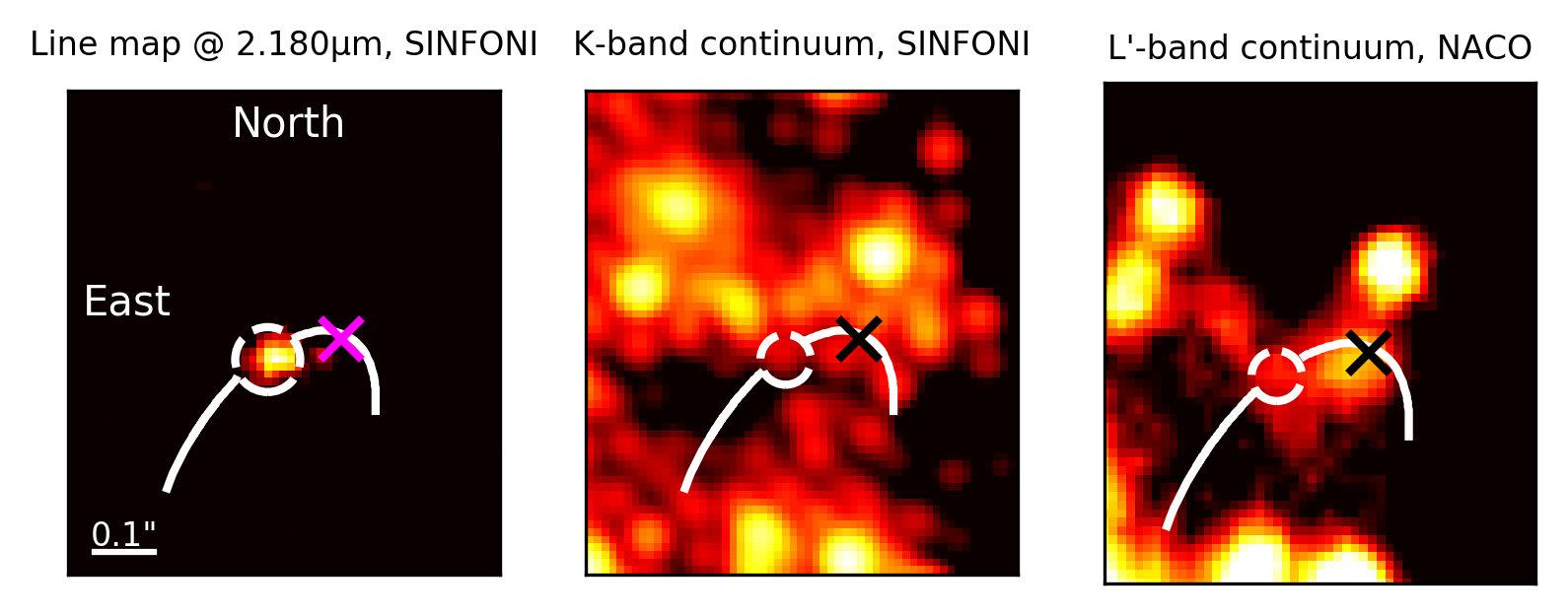}
\caption{Galactic center observed in the L- and H+K band with NACO and SINFONI in 2012. North is up, east is to the left. Here, the position of the DSO/G2 is marked with a dashed circle. Sgr A* is located at the position of the 'x'. The Keplerian orbital fit is based on the Br$\gamma$ line maps of the DSO/G2 (left image) between 2005 and 2016. The K-band continuum image is a background-subtracted high-pass filtered image with 10000 iterations. For the de/convolving process, a Gaussian-shaped APSF is used with an FWHM of 4.7px. The right image is a cutout of a NACO L-band mosaic observed in 2012 with a pixel scale of 27mas/px. The total integration of the NACO image is $\sim\,450$s, the on-source integration time of the SINFONI cube is $\sim\,610$min.}
\label{fig:naco_sinfo_dso}
\end{figure*}

\begin{figure*}[htbp!]
\centering
\includegraphics[width=0.8\textwidth]{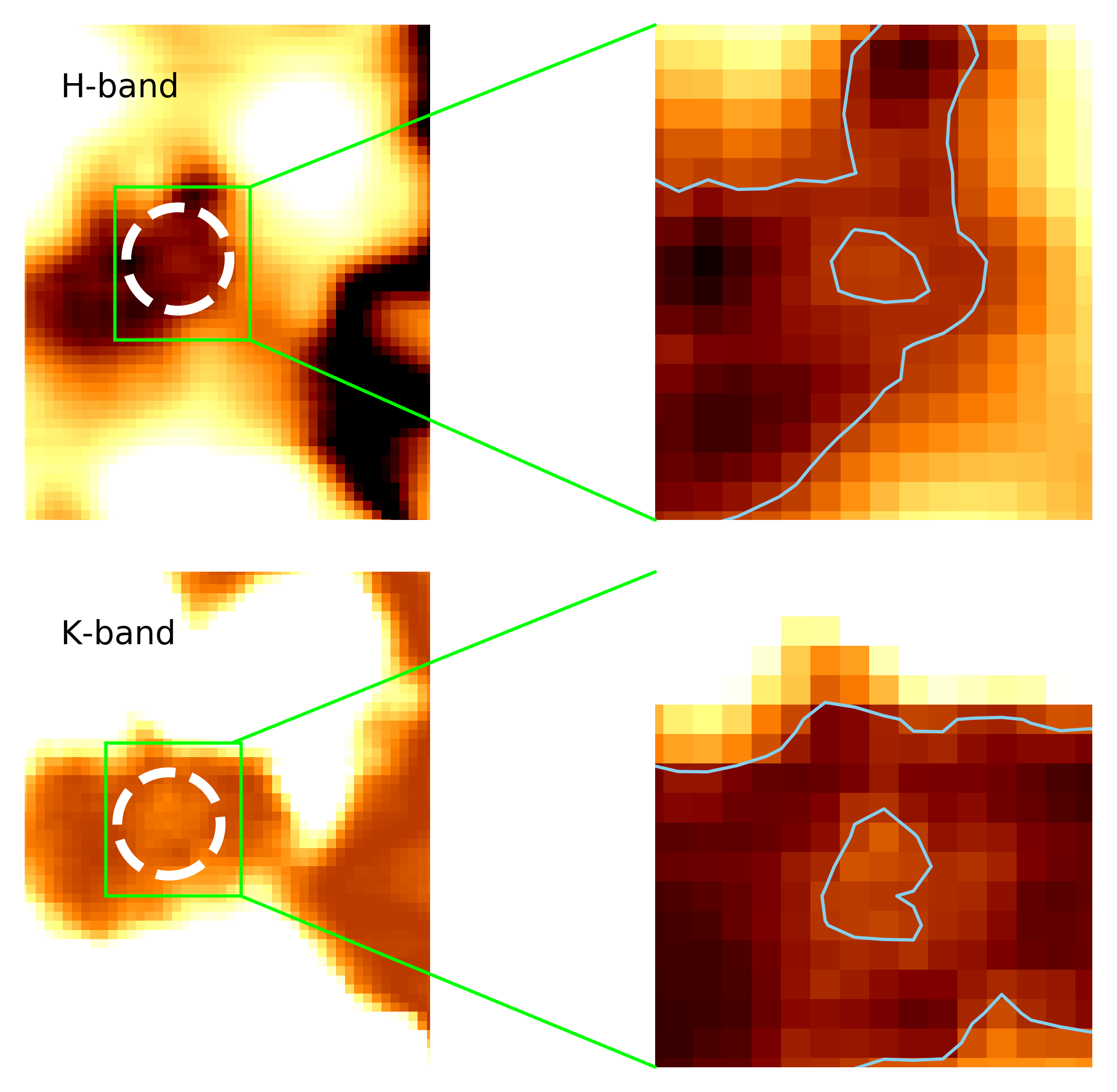}
\caption{H- and K-band detection of the DSO/G2 object. Here, we stack the continuum images that are used for the Lucy-Richardson detection of the DSO/G2 (see Appendix F, Fig. \ref{fig:hk_dso}). We apply for the K- and H-band detection the shifting vector from the Br$\gamma$ detection. The stacked continuum emission in K- and H-band is at the expected position considering the Br$\gamma$ and the Lucy-Richardson deconvolved images. In every image, north is up and east is to the left. The two overview figures to the left are $0".56\,\times\,0".68$. The two zoomed-in images to the right are $0".18\,\times\,0".21$. The contour levels are at 80$\%$ of the peak intensity of the object at the expected position of the DSO/G2 (right image).}
\label{fig:stackedKandHband}
\end{figure*}

The DSO/G2 is the focal point of many observations \citep[][]{Gillessen2012, Witzel2014, Valencia-S.2015, Pfuhl2015, Gillessen2019} and has a confirmed L$'$ counterpart that is observed with the NACO S27 camera in imaging mode.
Based on the H+K-band data-cubes, we find an H-, K-, and L-band continuum counterpart of the DSO/G2 that is presented in Fig. \ref{fig:naco_sinfo_dso} (middle figure), Fig. \ref{fig:stackedKandHband}, and Fig. \ref{fig:hk_dso} taken from Pei\ss{}ker et al. (in prep.). The position of the source is consistent with the line map detection and the L-band observation of the GC.
In the presented Br$\gamma$ channel maps as well as the H- and K-band detection, the emission that is associated with the DSO/G2 source indicates a rather compact origin (see also Fig. \ref{tab:fig1data}). 
In Fig. \ref{fig:naco_sinfo_dso}, we show that the L$'$-band detection is consistently compact. Besides the prominent Doppler-shifted Br$\gamma$ and HeI emission line detection in the K-band, we can not confirm additional features in that wavelength domain that could be connected to the DSO/G2 source. We also present stacked K- and H-band continuum images extracted from our data-cubes (Fig. \ref{fig:stackedKandHband}). The same continuum images are used for the Lucy-Richardson deconvolution that is presented in Fig. \ref{fig:hk_dso}. 

\subsubsection{D5}
The dusty source D5 is presented in \cite{EckartAA2013} but also mentioned in \cite{Meyer2014}. In Fig. \ref{spectrum}, we present a spectrum of this source extracted from the 2008 SINFONI data cube. Like D2 and D3, the proper motion is directed towards North-East in the projected direction of IRS16 (Fig. \ref{finding_chart}). Because of the on Sgr A* centered FOV of around $1".0$ and the trajectory of D5, we can not identify the source after 2009 in the SINFONI data sets. The LOS velocity of D5 can be determined to $60\,km\,s^{-1}$ with a red-shifted HeI line of $2.0590\,\mu m$ in 2008. However, the Doppler-shifted Br$\gamma$ line at $2.1676\,\mu m$ can be translated to a LOS velocity of $210\,km\,s^{-1}$. D5 is not showing signs of [FeIII] emission lines which is consistent with the spectral analysis presented in \cite{Meyer2014}.

\subsubsection{D9}
The source D9 is newly discovered and can be found North of S2 (see Fig. \ref{finding_chart_naco} for the position of the B2V star). Because of the extended wings of the star, the continuum information of D9 is difficult to obtain. However, we are able to trace the blue-shifted Br$\gamma$ and HeI line in the years 2008, 2010, and 2012 (see Fig. \ref{finding_chart}, \ref{spectrum}). We find an almost constant velocity of $150\,km\,s^{-1}$ with a related blue-shifted Br$\gamma$ emission line at $2.1649\mu m$ (Fig. \ref{finding_chart}). The blue-shifted HeI line of D9 at $2.05671\mu m$ is equivalent to $145\,km\,s^{-1}$ and matches the Br$\gamma$ line velocity.
Like G1, DSO/G2, and D5, D9 is not showing traces of the [FeIII] multiplet.
\begin{table*}[htbp!]
    \centering
    
    \begin{tabular}{ccccccc}
            \hline
            \hline
            & \multicolumn{2}{c}{D2} & \multicolumn{2}{c}{D23} & \multicolumn{2}{c}{D3}\\
            emission line [$\mu$m] & RV [km/s] & $\sigma$ [km/s] & RV [km/s] & $\sigma$ [km/s] & RV [km/s] & $\sigma$ [km/s]\\
            \hline
            2.145 & -227 & 24 & -248 & 10 & - & - \\
            2.218 & -264 & 21 & -251 & 9  & -676 & 14 \\
            2.243 & -357 & 34 & -393 & 17 & - & - \\
            2.348 & -265 & 28 & -280 & 10 & -511 & 63 \\
            \hline
    \end{tabular}
    \caption{Line-of-sight velocities based on the Doppler shifted [Fe III] line emission for D2, D23, and D3 in 2015. The $\sigma$ is the standard diviation. The ${}^3 G\,-\,{}^3 H$ multiplet is represented by the listed [Fe III] rest emission lines.}
    \label{tab:Fe_rv}
\end{table*}

\begin{figure*}[htbp!]
%\begin{sidewaysfigure*}[htbp!]
\centering
\includegraphics[width=1.\textwidth]{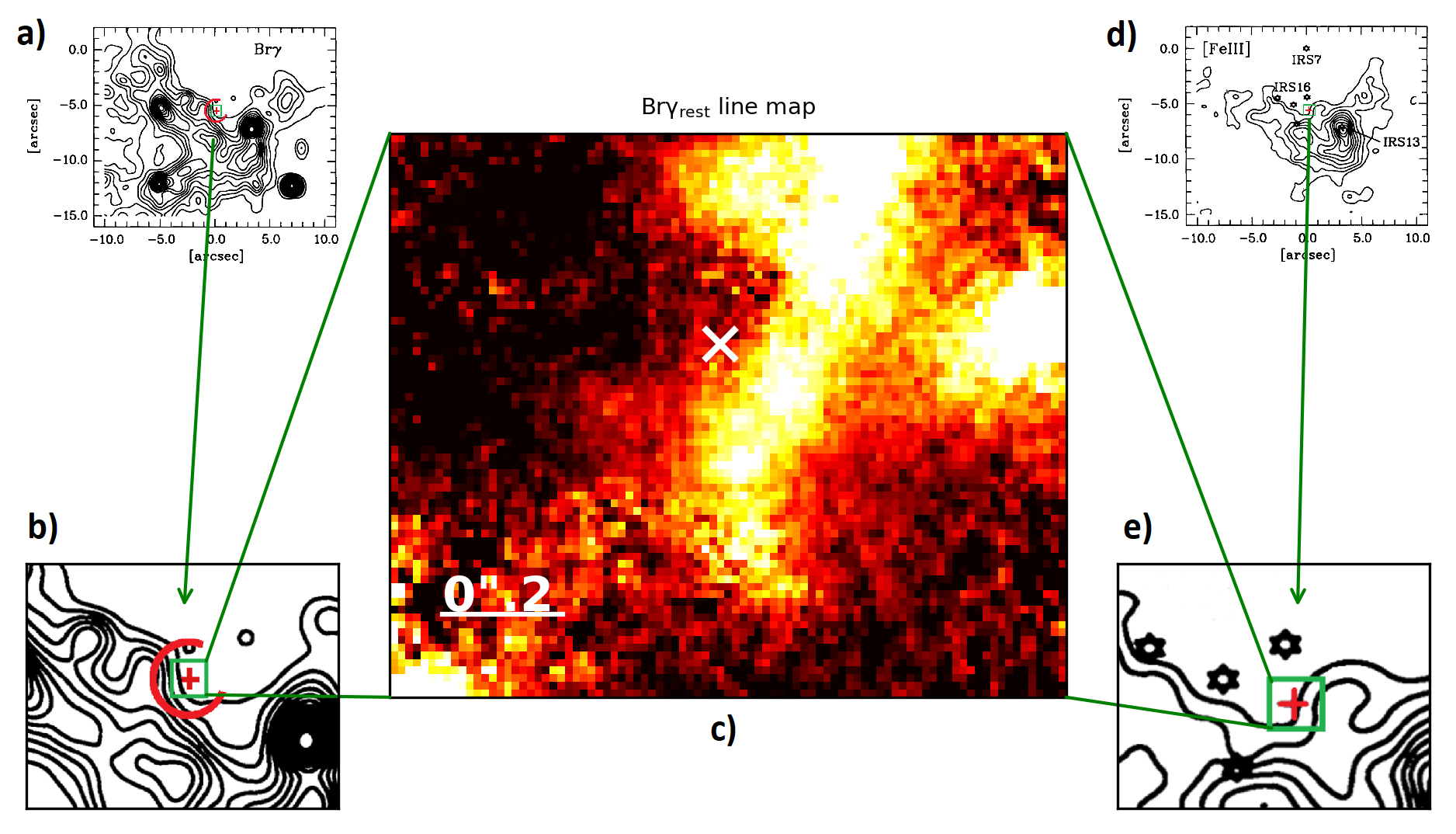}
\caption{[Fe III] and Br$\gamma$ line distribution in the Galactic center around Sgr A*. In every image, north is up and east to the left. The scaling of the figures is either mentioned or match the marked squares.
\newline
{\it Figure a):} We adapt the Br${\gamma}$ detection presented in \cite{Lutz1993} in the GC around SgrA*. The green box marks the size of the SINFONI FOV. The red circle with the open side towards the west indicates the position of the SgrA* bubble that is described in detail in \cite{Yusef-Zadeh2012}.
\newline
{\it Figure b):} A zoomed-in figure of {\it a)}. SgrA* is marked with a red cross.
\newline
{\it Figure c):} The Br$\gamma$ line map at around $2.1661\,\mu m$ extracted from our SINFONI data-cube in 2015. The bright north-south Br$\gamma$-bar can be detected in every data-cube and is approximately at the position of the open side of the SgrA* bubble.   
\newline
{\it Figure d):} [FeIII] distribution around SgrA*. The [FeIII] line-map emission of the dusty sources around SgrA* is shown in Fig. \ref{fig:feIIIdistriGC}.
\newline
{\it Figure e):} A zoomed-in cutout of d). The edge of the [FeIII] distribution coincides with the position of the detected Br$\gamma$-bar shown in c). From a) and b) we can conclude, that the [FeIII] emission is significantly decreased inside the SgrA* bubble. 
\newline
%\textbf{The upper line map shows the Br$\gamma$ emission at $2.1661\,\mu$m in the direct vicinity of SgrA*. The bright Br$\gamma$-bar can be detected in every SINFONI here investigated data-set. The contours are adapted from Fig. \ref{finding_chart} and show the Doppler-shifted Br$\gamma$ emission of the sources discussed in this work. The same contours are also included in the lower [FeIII] line map where we present the Doppler-shifted [FeIII] emission of the same sources. Additionally, the green vertical contour line marks the rough position of the boarder of the dense S-star cluster. In the lower line map, we include the emission of all sources that show Dopper-shifted [FeIII] lines. Coincidentally, they are all located west of SgrA*. %It could be speculated if the Br$\gamma$-bar or the low number of observed S-stars in that region leads to the emission of Dopper-shifted [FeIII] lines. 
The related velocities, uncertainties, and widths of the [FeIII] line emission are given in Tab. \ref{tab:fig16ata}. The here described phenomena are in line with \cite{Yusef-Zadeh2012} (see also Fig. 5 in the mentioned publication and Appendix E, Fig. \ref{fig:RadioNIRbubble}).}
\label{fig:fe_line}
%\end{sidewaysfigure*}
\end{figure*}
\begin{table*}[htbp!]
\tabcolsep=0.1cm
\centering
\begin{tabular}{|c|cc|cc|ccc|}
\hline
\hline
 [FeIII] line map properties & X7.1 & D3.1 & X7 & X8 & D2 & D23 & D3 \\
\hline          
Velocity in km/s & +383 & -550 & -488 & -122 & -264 & -251 & -676\\
\hline
FWHM in km/s & 216 & 216 & 162 & 148 & 175 & 94 & 110 \\
\hline 
Uncertainty in km/s & 33 & 33 & 20 & 20 & 21 & 9 & 14\\
\hline
\end{tabular}
\caption{[FeIII] line map properties related to Fig. \ref{fig:fe_line}. The blue- and red-shifted line maps are based on the ${}^3 G_{5}\,-\,{}^3 H_{6}$ [FeIII] emission line at $2.2178\mu m$.}
\label{tab:fig16ata}
\end{table*}
%X7.1: v1=2.2205 v2=2.2209 v3=2.2211 ; v_low= v_high=   
%v1 = 338.14 v2 = 392.24 v3 = 419.29  
%D3.1: v1=2.2136 v2=2.2140 v3=2.2142 ; v_low=2.2131 v_high=2.2147   
%v1 = 595.13 v2 = 541.02 v3 = 513.97  error

\subsection{Keplerian orbital fit}

Since D2 and D3 are the most prominent and bright dusty sources besides X7 in the vicinity around Sgr A*, we emphasize the analysis of these two objects by fitting a Keplerian orbit to the extracted data points (Tab. \ref{tab:k_band_pos} and Tab. \ref{tab:L-band-pos}). Nevertheless, during the analysis of D2 and D3, we collect additional information about the two close-by sources D23 and D3.1 (see Fig. \ref{D23D31_orbit} in the appendix).
\begin{figure*}[htbp!]
\centering
\includegraphics[width=1.\textwidth]{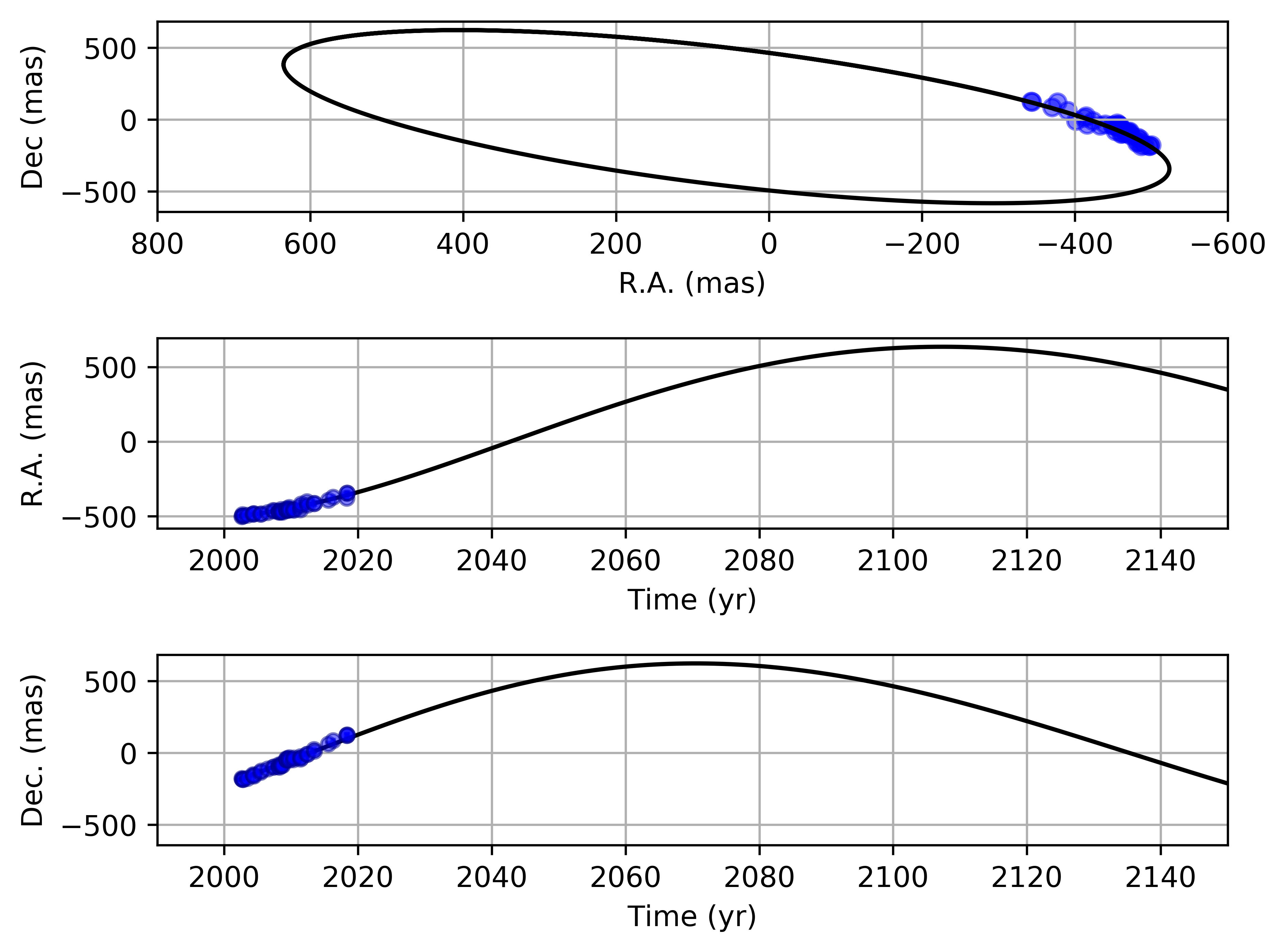}
\caption{Orbit of D2. The here used data is extracted from the spectrum extracted from the SINFONI data cube between 2002 and 2018. To enhance the accuracy of the Keplerian fit, we add the NACO L$'$- and K-band positions of D2.}
\label{D2_orbit}
\end{figure*}
\begin{figure*}[htbp!]
\centering
\includegraphics[width=1.\textwidth]{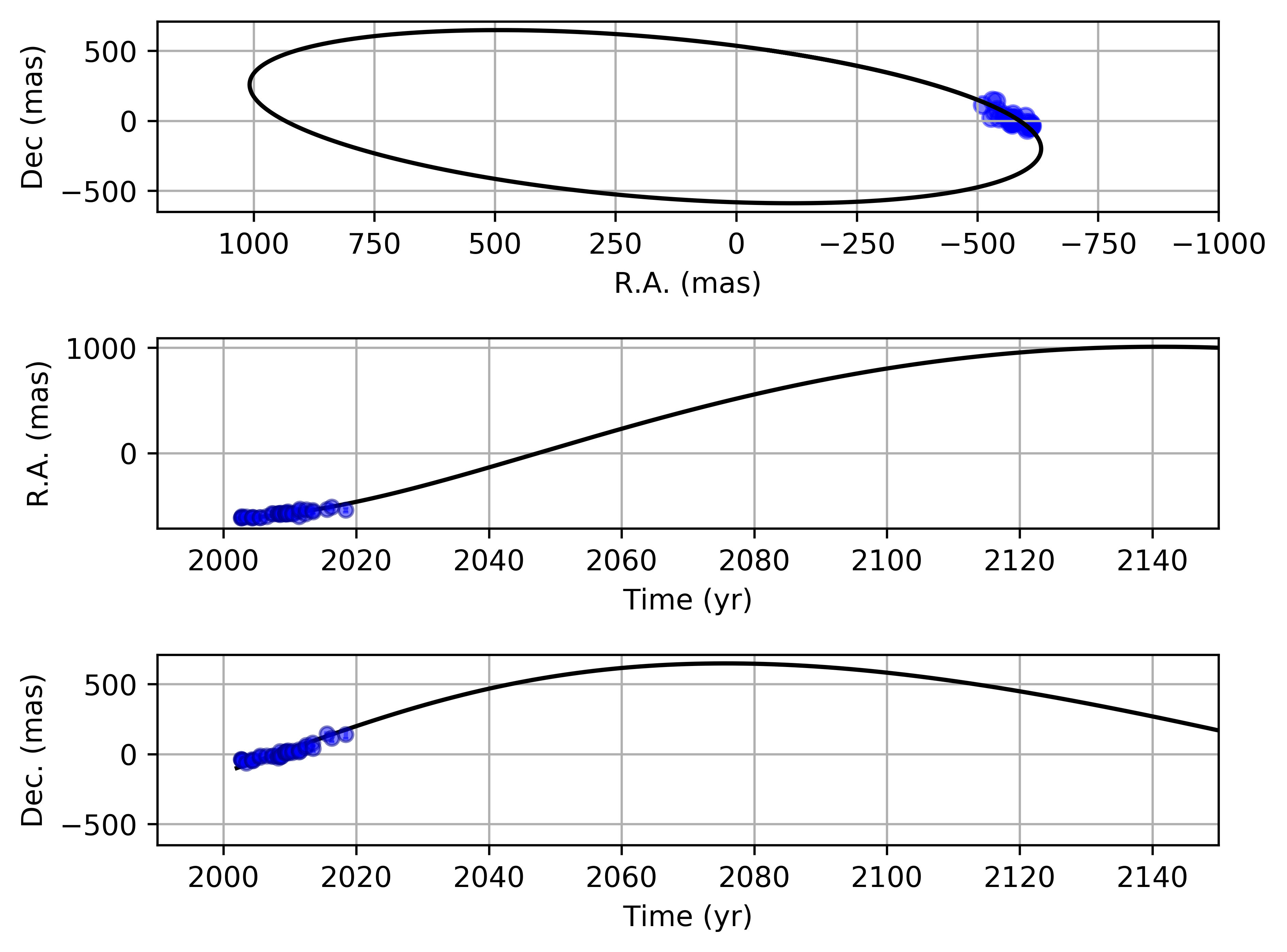}
\caption{Orbit of D3. This dusty source is part of the "D-complex" located west of Sgr A*. The used data is based on the NACO L$'$- and K-band identification of D2. The Doppler-shifted Br$\gamma$ and HeI LOS velocity is adapted from the spectrum presented in \cite{Meyer2014} and the spectroscopic analysis of the SINFONI data cube of 2015.}
\label{D3_orbit}
\end{figure*}
\begin{figure*}[htbp!]
\centering
\includegraphics[width=1.\textwidth]{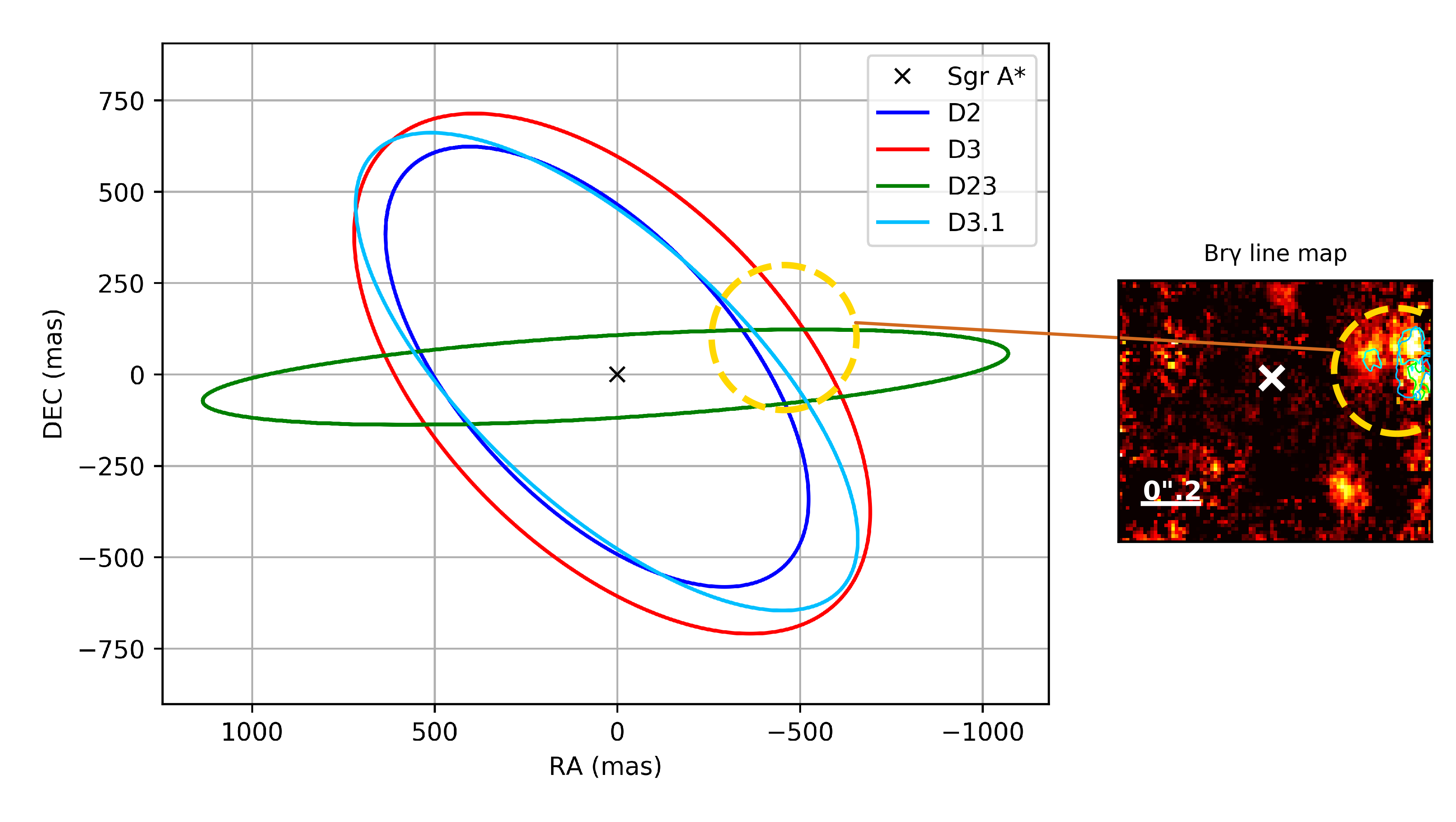}
\caption{Orbit of D2, D23, D3, and D3.1. The Keplerian fit is based on the data presented in Tab. \ref{tab:k_band_pos} and \ref{tab:L-band-pos}. The dashed circle indicates the investigated area in the SINFONI data cubes. The Doppler-shifted Br$\gamma$ line map shows the emission of D2, D23, and D3 (underlined by 50$\%$ peak intensity contour lines). In the line map, north is up, east to the left. We do not show the line map of D3.1 since it would dominate the emission of the other sources (for that, consider Fig. \ref{finding_chart}).}
\label{D2D23D3D31_orbit}
\end{figure*}
\begin{table*}[htbp!]
    \centering
    \begin{tabular}{ccccccc}
            \hline
            \hline
            Source & a [mpc] & e & i [$^\circ$] & $\Omega$ [$^\circ$]& $\omega$ [$^\circ$] & $t_{\rm closest}$ [years] \\
            \hline
            %D2 [3.00154241e+01 1.47113518e-01 1.00982203e+00 3.86468232e+00 7.65977818e-01 2.00216343e+03]  
            %      [0.03861378 0.02127108 0.02704456 0.00982139 0.02012749 0.02618904] 30.10.2019
            D2   & 30.01 $\pm$ 0.04 &  0.15 $\pm$ 0.02 & 57.86 $\pm$ 1.71 & 221.16 $\pm$ 1.43 & 44.11 $\pm$ 1.14 & 2002.16 $\pm$ 0.03 \\
            %D23  [4.42461751e+01 5.91986291e-02 1.67362479e+00 2.11377896e+00 1.62956035e+00 2.01599576e+03]
            %    [0.03366952 0.01153553 0.01572733 0.00868653 0.00555885 0.03382051] 30.10.2019
            D23  & 44.24 $\pm$ 0.03 &  0.06 $\pm$ 0.01 & 95.68 $\pm$ 1.14 & 120.89 $\pm$ 3.15 & 93.39 $\pm$ 1.89 & 2016.00 $\pm$ 0.03 \\
            %D3   [3.52003726e+01 2.40846133e-01 8.96200734e-01 3.60374156e+00 1.11653058e+00 2.00223780e+03]
            %    [0.01014247 0.01646685 0.0213452  0.03083599 0.03978047 0.00423226] 30.10.2019
            D3   & 35.20 $\pm$ 0.01&  0.24 $\pm$ 0.02 & 50.99 $\pm$ 1.20 & 206.26 $\pm$ 1.76 & 63.59 $\pm$ 2.29 & 2002.23 $\pm$ 0.01\\
            %D3.1[3.14687686e+01 5.89279844e-02 1.10296183e+00 8.36217538e-01 4.04003977e+00 2.00437844e+03]
            %   [0.00486478 0.0223755  0.01826476 0.01698417 0.02075249 0.00785427] 30.10.2019
            D3.1 & 31.46 $\pm$ 0.01&  0.06 $\pm$ 0.02 & 63.02 $\pm$ 1.04 & 47.89 $\pm$ 0.96 & 231.47 $\pm$ 1.18 & 2004.37 $\pm$ 0.008 \\
            \hline
    \end{tabular}
    \caption{Orbital parameters for the dusty sources D2, D23, D3, and D3.1. The given standard deviation is based on the spatial and detector uncertainties.}
    \label{tab:orbital_para}
\end{table*}
We find that the orbits for at least D2, D3, and D3.1 exhibit a similar orbital shape (see Fig. \ref{D2D23D3D31_orbit}). In contrast, the orientation of the D23 orbit is almost perpendicular to the other members of the "D-complex" and points towards IRS13N. It is also moving clockwise on its projected orbit whilst the other sources are moving counter-clockwise. Because of the distance, a connection between the IRS13 cluster and D23 can be excluded. It is evident that the absolute value of the eccentricity (Tab. \ref{tab:orbital_para}) and the LOS velocity (Tab. \ref{tab:LOS}) for D2, D3 and D3.1 are comparable. %In the next subsection, we will discuss a possible origin of these co-moving objects.

The derived errors (standard deviation) and the orbital parameters (mean values) in Tab. \ref{tab:orbital_para} are based on a variation of the positional data in R.A. and DEC. as well as the FWHM of the resolution of the velocity. In this way, we are including all known and unknown uncertainties.% The resulting orbital parameters define the given error range.}

\subsection{Abundance/Ionization parameter}
\label{sec.:metallicity}
\begin{figure*}[htbp!]
\centering
\includegraphics[width=1.\textwidth]{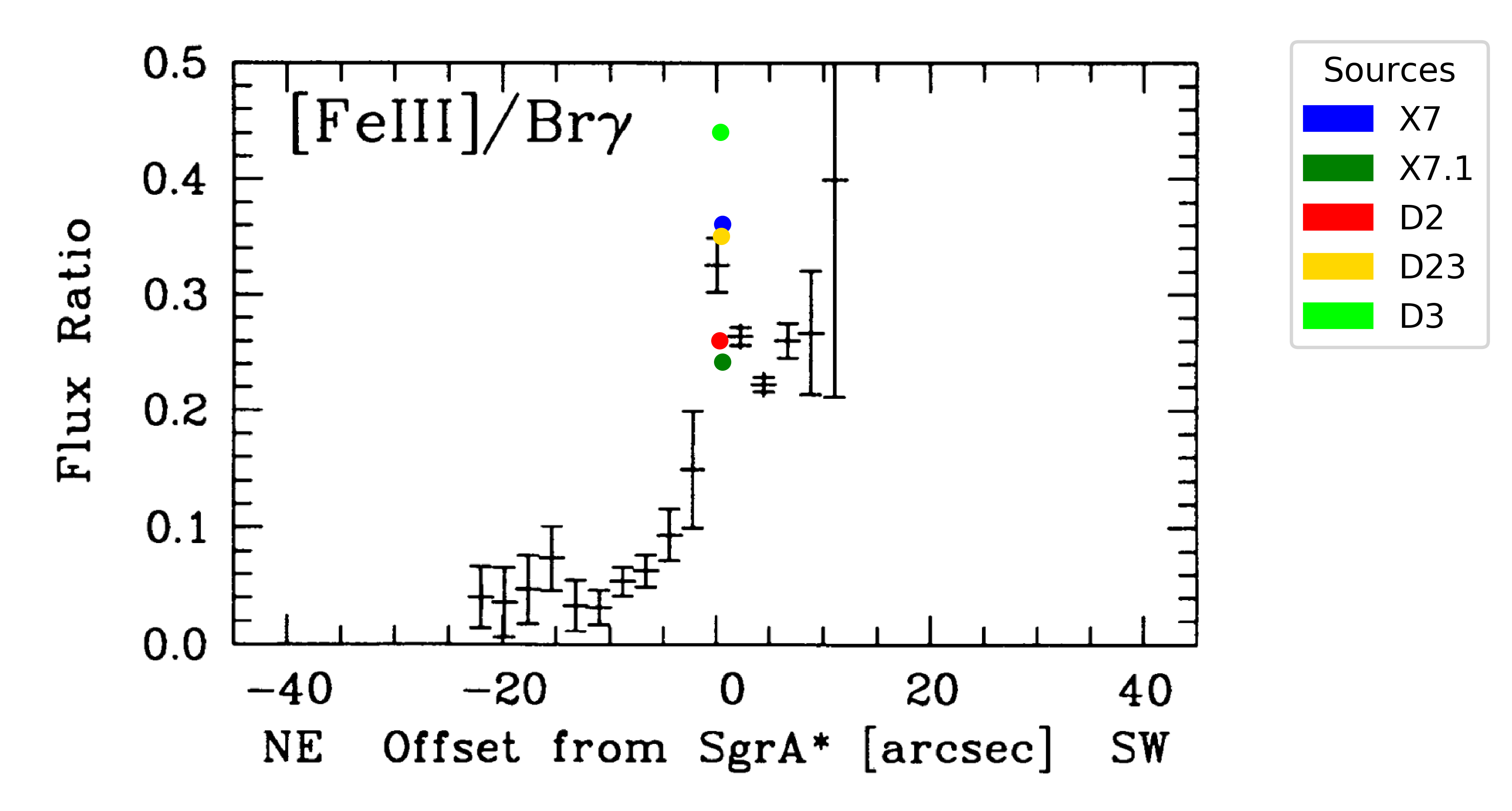}
\caption{The [Fe III]/Br$\gamma$ flux ratio as a function of the Sgr A* offset. The plot is adapted from \cite{Lutz1993}, the colored data points are extracted from the SINFONI data cube of 2015. We do not include X8 in this plot because its flux ratio is 0.67 and therefore outside the bonds of the adapted figure.}
\label{fig:flux_ratio}
\end{figure*}

We use the [Fe III]$\mathbin{/}Br{\gamma}$ line ratio as a combined abundance/ionization parameter (AIP) for the detected dusty sources that are located west of the Br$\gamma$ feature (Fig. \ref{fig:fe_line}). The orbital parameters of our estimated Keplerian orbit show, that D2, D3, and D3.1 have similar projected trajectories (see Fig. \ref{D2D23D3D31_orbit}). Nonetheless, we find at least for X7, X8, D2, D23, and D3 Doppler shifted [Fe III] lines (see Fig. \ref{spectrum}, \cite{Peissker2019}, and \citealp{Meyer2014}). For D3.1, we detect weak traces of the iron emission lines with a ratio of about $D23_{[Fe\,III]}$ / $D3.1_{[Fe\,III]} \,\sim$ 4.
\begin{table}[htbp!]
    \centering
    \begin{tabular}{cc}
        \hline
        \hline
        Source & AIP in [$\%$]\\
        \hline
        X7   & 36 \\
        X7.1 & 24 \\
        X8   & 67  \\
        D2   & 26 \\
        D23  & 35  \\
        D3   & 44  \\
        \hline
    \end{tabular}
    \caption{Abundance/Ionization parameter based on the [Fe III]$\mathbin{/}Br{\gamma}$ integrated line ratio in 2015.}
    \label{tab:metallicity}
\end{table}
Table \ref{tab:metallicity} shows the AIP ratio based on the [Fe III]$\mathbin{/}Br{\gamma}$ flux measurements. We find AIP values for D2, D23, D3, X7, X7.1, and X8 between 24$\%$ and 67$\%$. In this sample, X7.1 shows the lowest amount of metal abundance whilst the spectrum of X8 contains the strongest [Fe III] lines compared to its Br$\gamma$ emission (see Fig. \ref{fig:metallicity} in the appendix). Depending on the spectrum, we are using either the ${}^3 G_{5}\,-\,{}^3 H_{6} $ or ${}^3 G_{5}\,-\,{}^3 H_{5} $ transition of the Doppler shifted [Fe III] line (Fig. \ref{fig:metallicity}, appendix). By using the [Fe III]$\mathbin{/}Br{\gamma}$ ratio as a function of the Sgr A* offset plot from \cite{Lutz1993} and including the values for X7, X7.1, D2, D23, and D3, we find an increased AIP towards the Galactic center (see Fig. \ref{fig:flux_ratio}). 

\subsection{Magnitudes of dusty sources}

For the photometric analysis of D2, D23, and D3 in the NACO and SINFONI data between 2002 and 2018, we are using the calibrator stars S2, S7, and S10 (see Tab. \ref{tab:calib_stars}). With the method described in Sec. \ref{Analysis}, we derive a K$_S$-magnitude for the bright sources of the D-complex. For the H-band magnitude, we can estimate an upper limit. For D2, D23, and D3 sources, we find values of mag$_{H_{up}} = 22.98$, mag$_{H_{up}} = 22.22$, and mag$_{H_{up}} = 21.3$ for H-band upper limits, K-band magnitudes of  mag$_{K} = 18.1$, mag$_{K}=19.7$, and mag$_{K} = 19.8$, and L$'$-band magnitudes of mag$_{L'} = 13.5$, mag$_{L'} = 15.13$, and mag$_{L'}=13.8$, respectively.
Because of the limited SINFONI data sets that show D3 and the chance of interference effects for the flux measurements by the close source D3.1, we are adapting the H (upper limit), K, and L$'$-band magnitude values presented in \citet{EckartAA2013}. The magnitudes for D2, D23, and D3 can be found in Tab. \ref{tab:mag_D2_D3}. The L$'$-band magnitude of D23 is extrapolated from the D2 and D3 values by using the counts of the source. For that, we derive the ratio of the peak intensity from the emission area of D2 and D23. The resulting peak-count ratio can be used together with 
\begin{equation}
    \text{mag}_{L'} = -\text{mag}_{S2} + 2.5\times \log{(\text{ratio})}\,,
\end{equation}
where $mag_{S2}$ donates the extinction corrected S2 L$'$ magnitude of 11.5\footnote{Dereddened with A$_K$=2.7} \citealp{Clenet2001, Clenet2003} and 11.16 (Hosseini et al., in prep.; Dereddened with A$_K$=2.4). The L$'$ magnitude for D23 of mag$_{L'}$ = 15.14 is an average because of the slight variations of the S2 L$'$ magnitude.
\begin{table}[htbp!]
    \centering
    \begin{tabular}{cccc}
        \hline
        \hline
        Source & mag$_H$ & mag$_K$ & mag$_{L'}$ \\ %& Radius in [AU] & Temperate in [K]\\
        \hline
        D2  & 22.98 & 18.1$^{+0.3}_{-0.8}$& 13.5  \\ %& 1.8 $\pm$ 0.8 & 494 $\pm$ 27\\
        D23 & 22.22 & 19.7$^{+0.4}_{-0.3}$& 15.13 \\ %& 0.38 $\pm$ 0.11 & 585 $\pm$ 33 \\
        D3  &  21.3 &  19.8               & 13.8  \\ %& 2.2 $\pm$ 0.8 & 444 $\pm$ 32\\
        \hline
    \end{tabular}
    \caption{Magnitudes of D2, D23 and D3 in the H-, K$_S$-, and L$'$-band in 2004 and 2015.}
    \label{tab:mag_D2_D3}
\end{table}
Using these values and applying a one-component, single-temperature blackbody model, we can reproduce the observed and detected emission (Fig. \ref{fig_SED_model}) of D2, D23, and D3 that are presented in the upcoming section. 

\subsection{SED modeling of dusty sources}
\label{sec:sed}
\begin{figure*}
    \centering
    \includegraphics[width=\textwidth]{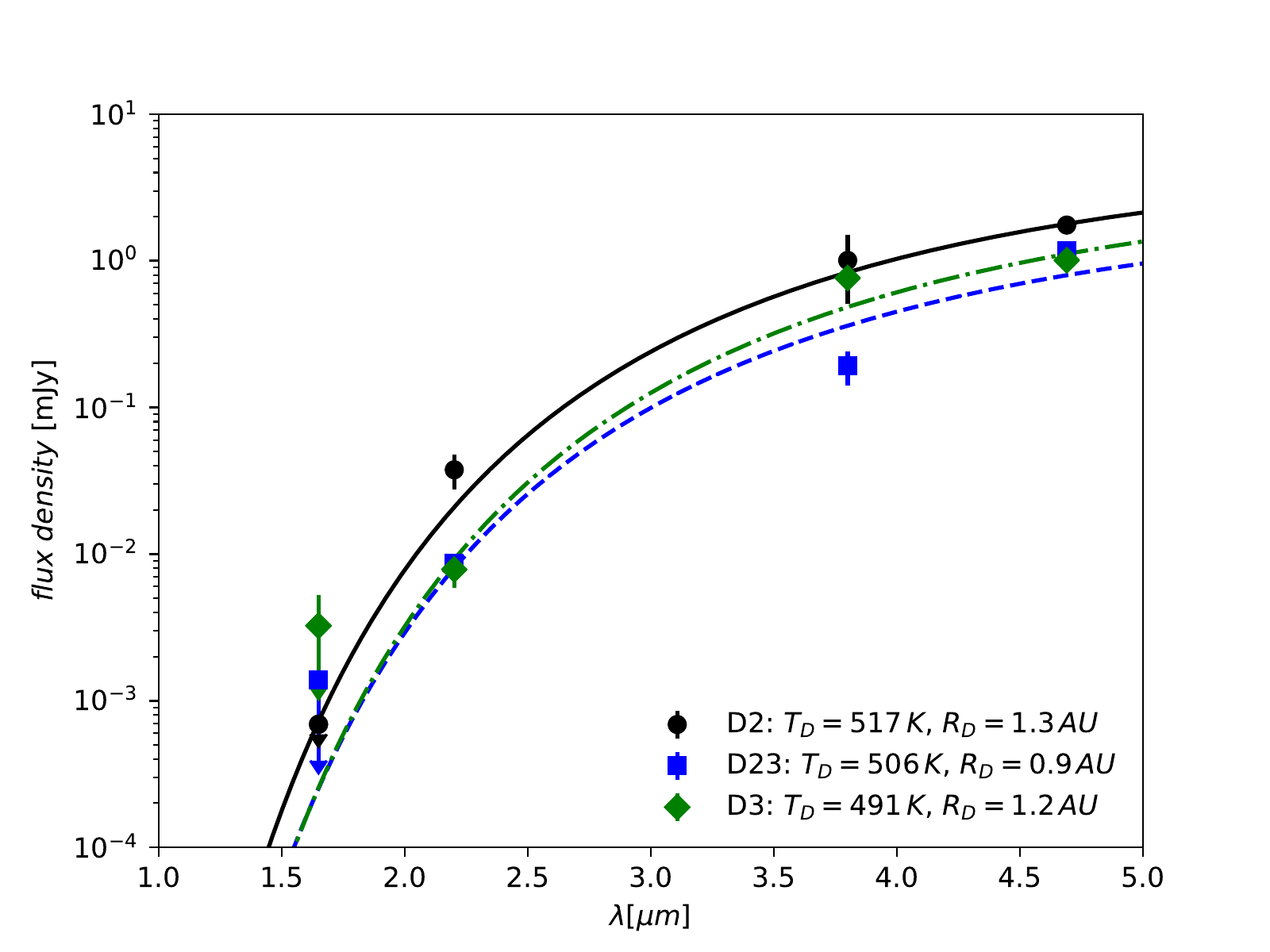}
    \caption{Single-temperature blackbody models of three dusty sources -- D2, D23, and D3 -- are represented by lines (black-solid, blue-dashed, green-dashdot, respectively) with the parameters according to the legend. The continuum flux densities are depicted by points according to the legend.}
    \label{fig_SED_model}
\end{figure*}

We use the flux densities of the studied dusty sources -- D2, D23, and D3 -- in H, K$_{\rm s}$, L$'$, and {\rm M bands} to construct their spectral energy distribution (SED) model. First, we fit a one-component, single-temperature blackbody model (BB) to their spectra using the flux-density prescription,

\begin{equation}
    S_{\nu}(R_{\rm D},T_{\rm D})=\left(R/d_{\rm GC}\right)^2 \pi B_{\nu}(T_{\rm D})\,,
    \label{eq_flux_blackbody}
\end{equation}

where R$_{\rm D}$ and T$_{\rm D}$ are the radius and the temperature of the optically thick photosphere of the dusty sources, respectively, and $B_{\nu}(T_{\rm D})$ is the Planck function. We set the Galactic centre distance to $d_{\rm GC}=8200\,{\rm pc}$, which is consistent with the recent result by the GRAVITY collaboration \citep{Gravity2019}. This value is close to the one inferred from S2 star observations with GRAVITY VLTI instrument in $K_{\rm s}$-band \citep{gravity2018}. The flux densities in mJy as well as the fitted models for the three dusty objects are depicted in Fig. \ref{fig_SED_model}. For all the three sources, the single-temperature BB model fits the continuum flux densities well, with the reduced $\chi^2_{\rm} r$ of $1.51$, $11.12$, and $5.76$ for D2, D23, and D3, respectively. The best-fit parameters along with their 1$\sigma$ uncertainties are listed in Tab. \ref{table_BB_parameters}. 

\begin{table*}[htbp!]
\centering
\begin{tabular}{|c|c|c|c|c|c|}
 \hline
 \hline
 Parameter & D2 & D23 & D3 & DSO/G2 (1C-fit) & DSO/G2 (2C-fit)\\
 \hline
 T$_{\rm D}\,[\rm K]$ & 517 $\pm$ 10  & 506 $\pm$ 44 & 491 $\pm$ 40 & 626 $\pm$ 87 & 514 $\pm$ 41\\
 R$_{\rm D}\, [\rm AU]$ & 1.26 $\pm$ 0.14  & 0.90 $\pm$ 0.39  & 1.17 $\pm$ 0.37 & 0.50 $\pm$ 0.31 & 0.97 $\pm$ 0.31 \\
slope KL$'$ (Lada YSO class) & 5.02 (I) & 4.69 (I)  & 7.38 (I) & 4.15 (I) & 4.15 (I) \\
 $\chi^2_{\rm r}$ & 1.51 & 11.12  & 5.76 & 6.20 & 1.00 \\
 \hline
\end{tabular}
\caption{Best-fit parameters $(T_{\rm D}, R_{\rm D})$ of the single-temperature BB model, including also the spectral slope (with the corresponding Lada YSO type), and the reduced $\chi^2$ of the BB model. The abbreviation 1C-fit stands for one component fit and 2C-fit represents two components -- star and dusty envelope.}
\label{table_BB_parameters}
\end{table*}

In general, the BB best-fit temperatures for all three sources are the same within the uncertainties $T_{\rm D} \sim 500\,{\rm K}$, which is most likely due to warm dust heated by the star from inside. The optically thick dust layer has the characteristic black-body length-scale of $R_{\rm D}\sim 1\,{\rm AU}$, which is also the same for all the objects within the uncertainties. 

\begin{figure*}
    \centering
    \includegraphics[width=0.48\textwidth]{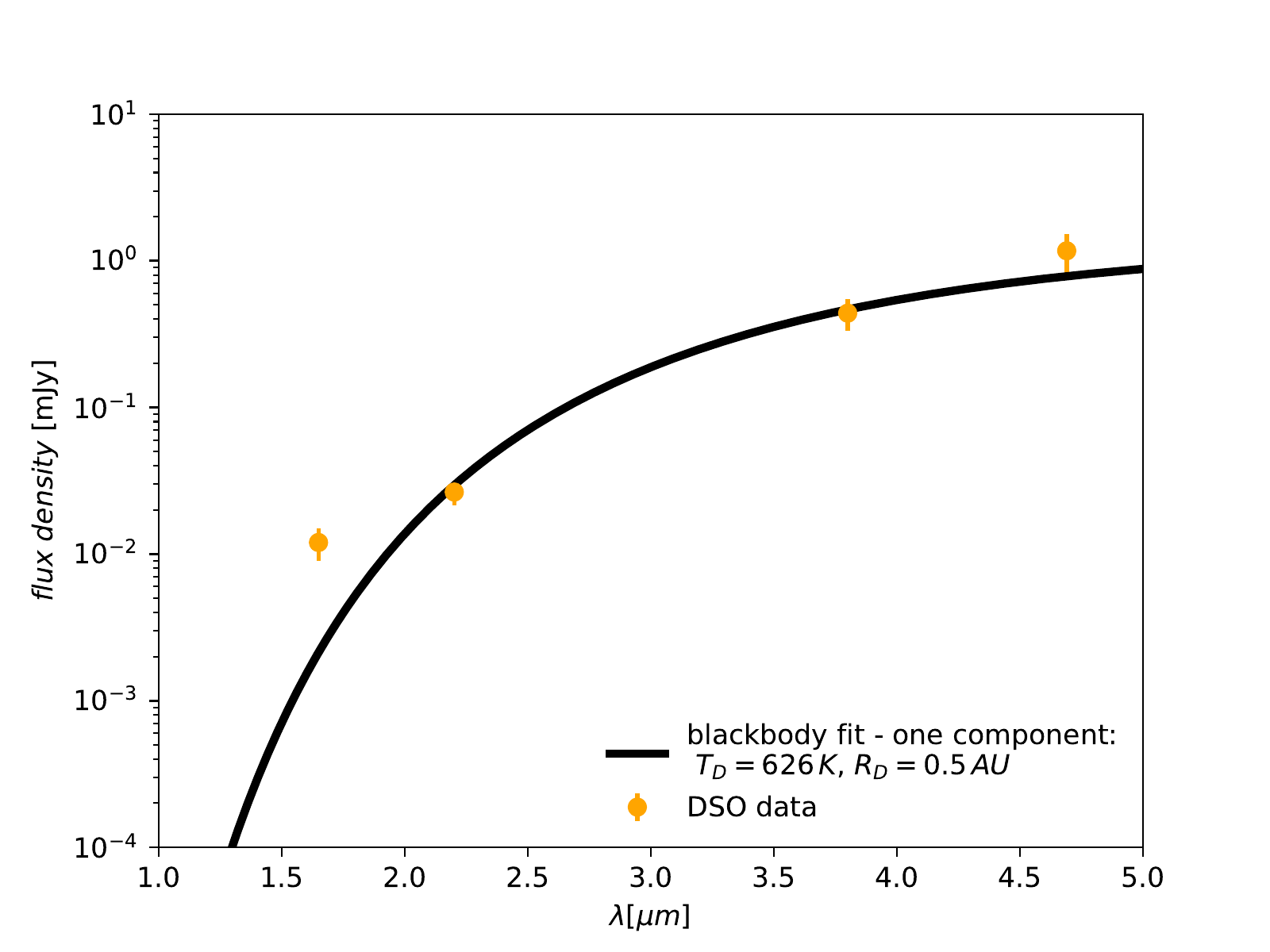}
    \includegraphics[width=0.48\textwidth]{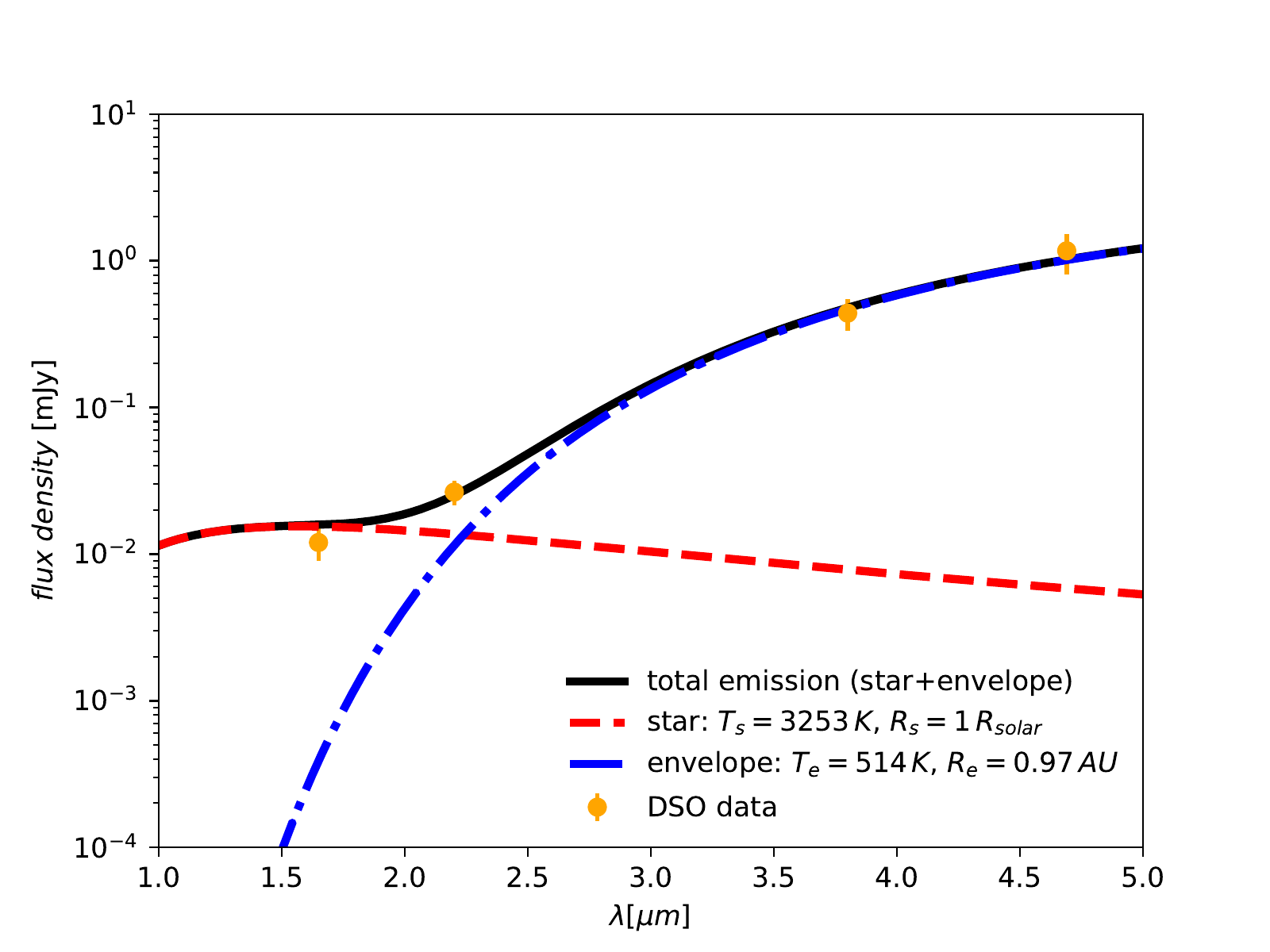}
    \caption{Blackbody SED model of the DSO/G2 infrared source. Left panel: One component blackbody fit with the parameters according to the legend. Measured flux densities are depicted as orange points with corresponding errorbars. Right panel: Two-component blackbody fit (star$+$ dusty envelope). The black solid represents the total emission, while the red dashed line stands for the stellar emission and the blue dot-dashed line represents the dusty envelope emission with the parameters according to the legend.}
    \label{fig_SED_model_DSO}
\end{figure*}

The temperatures of the investigated objects show a good agreement with the values for the DSO/G2 source (Fig. \ref{fig:naco_sinfo_dso}) derived in \cite{EckartAA2013} and \cite{Zajacek2017}. Here, we revisit the blackbody fit of the DSO, including the $H$-band detection for the first time, see Fig. \ref{fig_SED_model_DSO}. A single component blackbody fit yields the temperature of $T_{\rm DSO}=626\pm 87\,K$ and the radius of $R_{\rm DSO}=0.5\pm 0.3\,{\rm AU}$ with the reduced $\chi^2$ of $\chi^2_{\rm r}=6.2$, see also Table~\ref{table_BB_parameters} for the one component (1C) fit parameters. For this model fit, the $H$-band flux density is clearly offset and larger than the model-fit value. This motivates an improved, two-component model that consists of a star and its circumstellar envelope. We perform several trial fits, where we fix the SED of a star and search for the temperature and the radius of the surrounding circumstellar envelope. The best-fit with the reduced $\chi^2$ equal to unity is obtained for the star with the effective temperature of $3253\,{\rm K}$ and the stellar radius equal to one Solar radius. The corresponding best-fit envelope temperature is $T_{\rm D}=514\pm 41\,{\rm K}$ and its radius is $R_{\rm D}=0.97\pm 0.31\,{\rm AU}$. In summary, the two-component model including the star significantly improves the goodness of fit.

The standard classification of young stellar objects (YSOs) is based on the slope $\alpha$ of spectral energy distribution (SED) between $2.2\,{\rm \mu m}$ and $20\,{\rm \mu m}$ \citep{Lada1987}, using the prescription $\nu S_{\nu}\propto \lambda^{\alpha}$, which can be calculated between K and L$'$ bands as follows, 

\begin{equation}
    \alpha\equiv \frac{\log{(\nu_{L'}S_{L'}/\nu_{K}S_{K})}}{\log{(\lambda_{L'}/\lambda_{K})}}\,.
    \label{eq_slope_lada}
\end{equation}

Using Eq. \eqref{eq_slope_lada}, we calculate the spectral slope between K and L$'$ bands for the dusty sources, see Table \ref{table_BB_parameters}. According to \cite{Lada1987}, D2, D23, and D3 would be YSOs of type I, since their derived spectral slope is $\alpha\gg 0.3$. Thus, dusty sources studied in this paper could represent young stars embedded in optically thick dusty envelopes, which can explain the prominent NIR-excess between K$_{\rm s}$- and L$'$-bands, see Fig.~\ref{fig_SED_model}. This is underlined by the Keplerian orbit (Fig. \ref{D2D23D3D31_orbit}) and the not present evaporation effects caused by the high ionizing environment around Sgr A*. For the D23 and D3 source, the upper limit of the flux density in H$_{\rm s}$-band is larger than the corresponding fitted value, which could hint the contribution from the stellar photosphere in a similar way as for the DSO, see Fig.~\ref{fig_SED_model_DSO} (right panel). For the DSO, the $K-L'$-band slope is similarly inverted, $\alpha=4.15$, as for other D-sources which is quite distinct from the stellar SED slope of $\alpha\approx -2$. 

\subsection{Implications for the nature of D-objects}

\begin{figure}[h!]
    \centering
    \includegraphics[width=0.5\textwidth]{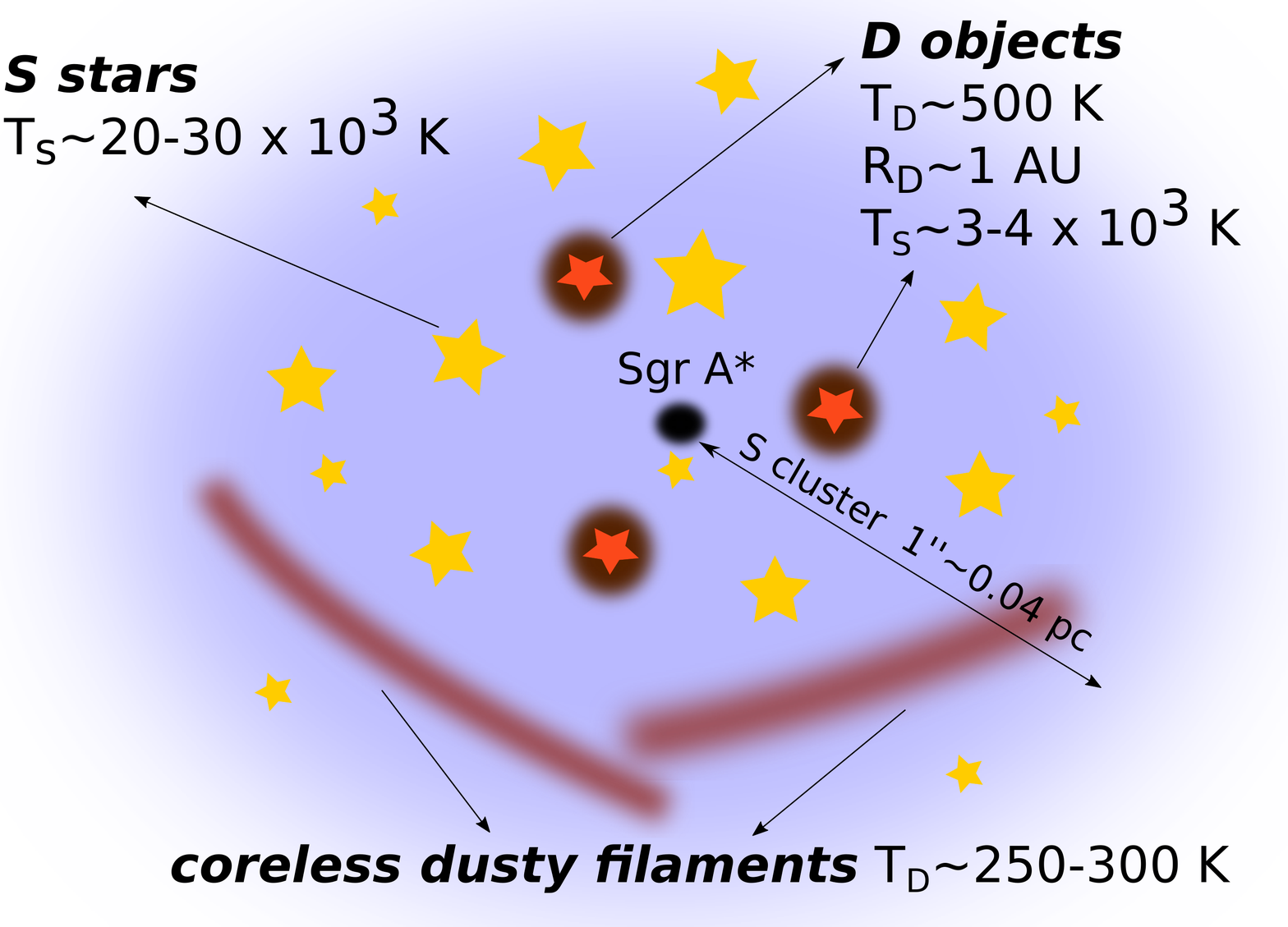}
    \caption{ Illustration of colder D-objects as dust-enshrouded stars inside the S cluster that consists predominantly of stars of spectral type B. D-objects have the characteristic temperature of $\sim 500\,{\rm K}$, which is larger than the temperature of other dusty structures in the region, namely the minispiral filaments of $200-300\,{\rm K}$.}
    \label{fig_dusty_illustration}
\end{figure}

D-sources are located inside the S-cluster with an approximate radius of $r_{\rm cluster}=1''\sim 0.04\,{\rm pc}$. Their blackbody temperature of $\sim 500\,{\rm K}$, which was determined in the previous section, is larger than the typical temperature of $\sim 200-300\,{\rm K}$ of dusty filaments in the innermost parsec, namely the minispiral arms and dust around infrared sources IRS3 and IRS7 \citep{1999ASPC..186..240C}, see also the illustration in Fig.~\ref{fig_dusty_illustration}. Given the number of S-stars -- around $N\sim 30$, the mean distance is expected to be $\overline{l}\sim r_{\rm cluster}/N^{1/3}\sim 0.01\,{\rm pc}=2700\,{\rm AU}$. If there is any dusty clump or filament, its temperature due to external irradiation is given by \citep{1987ApJ...320..537B},

\begin{equation}
    T_{\rm dust}=9.627\left[\left(\frac{L_{\rm UV}}{L_{\odot}}\right)\left(\frac{r_{\rm d}}{1\,\rm{pc}}\right)^{-2}\exp{(-\tau_{\rm UV})}\right]^{1/5.6}\,{\rm K}\,,
    \label{eq_dust_temperature}
\end{equation}
where $L_{\rm UV}$ is the UV luminosity of the central source expressed in Solar luminosity, $r_{\rm d}$ is the dust distance from the source, and $\tau_{\rm UV}$ is the optical depth. Considering the S2 star, its bolometric luminosity was inferred to be $L_{\rm S2}=10^{4.26}\,L_{\odot}$ \citep{Habibi2017}, from which we calculate the UV luminosity in the UV wavelength range of $\lambda=1-400\,{\rm nm}$, $L_{\rm UV}=10^{4.23}\,L_{\odot}$. For this luminosity, and using Eq.~\ref{eq_dust_temperature} in the optically thin case, we get the dust temperature of $T_{\rm dust}=260\,{\rm K}$, which is consistent with the warmest dust sources in the Galactic Center region \citep{1999ASPC..186..240C}.

Therefore the higher temperature of the D-sources implies the presence of an internal source of heating, that is a star with the UV luminosity of less than 10$\,L_{\odot}$, which gives the sublimation radius of $r_{\rm sub}\simeq 1.5\,{\rm AU}$. The stellar parameters that we inferred in the best model fit for the DSO -- $T_{\star}=3253\,{\rm K}$ and $R_{\star}=1R_{\odot}$ -- meet this criterion, since the overall stellar bolometric luminosity is $L_{\star}=4 \pi R_{\star}^2\sigma T_{\star}^4=0.1\,L_{\odot}$ and the UV luminosity is $L_{\rm UV}=4\times 10^{-4}L_{\odot}$. By inverting Eq.~\eqref{eq_dust_temperature}, we obtain the sublimation radius $r_{\rm sub}(T_{\rm dust}=1000\,{\rm K})=0.009\,{\rm AU}$ and the dust temperature of $T_{\rm dust}=500\,{\rm K}$ is reached at $r_{\rm d}(T_{\rm dust}=500\,{\rm K})=0.065\,{\rm AU}$, that is well within the overall extent of the DSO, $R_{\rm DSO}\sim 1\,{\rm AU}$.

The bolometric luminosity of the dusty envelope then is ${L_{\rm env}}=4\pi{R_{env}}^2\sigma{T_{env}}^4=2.6\,L_{\odot}$, hence it clearly dominates the overall luminosity of the DSO. In this context, the D-sources could be low-luminosity, light young stars of type I that are remnants of the most recent star-formation episode in the inner parsec of the Galactic center. It is not clear whether they manifest a single separate star-formation epoch, such as due to an infall of a cloud from the more distant regions in the Central Molecular Zone, or rather they formed from the material left-over from the formation of OB stars in an accretion disc $\sim 5$ million years ago \citep{Levin}, that is a secondary population of lighter objects formed from compressed wind-swept shells. The presence of low-mass stars still surrounded by primordial accretion material in the OB associations in the Galactic plane is observationally well-established \citep{2007prpl.conf..345B}. While massive OB stars evolve fast on the main sequence at 5 Myr, low-mass stars of $0.1$-$0.2\,M_{\odot}$ move still vertically along the Hayashi track \citep{1998A&A...337..403B}, being surrounded by accretion discs up to $10-20$ Myr. D-objects could in principle manifest this ''mass-separation'' of young stars in the direct vicinity of Sgr~A*, providing an opportunity for testing various star-formation models in the direct influence of the SMBH.

\section{Discussion}
\label{sec:discussion}
Here, we discuss our findings of the NACO and SINFONI analysis. We also give a brief outlook towards upcoming observations and possibilities with the Mid-Infrared Instrument (MIRI).
\subsection{NIR sources}
Our analysis of the infrared H-, K-, and L$'$-data combined with the one- and two-component blackbody modeling shows that the dusty sources may be part of the
S-cluster ($r<0.04$ pc) population. These dusty sources are possible young stellar objects that are embedded in a potentially non-spherical dusty envelope as shown for the DSO/G2 (Fig. \ref{fig_SED_model_DSO}). These envelopes may include a optically thick dusty section, bipolar cavities, as well as a bow shock. 
%\\
%Having young stars present in the strong gravitational field of a super-massive black hole can be explained. Young stars have been observed within the entire nuclear star cluster including the innermost region close to Sgr A*. 
\\
\cite{Jalali2014} have focused particularly on the IRS~13N located 0.1 pc away from Sgr A*. Its properties suggest that these objects form 
a group of young stars with about five embedded young stellar objects ($<$1 Myr). \cite{Jalali2014} performed three-dimensional
hydrodynamical simulations to follow the evolution of molecular clumps orbiting Sgr A*. Assuming that the molecular clumps are isothermal 50-100K containing $100\,M_{\odot}$ in a region of $\ll0.2$ pc radius. Hence, the authors were able to constrain the formation and the physical conditions of such groups of young stars. 
\\
The authors also argue that such molecular clumps can originate in the CND. Dissipation of orbital energy in the CND
result in highly eccentrical orbits with clumps experiencing a strong compression along the orbital radius vector and perpendicular to the orbital plane. This causes increased gas densities higher than the tidal density of Sgr A*, triggering the rapid formation of solar mass stars.

\begin{itemize}
\item \textup{D2, D23, D3, D3.1}\newline
In Fig. \ref{finding_chart_naco}, we show dusty sources D2, D3 with the newly discovered objects D23 and D3.1. In 2006, these sources are still isolated compared to 2015 where the chance of confusion is high. 
%
%The positions in mas} from the K-band NACO data presented in Tab. \ref{tab:k_band_pos} are extracted after a positional comparison %with the spectroscopic SINFONI line maps and the bright and dusty L'-band counterparts of the sources. The determined NACO K-band positions %are given with respect to Sgr A* (Tab. \ref{tab:k_band_pos}).

Since there is an overlap of D3 and D3.1 (see Fig. \ref{spectrum}), the shape of the spectrum shows multiple peaks. We identify the D3 related Doppler-shifted emission peaks with the SINFONI line maps. However, a bipolar morphology \citep{Yusef-Zadeh2017, Peissker2019}, that is indicated by a complex peak structure, could be speculated and is discussed in Sec. \ref{sec: special case of D3 and D3.1}. For the sake of completeness, it should be mentioned that the chance of confusion for the identification of D3.1 in the SINFONI data cubes is rather high in 2015. Spectroscopically, it is close to the Br$\gamma$ rest wavelength. However, we also find red-shifted HeI and [Fe III] emission lines at the approximate position of their related rest wavelength. Since the chosen location of the SINFONI FOV in various years is not covering the expected area of D3.1, we are not able to expand our analysis. 

In contrast, the D-complex members D2, D23, D3 can be traced an analyzed as presented in Sec. \ref{Results}. Based on our K- and L$'$-band data, we determine their magnitudes and give an upper limit for the H-band. Our single temperature BB model fits the flux emission of the dusty sources. In general, gas clouds/clumps in the vicinity of Sgr A* show temperatures much lower than our fitted model \cite[see also][]{Genzel2010}. Therefore, the external heating of the dusty-sources seems unlikely. Vice versa, the dusty sources D2, D23, and D3 could have an internal heating source that might be indeed of stellar nature.\newline

\item \textup{X7, X7.1, X8}\newline
The X-sources X7, X7.1, and X8 are grouped like the D-complex in the vicinity of Sgr A*. All three sources show Doppler-shifted HeI, Br$\gamma$, and [FeIII] emission lines and can be detected in the SINFONI line maps. From the observations, it is not clear if X7 and X7.1 share a similar history because the detection of the later one is limited to 2015. X7 can be detected in various bands with a comparable position angle (North-East) to X8 of around 45 deg. Because X7.1 is located too close to the edge of the SINFONI FOV, a analysis of the shape and its elongation is difficult. Since X7 and X8 show signs of a bow-shock \citep{muzic2010, Peissker2019}, we can just speculate if X7.1 might also share a similar nature. This, however, could be part of future observation programs. If a bow-shock could be confirmed for X7.1, it might add another source to the proposed wind-direction of East-North to South-West with respect to Sgr A* \citep{Yusef-Zadeh2012, Shahzamanian2015}. This would increase the importance of the X-sources since the proper motion of at least X7 is directed in a direction, where no wind should be active. Therefore, X7 could change its shape in the future.\newline

\item \textup{DSO, G1, D5, D9}\newline
Like the DSO/G2, G1 survived its peri-center passage without being disrupted and can be traced on a highly eccentric orbit (see Fig. \ref{finding_chart} for a detection of G1 after the pericenter passage in 2008). As \cite{Witzel2017} and \cite{Zajacek2017} points out, these two sources are more likely a YSO/star embedded in a dusty envelope. This is underlined by the detection of the DSO/G2 source as a compact object at peri-apse by \cite{Witzel2014} as well as the H- and K-band detection presented and analyzed in this work (see Fig. \ref{fig:hk_dso} and Fig. \ref{fig_SED_model_DSO}). The, in several publications described, tidal evolution of the gas emission \citep{Gillessen2013a, Gillessen2013b, Pfuhl2015, Plewa2017} can not be confirmed in our data-sets \citep[see][]{Valencia-S.2015}. In contrast, we are detecting a compact source close to periapse in 2013 (Fig. \ref{fig:naco_sinfo_dso}, \ref{fig:hk_dso}). We want to add, that the authors of \cite{Valencia-S.2015} show the sensitive connection between the subtracted background and the extracted spectrum of the source. Considering the report of a drag force for the DSO/G2 by \cite{Gillessen2019}, it would be interesting to investigate the effect of the background subtraction.%However, \cite{Gillessen2019} report a drag force that interacts with the DSO/G2 source.

In the case of the DSO/G2, dust-dust scattering could be accounted for the observed emission \citep{Zajacek2017}. Even though G1 and the DSO/G2 share some similarities, \cite{Witzel2017} conclude that both sources are not part of a gas streamer in the way, that was proposed by \cite{Pfuhl2015}. The authors propose, that the emission of the two sources is rather of non-stellar origin. Considering also the polarization results of \cite{Shahzamanian2016} for the DSO/G2, the observational evidence presented by the authors strongly favors an explanation that goes beyond a pure gas-cloud model.

Some sources are moving out of the available SINFONI FOV because of their proper motion. D5 (\citep{EckartAA2013} and Fig. \ref{finding_chart}) for example can be barely observed after 2008. More observations are needed to expand the analysis in order to evaluate the nature of the object. This can also be applied to D9 that is located north of S2.

%As mentioned before, \cite{Jalali2014} proposed the idea, that gas clumps from the CND could have moved towards Sgr A*. These $100\,M_\odot$ %clumps at 50-100K could have formed dusty sources through the gravitational influence from the SMBH. This process could be responsible for %forming YSOs. The observations fit the SED, that is based on a pre-main sequence star embedded in a dust envelope, presented in %\cite{Zajacek2017}.
 
\end{itemize}

\subsection{The case of D3 and D3.1}
\label{sec: special case of D3 and D3.1}
Figure \ref{finding_chart} and Fig. \ref{finding_chart_naco} both show the sources D3 and D3.1 in the 2006 NACO dataset and the 2015 SINFONI cubs. Over 10 years, both sources moved with a very similar projected proper motion direction towards North-East (see Tab. \ref{tab:L-band-pos} and Fig. \ref{D2D23D3D31_orbit}).
They show Doppler-shifted Br$\gamma$, HeI, and [Fe III] emission that clearly distinguishes these sources against the background emission. As indicated by Fig. \ref{spectrum}, the bright Br$\gamma$ emission line of both sources is present in the spectrum as a double peak that is blue and red-shifted. As shown by \cite{Yusef-Zadeh2017}, star formation in the form of Bipolar outflow sources is still ongoing close to the Galactic center. For D3 and D3.1, it can just be speculated if the observed properties indicate the presence of a bipolar scenario. Judging by our orbit overview presented in Fig. \ref{D2D23D3D31_orbit}, the possibility for a common origin for both sources is rather low. Currently, they are not targeted by a possible wind that probably originates at the position of Sgr A* \citep[see][for more information]{Lutz1993,muzic2010,Yusef-Zadeh2012,Peissker2019,Yusef-Zadeh2017-ALMAVLA}. If the wind coming from Sgr A* causes a shock or the elongated shape that we observe for X7 and X8, the objects D3 and D3.1 but also D2 are the perfect candidates to prove its existence. For that matter, the speculated shielding from the ionizing X-ray radiation could lead to disappearing [Fe III] emission lines when these sources cross the discussed Br$\gamma$-bar (Fig. \ref{fig:fe_line}, see Sec. \ref{sec:brgamma_bar}) in the Galactic center. However, an identification of these sources is afflicted with a high chance of confusion when they move towards the more crowded S-cluster. For a clear identification in the objects in the vicinity of Sgr A*, spectroscopic information is crucial. Because of their dust, mid-infrared observations (with, e.g., MIRI) combined with a spectroscopic analysis is crucial to get a deeper understanding of these sources.

\subsection{[Fe III] emission}
\label{sec.:iron_emission_lines}
In contrast to the already indicated but not further commented line detection of ionized iron for D2 and D3 in \cite{Meyer2014}, we report in addition to the blue-shifted Br$\gamma$ and He$I$ a Doppler-shifted [FeIII] ${}^3 G\,-\,{}^3 H$ multiplet for the investigated sources (see Fig. \ref{spectrum} and Sec. \ref{Results}). A confusion, as mentioned by \cite{DepoyPogge1994}, with other near-infrared emission lines can be expelled since the observed [FeIII] lines are Doppler-shifted and can be thereby distinguished from other lines. 
For the X7, X7.1, X8, D2, D23, and D3 sources, a Doppler shifted [Fe III] line emission is detected in the SINFONI data cubes (Fig. \ref{finding_chart}, \ref{spectrum}). We find also a weak Doppler-shifted [FeIII] emission at the position of D3.1 that distinguishes the source from the rest wavelength emission. 
When it comes to the charge transfer reaction of [Fe II] and [Fe III], the authors of \cite{Lutz1993} discuss the ratio of the ionized iron lines with
\begin{equation}
      Fe^{2+}\,\,+\,\,H\,\,\rightarrow\,\,Fe^{+}\,\, +\,\, H^{+}\,\, .
      \label{equation1}
\end{equation}
As pointed out by the authors, because of the missing information about the efficiency of the charge transfer reaction, it is difficult to evaluate the importance of the [FeIII] line compared to the total iron budget. Even though the amount of the, in this work analyzed sources, is low for a statistical evaluation, the findings indicate, that the chance is rather high to detect Doppler-shifted [FeIII] lines in dusty sources west of Sgr A*. So far, none of the observed sources east of the Br$\gamma$ emission bar (Fig. \ref{fig:fe_line}) are showing signs of Doppler-shifted [FeIII] lines. %In the following subsection, we want to discuss the possibility to draw a connection between the [FeIII] emission and the nature of the investigated objects.} 
%\subsection{Doppler-shifted iron emission lines}
%The presence of ionized [Fe III] lines west of the Br$\gamma$ feature and the abundance of the iron multiplet in the spectrum of the observed objects east of the dividing emission 
Even without the efficiency information, this could be explained by the charge transfer radiation equation (Eq. \ref{equation1}). It can be speculated, that the origin of these lines is most probably the strong X-ray radiation of Sgr A*. A possible wind, that originates at the position of the SMBH produces a shock front in the ambient medium in the form of the observed Br$\gamma$-bar feature (Fig. \ref{fig:fe_line}). %This Br$\gamma$-bar is detected in every SINFONI data set between 2005 and 2016. However, the continuum does not show emission that could be linked to the line map detection (Fig. \ref{fig:fe_line}).}

\subsection{Br$\gamma$ emission bar}
\label{sec:brgamma_bar}
The here reported ionized and Doppler-shifted iron lines can be detected throughout almost all sources that are located west of the Br$\gamma$-bar that is observed in the SINFONI line maps at $2.1661\mu m$ (see Fig. \ref{fig:fe_line}, middle) that spans from north to south with a projected distance to Sgr A* of $\sim\,0".08$ in 2006 and $\sim\,0".15$ in 2015. This structural detail of the inner parsec can be observed in every available SINFONI data-set.

Keeping this finding in mind, we shortly want to summarize the analysis of features at the position of Sgr A* that could be connected to the Br$\gamma$-bar and speculate about a possible connection to the recent publication of \cite{Murchikova2019}. 

\begin{itemize}
    
\item 
\cite{Lutz1993} reports a [FeIII] bubble around Sgr A*. It is speculated that the emission originates either in stellar winds or a jet-driven outflow from Sgr A*. The authors also present a V-shaped Doppler-shifted [Fe III] feature directed towards the mini-cavity with a bow-shock morphology (see Fig. \ref{finding_chart} and \ref{fig:fe_line}) and a LOS velocity of around $40\,km s^{-1}$. The position angle of the emission with respect to Sgr A* is around $45\deg$.

\item
Followed by this, \cite{Yusef-Zadeh2012} reports a teardrop-shaped X-ray feature roughly at the position of Sgr A*. The authors speculate that it is probably produced by a synchrotron jet or shocked winds, that are created by mass losing stars and seems to be consistent with the [FeIII] bubble shown in \cite{Lutz1993}. The ram pressure of the winds could then push the ionized bar away and produce the teardrop-shaped X-ray feature. However, the VLA data show, that the bright X-ray feature is open towards the western side. It is unclear if this open side of the teardrop can be linked to the Br$\gamma$-bar shown in Fig. \ref{fig:fe_line}. 
\end{itemize}

If future observations of the Br$\gamma$-bar could confirm a connection between [FeIII] bubble and the teardrop-shaped X-ray feature, it might indicate the transition of a region that is dominated by a strong wind to a region with a strong ionizing X-ray emission. This could explain the Doppler shifted and ionized [Fe III] ${}^3 G\,-\,{}^3 H$ multiplet that can be observed in almost every dusty source west of the Br$\gamma$ transition feature (see Fig. \ref{spectrum} and \ref{fig:fe_line}).
Recently, \cite{Murchikova2019} reported a blue- and red-shifted Br$\gamma$ emission around Sgr A* with a line width of over $2000\,km\,s^{-1}$ on scales that are comparable to the here presented SINFONI data. Based on the presented data of the authors, there is a small but noticeable gap of the Doppler-shifted Br$\gamma$ feature. The authors report a displacement of the red-shifted Br$\gamma$ side of around 0.11 as or 110 mas. Based on our SINFONI data-cube, the here reported Br$\gamma$-bar shows a measured size of 0.10-0.12 as and therefore could be connected to the reported Doppler-shifted Br$\gamma$ feature by \cite{Murchikova2019}.

\subsection{[FeIII] distribution in the Galactic center around Sgr A*}

We already discussed the [FeIII]-bubble analyzed by \cite{Lutz1993}. In addition, we want to draw a connection between the large scale [FeIII] distribution from the same publication (see Fig. \ref{finding_chart}) that is located south-east of Sgr A* at a distance of a few arcsec. This distribution is arranged in a V-shape and at the position of the mini-cavity. The IRS13 complex can also be found inside of this bow-shock like feature that contains many dusty sources with iron emission lines (\citep{muzic2008}). 
We show in Fig. \ref{finding_chart}, that the iron distribution is in close contact to the position of Sgr A*. As mentioned before, it is plausible, that iron in the Galactic center is not only created by a continuous distribution, but also by compact sources like in, e.g., IRS13N (\citealp{muzic2008}) or the D-complex (this work). 
Based on our spectral analysis, we speculate that the chance of finding dusty sources with iron emission lines is higher inside the V-shaped [Fe III] feature compared to dusty sources outside of the iron distribution. We discussed, that the detection of the here discussed sources can be divided by the bright and prominent Br$\gamma$-bar (Fig. \ref{finding_chart} and \ref{fig:fe_line}). 

As mentioned before, \cite{Eckart1992}, \cite{Lutz1993}, and \cite{Yusef-Zadeh2017} report a bubble of hot gas that is found around Sgr A*. 
It is still unclear, why the bubble around Sgr A* is open towards the western direction. It just can be speculated that maybe the S-stars or the here reported Br$\gamma$-bar are shielding the dust grains against depletion of [FeIII] (\cite{Jones2000}, \citealp{Zhukovska2018}) that leads to the presence of large amounts of excited iron in the gas phase. At the transition edge (Fig. \ref{finding_chart} and \ref{fig:fe_line}), hydrogen gets excited and the depletion of [FeIII] from the dust grains of the dusty sources is increased. Upcoming observations with MIRI and long term monitoring with SINFONI on larger scales of these sources and the GC will probably reveal more interesting details about the connection of the Sgr A* bubble, the wind, the in \cite{Murchikova2019} presented Doppler-shifted Br$\gamma$ emission, and the presented north- to south Br$\gamma$-bar.  
These observations becoming more important in the future since some dusty sources, e.g., D2 and D3 are moving in projection towards regions, where no wind features are present. We expect, that the Doppler-shifted [Fe III] emission lines could be dampened or vanish completely. This is also reflected in Fig. \ref{fig:flux_ratio}. Partially, this plot is adapted from \cite{Lutz1993} and complemented with our measurements. The plot suggests, that a possible ionization wind shows an increased impact towards Sgr A* from the north-east direction. This trend continues towards south-west till the V-shaped bow shock like feature at the position of the mini-cavity (Fig. \ref{finding_chart} and \ref{fig:fe_line}). 

\subsection{Indications from the [FeIII] findings}

\subsubsection{Iron distribution in the GC}

We assume the dust-growth ratios presented in \cite{Zhukovska2018} where the authors discuss the spallation of iron from dust grains into the gas-phase. Combined with the LOS, our results (Fig. \ref{fig:flux_ratio}) are in agreement with the in \cite{Jones2000} presented connection between shock velocity and dust-depletion of Fe. The [FeIII] flux presented in Fig. \ref{fig:flux_ratio} originates in the mini-cavity (black data points) and could be associated with an ionized V-shaped structure that shows a bow-shock morphology \citep{Lutz1993, Yusef-Zadeh2012}. The detection of compacted sources close to Sgr A* could indicate, that the [FeIII] distribution in the mini-cavity could be influenced strongly by single dusty-sources \citep[this work;][]{Moultaka2006, muzic2008}.

\subsubsection{Iron as an indication for YSOs}

As mentioned in Sec. \ref{sec.:iron_emission_lines}, we want to discuss the significance of the Doppler-shifted [Fe III] emission line compared to the total iron budget. As shown by \cite{Depoy1992}, the derived [Fe II]$\mathbin{/}Br{\gamma}$ ratio is between 0.5 (5" south of IRS7) and 0.7 (10" south of IRS7). IRS7 is located around 5".5 north of Sgr A*. Therefore, it is reasonable to assume, that a ratio of around $0.55\,\pm\,0.1$ is in good agreement with the derived values. Taking into account our measurements (Tab. \ref{tab:metallicity} and Fig. \ref{fig:flux_ratio}), we find an averaged value for the amount of [Fe III] to [Fe II] of around 0.7. This naive approach does not take different densities or temperatures into account. However, \cite{Lutz1993} derive with values from \cite{Depoy1992} for the [Fe II]/[Fe I] ratio a result of $\approx$ 1 in the GC. Additionally, our averaged AIP value of 0.38 is in excellent agreement with the common derived [Fe III]$\mathbin{/}Br{\gamma}$ ratios \citep{Depoy1992,Lutz1993, Luck1998}. Therefore, the presence of the detected [FeIII] emission line of the sources could be used as an indicator for the metallicity \citep{Kamath2014} or at least for the high abundance in metals. Moreover, it emphasizes the dusty nature of the objects \citep{Zhukovska2018}. \cite{Oliveira2013}, \cite{Shimonishi2016}, and most recently \cite{Sciortino2019} underline in their studies iron and metallicity as an important star formation parameter. The hypothesis, that the nature of the dusty sources can be described as recently formed YSOs \citep{Yusef-Zadeh2017, Zajacek2017} becomes, therefore, more acceptable and will be briefly summarized in the following.% and will be discussed in detail in Sec. \ref{sec:sed}.

\subsection{Origin of the NIR excess sources}

When it comes to the origin of the here discussed Doppler-shifted Br$\gamma$ sources that surround Sgr A* on highly eccentric orbits, we speculate a stellar nature. 
We associate dust-enshrouded stars that may, in fact, could have been formed very recently through the mechanism put forward by \cite{Jalali2014}. These mechanisms are also part of the discussion presented by \cite{tsuboi2016} and \cite{Trani2016}. Based on the fact that D2, D23, and D3 show the same spectral features and have similar magnitudes, it seems to be plausible, that they are part of one population when we consider the mentioned publication. They might be dust embedded pre-main-sequence stars presumably forming a small bow shock \citep[][]{Shahzamanian2016, Zajacek2017}. \cite{Yoshikawa2013}, \cite{Yusef-Zadeh2017-ALMAVLA}, and \cite{Naoz2018} discuss and show several plausible scenarios that underline the presence of binaries, YSOs, and intrinsically polarized low-mass stars as well as bow-shock sources in the GC close to Sgr A*.

In addition, \cite{muzic2007} and \cite{muzic2008} present trajectories of dusty sources that are co-moving in groups. This is already separately analyzed in \cite{EckartAA2013} where the authors discuss, that randomly located dusty sources in the GC can be found with a likelihood of around $25\%$ per resolution element and observing epoch. They derive, that co-moving objects in the GC can be observed with a probability of several percents. These two observational and statistical findings can be directly applied to the found sources D2, D23, D3, D3.1, X8, X7, and X7.1. These sources are either co-moving (see Fig. \ref{D2D23D3D31_orbit}) or can be found in groups (see Fig. \ref{finding_chart_naco}).

If upcoming observations could confirm the in this work proposed nature of the sources, the identification of the [FeIII] multiplet in unknown dusty sources could be used to underline their stellar character. However, the DSO/G2 spectrum does not show any signs of a [FeIII] line. But as shown in Fig. \ref{fig_SED_model_DSO}, the presented SED model based on our H- and K-band detection (Fig. \ref{fig:hk_dso}) fits a stellar core with a dusty envelope.
%In summary, the discussed scenarios always consider recent star formation processes that seem to be in line with the proposed model} of \cite{Jalali2014}.

%\subsection{Apparent co-motion and clustering}
%
%\cite{muzic2007} and \cite{muzic2008} present trajectories of dusty sources that are co-moving in groups. This %is already separately analyzed in \cite{EckartAA2013} where the authors discuss, that randomly located dusty %sources in the GC can be found with a likelihood of around $25\%$ per resolution element and observing epoch. %They derive, that co-moving objects in the GC can be observed with a probability of several percents. These two %observational and statistical findings can be directly applied to the found sources D2, D23, D3, D3.1, X8, X7, %and X7.1. These sources are either co-moving (see Fig. \ref{D2_orbit} and Fig. \ref{D3_orbit}) or can be found %in groups (see Fig. \ref{finding_chart_naco}). This finding is already modeled in \cite{Jalali2014} where the %authors propose, that clumps in the Circum Nuclear Disk (CND, see \citep{Hsieh2017,Tsuboi2018}) collide and %move towards Sgr A*. Because of the gravitational interaction between Sgr A* and these clouds, that can be as %big as 100$M_{\odot}$ with temperatures around 50-100K, star formation can be triggered. The described process %of star formation produces groups of stellar companions. The resulting members of these groups (YSOs and %pre-main-sequence stars) can lose angular momentum through collisions and then will naturally form in co-moving %groups. 

\subsection{Upcoming opportunities with MIRI}

Here, we want to discuss the impact of the mid-infrared instrument MIRI of the James Webb Space Telescope \citep[see][for further information]{Bouchet_2015, Ressler_2015, Rieke_2015}. Compared to the presented data in this work, MIRI will operate in the M-band. This will allow us to observe the here discussed sources with higher precision because of a more stable PSF. Since the SINFONI introduces an elongated PSF with changing x- and y-values for the FWHM, an optimized PSF for the high-pass filtering is necessary. The wings of close-by sources do interact in a way, that artificial signals could be created \citep{Sabha2012, EckartAA2013}. Additionally, sky emission lines can have an unavoidable influence on the spectrum that is described in \cite{Davies2007}. Hence, the investigation of dusty sources with MIRI in the central stellar cluster will greatly benefit from the up-cumming improved observational possibilities. Foremost, we have the JWST instrumentation that will allow us to obtain sensitive spectroscopy without the mentioned atmospheric and instrumental effects of individual sources in this densely populated region. In particular, the IFU observations with MIRI will be helpful to obtain broadband spectroscopy covering a few to several 10 micrometers. While spatial deconvolution of the resulting data cubes at longer wavelengths will be essential, the sources should readily be separable at the short-wavelength end of the spectrometers.
Highest angular resolution and/or diagnostic spectral line tracers available in MIRI are essential to make further progress in this investigation of the dusty sources. Especially when it comes to the local and general line of sight ISM towards them through ice absorption features like for example $H_2 O, CH_4, CO, N H_3$. Also, emission features from gaseous molecules like $ H_2, CO, H_2 O, CH_4, C_2 H_2, HCN, OH, SiO$ could provide a new insight of the nature and origin of the observed objects. Some of these emission lines like $ CH_4, C_2 H_2$, that can be found in gas and ice, can be used as probes for stellar disks. For the Galactic Center, we have already shown in \cite{Moultaka2006, Moultaka2009} the importance of the $H_2 O$ line emission and the CO gas and ice absorption to study the extinction towards the Galactic Center and the properties of the GC interstellar medium.

\section{Summary and Conclusions}
\label{Conclusion}

We present a detailed analysis of two of the dusty NIR excess sources
within the central arcseconds of the Galactic Nucleus, close to the 
supermassive black hole Sgr A*. These dust sources are of special interest
as their existence had not been predicted and may be indicative 
for ongoing star formation in this complex region.

\begin{itemize}
\item{Properties of the NIR excess  sources}\newline
We successfully fitted Keplerian orbits to both sources, D2 and D3. Furthermore, we found new sources that we named D23 and D3.1. 
D2, D3, and D3.1 have an elliptical orbit with similar orbital parameters. They also show Doppler-shifted Br$\gamma$, HeI and [FeIII] line emission exhibiting comparable line-of-sight velocities indicating that they arise from the same source. Both, the Br$\gamma$ and [Fe III] line emission, occur from high excitation shocked regions (see e.g. \citet{Tielens1994, Zhukovska2018}). Since D23 shows a clockwise orbital trajectory, this D-source might be part of other, not discovered groups of dusty emission objects. 

The obtained \textit{K$_S$}-band magnitudes are 18.1$^{+0.3}_{-0.8}$ for D2 and 19.7 $^{+0.4}_{-0.3}$ for D3. In the \textit{H}-band, we derived upper limits to their magnitudes of 22.98 for D2 and 22.22 for D3, respectively. 

\cite{Valencia-S.2015} and \cite{Zajacek2017} show that for a 1-2 ~ M$_{\odot}$ ~ embedded pre-main sequence star, hot accretion streams of matter very close to the star can fully account for its observed Br$\gamma$ luminous in combination with the line widths covering a range from 200 $km\, s^{-1}$ to 700 $km\, s^{-1}$. These streams can possibly be in combination with disk winds.

As indicated by pre-main-sequence evolutionary tracks of low- and intermediate-mass stars \citep[see, e.g., ][]{Siess2000}
it is entirely conceivable that after an initial phase of less than 10$^{6}$ years the 1-2$M_{\odot}$ stars present themselves in a T~Tauri stage with a luminosity that does not exceed $10M_{\odot}$ \citep[see also][]{chen2014}.

Comparing the here derived flux values for the DSO/G2 with a two component blackbody model and detecting the object in the H- and K-band continuum, show that the source might be indeed a pre-main-sequence star emdedded in a dusty envelope as already discussed by \cite{Shahzamanian2016} and \cite{Zajacek2017}. We sucessfully fitted also a single component BB model to the observed emission of D2, D23, and D3. Since the gas temperatures in the vicinity do not exceed 50 K--200 K \citep{Genzel2010}, it is obvious that a simple gas-cloud nature of the sources can be excluded. In combination with their Keplerian orbital motion, we see a strong tendency towards a stellar scenario. This is also supported by the H-band detection of the DSO/G2, which clearly favours the combined star+envelope model over the pure core-less cloud model. Young stellar objects of type I \citep{Lada1987} could be a plausible explanation. Adjustment of several variables, e.g. the accretion rate of the YSOs, the viewing angle and the dust temperature, might improve the SED of the near-infrared and mid-infrared fluxes. The intrinsic colors are consistent with a dust temperature of 500~K. The surrounding shell/disk, as well as the accretion material, can provide enough extinction to explain the observed infrared colors \citep{EckartAA2013}.
\\
\item{Linkage to flaring activity}\newline

The data obtained for the DSO are consistent with a very compact source on an elliptical orbit. It is sufficiently compact that it has not influenced the flaring activity of the black hole in any statistically convincingly way. That does however not exclude the possibility that an increase in accretion activity of Sgr A* may still be upcoming. However, the crossing time scales for the region in which the DSO passed by Sgr A* is only a few years, even if one includes effects from a possible wind from Sgr A*. Any Sgr A* activity occurring at times more than a few years from now become increasingly more difficult to be associated with the DSO passage close to Sgr A* when one considers in-spiral time scales of several decades. Distinguishing between the DSO/G2 passage and for example, the G1 source might be challenging. Both sources passed Sgr A* at a comparatively distance at $a_{min}\,=\,200\,-\,300$ a.u. \citep{Witzel2017}. As discussed, it becomes increasingly likely that the dust embedded sources are associated rather with young accreting stars than core-less gas and dust clouds. An identification of the DSO, D2, D23, and D3 with a dust embedded stars also puts the necessity of a common history of the DSO and G1 source at risk \citep{Pfuhl2015}. Due to the higher mass ($\sim 1-2\,M_{\odot}$), the action of drag forces on these objects is also more than questionable \citep{Gillessen2019}.
\\
\item{Bow shock stars}\newline

The here detected Br$\gamma$ and  [Fe III] lines might arise from collisional excitation in the inner bow shock. Also, accretion processes onto a central star is a considerable option. Furthermore, it is presumed that the Br$\gamma$ emission lines originate in photoionization of a UV radiation field. [Fe III] emission requires a high iron abundance gas, which can be created in shocks. The high abundance gas then has to be ionized, either by UV photons or by collisions. 

A possible bow shock might either be caused by strong stellar winds coming from, e.g., IRS13 or by an outflow from the Galactic Center. Moreover, the bow shock could arise from the potential supersonic motion of D2 and D3.

\cite{Shahzamanian2016} and \cite{Zajacek2017} constrain the nature and the geometry of the DSO. The authors have compared 3D radiative transfer models of the DSO with the near-infrared (NIR) continuum data including NIR K-band polarimetry. They present a composite model of the DSO with a dust-enshrouded star that consists of a stellar source, dusty, optically thick envelope, bipolar cavities, and a bow shock. 
\\
\item{An alternative to the thermal young stellar object model}\newline

An alternative scheme can explain the continuum emission via the non-thermal contribution of young neutron stars and their wind nebulae \citep[][]{Zajacek2014, Zajacek2017}. Their proposed model can match the NIR fluxes as well as polarized properties of the observed spectral energy distribution. The authors also show that, in principle, the SED may be explained by a young pulsar wind nebula (PWN). However, the analysis of the neutron star model shows that young, energetic neutron stars should be detected as near-infrared-excess sources which also exhibit themselves by the magnetospheric synchrotron emission. Both the thermal and the non-thermal models are thus consistent with the observed properties of these sources, i.e., their compactness, the total and (for the DSO) the polarized continuum emission. Nevertheless, there are as yet no such detections of X-ray and radio counterparts of the dusty sources.

\cite{Naoz2018} discuss, that binaries could play a major role by contributing and explaining observed properties of the GC disk. With that, the number of similar objects as discussed here could be much higher but also expected. The LOS velocities of the investigated objects are consistent with the discussed binary velocities $v_{Z}$ as a function of the semi-major axis of the GC disk.

Recently, \cite{Yusef-Zadeh2017-ALMAVLA} reported the detection of bipolar outflow sources with ALMA. The authors find evidence for low-mass star formation even though considering the difficulties, distinguishing between low-mass YSOs from starless dusty cores \citep{Gillessen2012, EckartAA2013}. However, the findings of \cite{Yusef-Zadeh2017-ALMAVLA} are in line with \cite{Yoshikawa2013} where the authors report, that intrinsically polarized stars could be associated with YSOs. At least for the DSO/G2, \cite{Shahzamanian2016} finds a polarization degree of $>\,20\%$. A future polarization analysis of the remaining here discussed sources could contribute to a more detailed discussion about the nature of the dusty-sources in the GC.
\end{itemize}

\section*{Acknowledgements}%We highly appreciate the comments of the anonymous referee that helped to improve this paper.
This work was supported in part by the
Deutsche Forschungsgemeinschaft (DFG) via the Cologne
Bonn Graduate School (BCGS), the Max Planck Society
through the International Max Planck Research School
(IMPRS) for Astronomy and Astrophysics, as well as special
funds through the University of Cologne and
SFB 956 – Conditions and Impact of Star Formation. Part of this
work was supported by fruitful discussions with members of
the European Union funded COST Action MP0905: Black
Holes in a Violent Universe and the Czech Science Foundation
-- DFG collaboration (No.\ 19-01137J). MZ acknowledges the support by the National
Science Centre in Poland (Maestro 9 grant No.\ 2017/26/A/ST9/00756). We also would like to 
thank the members of the SINFONI/NACO and ESO's Paranal/Chile team for their support and collaboration.

\bibliographystyle{aa}
\bibliography{bib.bib}

\begin{thebibliography}{91}
\expandafter\ifx\csname natexlab\endcsname\relax\def\natexlab#1{#1}\fi

\bibitem[{Alig {et~al.}(2013)Alig, Schartmann, Burkert, \& Dolag}]{alig2013}
Alig, C., Schartmann, M., Burkert, A., \& Dolag, K. 2013, Proceedings of the
  International Astronomical Union, 9, 185–187

\bibitem[{{Baraffe} {et~al.}(1998){Baraffe}, {Chabrier}, {Allard}, \&
  {Hauschildt}}]{1998A&A...337..403B}
{Baraffe}, I., {Chabrier}, G., {Allard}, F., \& {Hauschildt}, P.~H. 1998, \aap,
  337, 403

\bibitem[{{Barvainis}(1987)}]{1987ApJ...320..537B}
{Barvainis}, R. 1987, \apj, 320, 537

\bibitem[{{Bonnet} {et~al.}(2004){Bonnet}, {Abuter}, {Baker}, {Bornemann},
  {Brown}, {Castillo}, {Conzelmann}, {Damster}, {Davies}, {Delabre},
  {Donaldson}, {Dumas}, {Eisenhauer}, {Elswijk}, {Fedrigo}, {Finger},
  {Gemperlein}, {Genzel}, {Gilbert}, {Gillet}, {Goldbrunner}, {Horrobin}, {Ter
  Horst}, {Huber}, {Hubin}, {Iserlohe}, {Kaufer}, {Kissler-Patig}, {Kragt},
  {Kroes}, {Lehnert}, {Lieb}, {Liske}, {Lizon}, {Lutz}, {Modigliani}, {Monnet},
  {Nesvadba}, {Patig}, {Pragt}, {Reunanen}, {R{\"o}hrle}, {Rossi}, {Schmutzer},
  {Schoenmaker}, {Schreiber}, {Stroebele}, {Szeifert}, {Tacconi}, {Tecza},
  {Thatte}, {Tordo}, {van der Werf}, \& {Weisz}}]{Bonnet2004}
{Bonnet}, H., {Abuter}, R., {Baker}, A., {et~al.} 2004, The Messenger, 117, 17

\bibitem[{Bouchet {et~al.}(2015)Bouchet, Garc{\'{\i}}a-Mar{\'{\i}}n, Lagage,
  Amiaux, Augu{\'{e}}res, Bauwens, Blommaert, Chen, Detre, Dicken, Dubreuil,
  Galdemard, Gastaud, Glasse, Gordon, Gougnaud, Guillard, Justtanont, Krause,
  Leboeuf, Longval, Martin, Mazy, Moreau, Olofsson, Ray, Rees, Renotte,
  Ressler, Ronayette, Salasca, Scheithauer, Sykes, Thelen, Wells, Wright, \&
  Wright}]{Bouchet_2015}
Bouchet, P., Garc{\'{\i}}a-Mar{\'{\i}}n, M., Lagage, P.-O., {et~al.} 2015,
  Publications of the Astronomical Society of the Pacific, 127, 612

\bibitem[{{Brice{\~n}o} {et~al.}(2007){Brice{\~n}o}, {Preibisch}, {Sherry},
  {Mamajek}, {Mathieu}, {Walter}, \& {Zinnecker}}]{2007prpl.conf..345B}
{Brice{\~n}o}, C., {Preibisch}, T., {Sherry}, W.~H., {et~al.} 2007, in
  Protostars and Planets V, ed. B.~{Reipurth}, D.~{Jewitt}, \& K.~{Keil}, 345

\bibitem[{{Burkert} {et~al.}(2012){Burkert}, {Schartmann}, {Alig}, {Gillessen},
  {Genzel}, {Fritz}, \& {Eisenhauer}}]{Burkert2012}
{Burkert}, A., {Schartmann}, M., {Alig}, C., {et~al.} 2012, \apj, 750, 58

\bibitem[{{Calder{\'o}n} {et~al.}(2018){Calder{\'o}n}, {Cuadra}, {Schartmann},
  {Burkert}, {Plewa}, {Eisenhauer}, \& {Habibi}}]{Calderon2018}
{Calder{\'o}n}, D., {Cuadra}, J., {Schartmann}, M., {et~al.} 2018, \mnras, 478,
  3494

\bibitem[{{Chen} {et~al.}(2014){Chen}, {Girardi}, {Bressan}, {Marigo},
  {Barbieri}, \& {Kong}}]{chen2014}
{Chen}, Y., {Girardi}, L., {Bressan}, A., {et~al.} 2014, \mnras, 444, 2525

\bibitem[{{Cl{\'e}net} {et~al.}(2001){Cl{\'e}net}, {Rouan}, {Gendron},
  {Montri}, {Rigaut}, {L{\'e}na}, \& {Lacombe}}]{Clenet2001}
{Cl{\'e}net}, Y., {Rouan}, D., {Gendron}, E., {et~al.} 2001, \aap, 376, 124

\bibitem[{{Cl{\'e}net} {et~al.}(2005){Cl{\'e}net}, {Rouan}, {Gratadour},
  {Marco}, {L{\'e}na}, {Ageorges}, \& {Gendron}}]{Clenet2005}
{Cl{\'e}net}, Y., {Rouan}, D., {Gratadour}, D., {et~al.} 2005, \aap, 439, L9

\bibitem[{{Cl{\'e}net} {et~al.}(2003){Cl{\'e}net}, {Rouan}, {Lacombe},
  {Gendron}, \& {Gratadour}}]{Clenet2003}
{Cl{\'e}net}, Y., {Rouan}, D., {Lacombe}, F., {Gendron}, E., \& {Gratadour}, D.
  2003, Astronomische Nachrichten Supplement, 324, 327

\bibitem[{{Cotera} {et~al.}(1999){Cotera}, {Morris}, {Ghez}, {Becklin},
  {Tanner}, {Werner}, \& {Stolovy}}]{1999ASPC..186..240C}
{Cotera}, A., {Morris}, M., {Ghez}, A.~M., {et~al.} 1999, in Astronomical
  Society of the Pacific Conference Series, Vol. 186, The Central Parsecs of
  the Galaxy, ed. H.~{Falcke}, A.~{Cotera}, W.~J. {Duschl}, F.~{Melia}, \&
  M.~J. {Rieke}, 240

\bibitem[{{Davies}(2007)}]{Davies2007}
{Davies}, R.~I. 2007, \mnras, 375, 1099

\bibitem[{{Depoy}(1992)}]{Depoy1992}
{Depoy}, D.~L. 1992, \apj, 398, 512

\bibitem[{{Depoy} \& {Pogge}(1994)}]{DepoyPogge1994}
{Depoy}, D.~L. \& {Pogge}, R.~W. 1994, \apj, 433, 725

\bibitem[{{Dodds-Eden} {et~al.}(2011){Dodds-Eden}, {Gillessen}, {Fritz},
  {Eisenhauer}, {Trippe}, {Genzel}, {Ott}, {Bartko}, {Pfuhl}, {Bower},
  {Goldwurm}, {Porquet}, {Trap}, \& {Yusef-Zadeh}}]{Dodds-Eden2011}
{Dodds-Eden}, K., {Gillessen}, S., {Fritz}, T.~K., {et~al.} 2011, \apj, 728, 37

\bibitem[{{Eckart} \& {Genzel}(1996)}]{Eckart1996a}
{Eckart}, A. \& {Genzel}, R. 1996, 102, 196

\bibitem[{{Eckart} {et~al.}(1992){Eckart}, {Genzel}, {Krabbe}, {Hofmann}, {van
  der Werf}, \& {Drapatz}}]{Eckart1992}
{Eckart}, A., {Genzel}, R., {Krabbe}, A., {et~al.} 1992, \nat, 355, 526

\bibitem[{{Eckart} {et~al.}(2017){Eckart}, {H{\"u}ttemann}, {Kiefer},
  {Britzen}, {Zaja{\v{c}}ek}, {L{\"a}mmerzahl}, {St{\"o}ckler}, {Valencia-S},
  {Karas}, \& {Garc{\'\i}a-Mar{\'\i}n}}]{2017FoPh...47..553E}
{Eckart}, A., {H{\"u}ttemann}, A., {Kiefer}, C., {et~al.} 2017, Foundations of
  Physics, 47, 553

\bibitem[{{Eckart, A.} {et~al.}(2013){Eckart, A.}, {Muzi\'{}c, K.}, {Yazici,
  S.}, {Sabha, N.}, {Shahzamanian, B.}, {Witzel, G.}, {Moser, L.},
  {Garcia-Marin, M.}, {, M. Valencia-S.}, {Jalali, B.}, {Bremer, M.},
  {Straubmeier, C.}, {Rauch, C.}, {Buchholz, R.}, {Kunneriath, D.}, \&
  {Moultaka, J.}}]{EckartAA2013}
{Eckart, A.}, {Muzi\'{}c, K.}, {Yazici, S.}, {et~al.} 2013, A\&A, 551, A18

\bibitem[{{Eisenhauer} {et~al.}(2003){Eisenhauer}, {Abuter}, {Bickert},
  {Biancat-Marchet}, {Bonnet}, {Brynnel}, {Conzelmann}, {Delabre}, {Donaldson},
  {Farinato}, {Fedrigo}, {Genzel}, {Hubin}, {Iserlohe}, {Kasper},
  {Kissler-Patig}, {Monnet}, {Roehrle}, {Schreiber}, {Stroebele}, {Tecza},
  {Thatte}, \& {Weisz}}]{Eisenhauer2003}
{Eisenhauer}, F., {Abuter}, R., {Bickert}, K., {et~al.} 2003, in \procspie,
  Vol. 4841, Instrument Design and Performance for Optical/Infrared
  Ground-based Telescopes, ed. M.~{Iye} \& A.~F.~M. {Moorwood}, 1548--1561

\bibitem[{{Eisenhauer} {et~al.}(2005){Eisenhauer}, {Genzel}, {Alexander},
  {Abuter}, {Paumard}, {Ott}, {Gilbert}, {Gillessen}, {Horrobin}, {Trippe},
  {Bonnet}, {Dumas}, {Hubin}, {Kaufer}, {Kissler-Patig}, {Monnet},
  {Str{\"o}bele}, {Szeifert}, {Eckart}, {Sch{\"o}del}, \&
  {Zucker}}]{Eisenhauer2005}
{Eisenhauer}, F., {Genzel}, R., {Alexander}, T., {et~al.} 2005, apj, 628, 246

\bibitem[{{Fritz} {et~al.}(2011){Fritz}, {Gillessen}, {Dodds-Eden}, {Lutz},
  {Genzel}, {Raab}, {Ott}, {Pfuhl}, {Eisenhauer}, \& {Yusef-Zadeh}}]{Fritz2011}
{Fritz}, T.~K., {Gillessen}, S., {Dodds-Eden}, K., {et~al.} 2011, ApJ, 737, 73

\bibitem[{{Genzel} {et~al.}(2010){Genzel}, {Eisenhauer}, \&
  {Gillessen}}]{Genzel2010}
{Genzel}, R., {Eisenhauer}, F., \& {Gillessen}, S. 2010, Reviews of Modern
  Physics, 82, 3121

\bibitem[{{Ghez} {et~al.}(2003){Ghez}, {Duch{\^e}ne}, {Matthews}, {Hornstein},
  {Tanner}, {Larkin}, {Morris}, {Becklin}, {Salim}, {Kremenek}, {Thompson},
  {Soifer}, {Neugebauer}, \& {McLean}}]{Ghez2003}
{Ghez}, A.~M., {Duch{\^e}ne}, G., {Matthews}, K., {et~al.} 2003, \apjl, 586,
  L127

\bibitem[{{Ghez} {et~al.}(1998){Ghez}, {Klein}, {Morris}, \&
  {Becklin}}]{Ghez1998}
{Ghez}, A.~M., {Klein}, B.~L., {Morris}, M., \& {Becklin}, E.~E. 1998, \apj,
  509, 678

\bibitem[{{Gillessen} {et~al.}(2009){Gillessen}, {Eisenhauer}, {Trippe},
  {Alexander}, {Genzel}, {Martins}, \& {Ott}}]{Gillessen2009}
{Gillessen}, S., {Eisenhauer}, F., {Trippe}, S., {et~al.} 2009, \apj, 692, 1075

\bibitem[{{Gillessen} {et~al.}(2013{\natexlab{a}}){Gillessen}, {Genzel},
  {Fritz}, {Eisenhauer}, {Pfuhl}, {Ott}, {Schartmann}, {Ballone}, \&
  {Burkert}}]{Gillessen2013a}
{Gillessen}, S., {Genzel}, R., {Fritz}, T.~K., {et~al.} 2013{\natexlab{a}},
  \apj, 774, 44

\bibitem[{{Gillessen} {et~al.}(2013{\natexlab{b}}){Gillessen}, {Genzel},
  {Fritz}, {Eisenhauer}, {Pfuhl}, {Ott}, {Schartmann}, {Ballone}, \&
  {Burkert}}]{Gillessen2013b}
{Gillessen}, S., {Genzel}, R., {Fritz}, T.~K., {et~al.} 2013{\natexlab{b}},
  \apj, 774, 44

\bibitem[{{Gillessen} {et~al.}(2012){Gillessen}, {Genzel}, {Fritz}, {Quataert},
  {Alig}, {Burkert}, {Cuadra}, {Eisenhauer}, {Pfuhl}, {Dodds-Eden}, {Gammie},
  \& {Ott}}]{Gillessen2012}
{Gillessen}, S., {Genzel}, R., {Fritz}, T.~K., {et~al.} 2012, \nat, 481, 51

\bibitem[{{Gillessen} {et~al.}(2019){Gillessen}, {Plewa}, {Widmann}, {von
  Fellenberg}, {Schartmann}, {Habibi}, {Jimenez Rosales}, {Baub{\"o}ck},
  {Dexter}, {Gao}, {Waisberg}, {Eisenhauer}, {Pfuhl}, {Ott}, {Burkert}, {de
  Zeeuw}, \& {Genzel}}]{Gillessen2019}
{Gillessen}, S., {Plewa}, P.~M., {Widmann}, F., {et~al.} 2019, \apj, 871, 126

\bibitem[{{Gravity Collaboration} {et~al.}(2018){Gravity Collaboration},
  {Abuter}, {Amorim}, {Anugu}, {Baub{\"o}ck}, {Benisty}, {Berger}, {Blind},
  {Bonnet}, {Brandner}, {Buron}, {Collin}, {Chapron}, {Cl{\'e}net}, {Coud{\'e}
  Du Foresto}, {de Zeeuw}, {Deen}, {Delplancke-Str{\"o}bele}, {Dembet},
  {Dexter}, {Duvert}, {Eckart}, {Eisenhauer}, {Finger}, {F{\"o}rster
  Schreiber}, {F{\'e}dou}, {Garcia}, {Garcia Lopez}, {Gao}, {Gendron},
  {Genzel}, {Gillessen}, {Gordo}, {Habibi}, {Haubois}, {Haug}, {Hau{\ss}mann},
  {Henning}, {Hippler}, {Horrobin}, {Hubert}, {Hubin}, {Jimenez Rosales},
  {Jochum}, {Jocou}, {Kaufer}, {Kellner}, {Kendrew}, {Kervella}, {Kok},
  {Kulas}, {Lacour}, {Lapeyr{\`e}re}, {Lazareff}, {Le Bouquin}, {L{\'e}na},
  {Lippa}, {Lenzen}, {M{\'e}rand}, {M{\"u}ler}, {Neumann}, {Ott}, {Palanca},
  {Paumard}, {Pasquini}, {Perraut}, {Perrin}, {Pfuhl}, {Plewa}, {Rabien},
  {Ram{\'\i}rez}, {Ramos}, {Rau}, {Rodr{\'\i}guez-Coira}, {Rohloff}, {Rousset},
  {Sanchez-Bermudez}, {Scheithauer}, {Sch{\"o}ller}, {Schuler}, {Spyromilio},
  {Straub}, {Straubmeier}, {Sturm}, {Tacconi}, {Tristram}, {Vincent}, {von
  Fellenberg}, {Wank}, {Waisberg}, {Widmann}, {Wieprecht}, {Wiest},
  {Wiezorrek}, {Woillez}, {Yazici}, {Ziegler}, \& {Zins}}]{gravity2018}
{Gravity Collaboration}, {Abuter}, R., {Amorim}, A., {et~al.} 2018, \aap, 615,
  L15

\bibitem[{{Gravity Collaboration} {et~al.}(2019){Gravity Collaboration},
  {Abuter}, {Amorim}, {Baub{\"o}ck}, {Berger}, {Bonnet}, {Brand ner},
  {Cl{\'e}net}, {Coud{\'e} Du Foresto}, {de Zeeuw}, {Dexter}, {Duvert},
  {Eckart}, {Eisenhauer}, {F{\"o}rster Schreiber}, {Garcia}, {Gao}, {Gendron},
  {Genzel}, {Gerhard}, {Gillessen}, {Habibi}, {Haubois}, {Henning}, {Hippler},
  {Horrobin}, {Jim{\'e}nez-Rosales}, {Jocou}, {Kervella}, {Lacour},
  {Lapeyr{\`e}re}, {Le Bouquin}, {L{\'e}na}, {Ott}, {Paumard}, {Perraut},
  {Perrin}, {Pfuhl}, {Rabien}, {Rodriguez Coira}, {Rousset}, {Scheithauer},
  {Sternberg}, {Straub}, {Straubmeier}, {Sturm}, {Tacconi}, {Vincent}, {von
  Fellenberg}, {Waisberg}, {Widmann}, {Wieprecht}, {Wiezorrek}, {Woillez}, \&
  {Yazici}}]{Gravity2019}
{Gravity Collaboration}, {Abuter}, R., {Amorim}, A., {et~al.} 2019, \aap, 625,
  L10

\bibitem[{{Habibi} {et~al.}(2017){Habibi}, {Gillessen}, {Martins},
  {Eisenhauer}, {Plewa}, {Pfuhl}, {George}, {Dexter}, {Waisberg}, {Ott}, {von
  Fellenberg}, {Baub{\"o}ck}, {Jimenez-Rosales}, \& {Genzel}}]{Habibi2017}
{Habibi}, M., {Gillessen}, S., {Martins}, F., {et~al.} 2017, \apj, 847, 120

\bibitem[{{Jalali} {et~al.}(2014){Jalali}, {Pelupessy}, {Eckart}, {Portegies
  Zwart}, {Sabha}, {Borkar}, {Moultaka}, {Mu{\v z}i{\'c}}, \&
  {Moser}}]{Jalali2014}
{Jalali}, B., {Pelupessy}, F.~I., {Eckart}, A., {et~al.} 2014, \mnras, 444,
  1205

\bibitem[{{Jones}(2000)}]{Jones2000}
{Jones}, A.~P. 2000, \jgr, 105, 10257

\bibitem[{{Kamath} {et~al.}(2014){Kamath}, {Wood}, \& {Van
  Winckel}}]{Kamath2014}
{Kamath}, D., {Wood}, P.~R., \& {Van Winckel}, H. 2014, \mnras, 439, 2211

\bibitem[{{Lada}(1987)}]{Lada1987}
{Lada}, C.~J. 1987, in IAU Symposium, Vol. 115, Star Forming Regions, ed.
  M.~{Peimbert} \& J.~{Jugaku}, 1

\bibitem[{{Lenzen} {et~al.}(2003){Lenzen}, {Hartung}, {Brandner}, {Finger},
  {Hubin}, {Lacombe}, {Lagrange}, {Lehnert}, {Moorwood}, \&
  {Mouillet}}]{Lenzen2003}
{Lenzen}, R., {Hartung}, M., {Brandner}, W., {et~al.} 2003, in \procspie, Vol.
  4841, Instrument Design and Performance for Optical/Infrared Ground-based
  Telescopes, ed. M.~{Iye} \& A.~F.~M. {Moorwood}, 944--952

\bibitem[{{Levin} \& {Beloborodov}(2003)}]{Levin}
{Levin}, Y. \& {Beloborodov}, A.~M. 2003, \apjl, 590, L33

\bibitem[{{Luck} {et~al.}(1998){Luck}, {Moffett}, {Barnes}, \&
  {Gieren}}]{Luck1998}
{Luck}, R.~E., {Moffett}, T.~J., {Barnes}, Thomas~G., I., \& {Gieren}, W.~P.
  1998, \aj, 115, 605

\bibitem[{{Lucy}(1974)}]{Lucy1974}
{Lucy}, L.~B. 1974, \aj, 79, 745

\bibitem[{{Lutz} {et~al.}(1993){Lutz}, {Krabbe}, \& {Genzel}}]{Lutz1993}
{Lutz}, D., {Krabbe}, A., \& {Genzel}, R. 1993, \apj, 418, 244

\bibitem[{{Meyer} \& {Meyer-Hofmeister}(2012)}]{Meyer2012}
{Meyer}, F. \& {Meyer-Hofmeister}, E. 2012, aap, 546, L2

\bibitem[{{Meyer} {et~al.}(2014){Meyer}, {Ghez}, {Witzel}, {Do}, {Phifer},
  {Sitarski}, {Morris}, {Boehle}, {Yelda}, {Lu}, \& {Becklin}}]{Meyer2014}
{Meyer}, L., {Ghez}, A.~M., {Witzel}, G., {et~al.} 2014, 303, 264

\bibitem[{{Modigliani} {et~al.}(2007){Modigliani}, {Hummel}, {Abuter}, {Amico},
  {Ballester}, {Davies}, {Dumas}, {Horrobin}, {Neeser}, {Kissler-Patig},
  {Peron}, {Rehunanen}, {Schreiber}, \& {Szeifert}}]{Modigliani2007}
{Modigliani}, A., {Hummel}, W., {Abuter}, R., {et~al.} 2007, ArXiv Astrophysics
  e-prints [\eprint{astro-ph/0701297}]

\bibitem[{{Morris} \& {Serabyn}(1996)}]{Morris1996}
{Morris}, M. \& {Serabyn}, E. 1996, \araa, 34, 645

\bibitem[{{Moultaka} {et~al.}(2009){Moultaka}, {Eckart}, \&
  {Sch{\"o}del}}]{Moultaka2009}
{Moultaka}, J., {Eckart}, A., \& {Sch{\"o}del}, R. 2009, \apj, 703, 1635

\bibitem[{{Moultaka} {et~al.}(2006){Moultaka}, {Eckart}, {Viehmann}, \&
  {Sch{\"o}del}}]{Moultaka2006}
{Moultaka}, J., {Eckart}, A., {Viehmann}, T., \& {Sch{\"o}del}, R. 2006, in
  Journal of Physics Conference Series, Vol.~54, Journal of Physics Conference
  Series, ed. R.~{Sch{\"o}del}, G.~C. {Bower}, M.~P. {Muno}, S.~{Nayakshin}, \&
  T.~{Ott}, 57--61

\bibitem[{{Murchikova} {et~al.}(2019){Murchikova}, {Phinney}, {Pancoast}, \&
  {Blandford}}]{Murchikova2019}
{Murchikova}, E.~M., {Phinney}, E.~S., {Pancoast}, A., \& {Blandford}, R.~D.
  2019, \nat, 570, 83

\bibitem[{{Mu{\v z}i{\'c}} {et~al.}(2010){Mu{\v z}i{\'c}}, {Eckart},
  {Sch{\"o}del}, {Buchholz}, {Zamaninasab}, \& {Witzel}}]{muzic2010}
{Mu{\v z}i{\'c}}, K., {Eckart}, A., {Sch{\"o}del}, R., {et~al.} 2010, \aap,
  521, A13

\bibitem[{{Mu{\v z}i{\'c}} {et~al.}(2007){Mu{\v z}i{\'c}}, {Eckart},
  {Sch{\"o}del}, {Meyer}, \& {Zensus}}]{muzic2007}
{Mu{\v z}i{\'c}}, K., {Eckart}, A., {Sch{\"o}del}, R., {Meyer}, L., \&
  {Zensus}, A. 2007, \aap, 469, 993

\bibitem[{{Mu{\v{z}}i{\'c}} {et~al.}(2008){Mu{\v{z}}i{\'c}}, {Sch{\"o}del},
  {Eckart}, {Meyer}, \& {Zensus}}]{muzic2008}
{Mu{\v{z}}i{\'c}}, K., {Sch{\"o}del}, R., {Eckart}, A., {Meyer}, L., \&
  {Zensus}, A. 2008, \aap, 482, 173

\bibitem[{{Naoz} {et~al.}(2018){Naoz}, {Ghez}, {Hees}, {Do}, {Witzel}, \&
  {Lu}}]{Naoz2018}
{Naoz}, S., {Ghez}, A.~M., {Hees}, A., {et~al.} 2018, \apjl, 853, L24

\bibitem[{{Oliveira} {et~al.}(2013){Oliveira}, {van Loon}, {Sloan},
  {Sewi{\l}o}, {Kraemer}, {Wood}, {Indebetouw}, {Filipovi{\'c}}, {Crawford},
  {Wong}, {Hora}, {Meixner}, {Robitaille}, {Shiao}, \& {Simon}}]{Oliveira2013}
{Oliveira}, J.~M., {van Loon}, J.~T., {Sloan}, G.~C., {et~al.} 2013, \mnras,
  428, 3001

\bibitem[{{Parsa} {et~al.}(2017){Parsa}, {Eckart}, {Shahzamanian}, {Karas},
  {Zaja{\v c}ek}, {Zensus}, \& {Straubmeier}}]{Parsa2017}
{Parsa}, M., {Eckart}, A., {Shahzamanian}, B., {et~al.} 2017, \apj, 845, 22

\bibitem[{{Pei{\ss}ker} {et~al.}(2019){Pei{\ss}ker}, {Zaja{\v{c}}ek}, {Eckart},
  {Sabha}, {Shahzamanian}, \& {Parsa}}]{Peissker2019}
{Pei{\ss}ker}, F., {Zaja{\v{c}}ek}, M., {Eckart}, A., {et~al.} 2019, \aap, 624,
  A97

\bibitem[{{Pfuhl} {et~al.}(2015){Pfuhl}, {Gillessen}, {Eisenhauer}, {Genzel},
  {Plewa}, {Ott}, {Ballone}, {Schartmann}, {Burkert}, {Fritz}, {Sari},
  {Steinberg}, \& {Madigan}}]{Pfuhl2015}
{Pfuhl}, O., {Gillessen}, S., {Eisenhauer}, F., {et~al.} 2015, \apj, 798, 111

\bibitem[{{Plewa} {et~al.}(2015){Plewa}, {Gillessen}, {Eisenhauer}, {Ott},
  {Pfuhl}, {George}, {Dexter}, {Habibi}, {Genzel}, {Reid}, \&
  {Menten}}]{Plewa2015}
{Plewa}, P.~M., {Gillessen}, S., {Eisenhauer}, F., {et~al.} 2015, \mnras, 453,
  3234

\bibitem[{{Plewa} {et~al.}(2017){Plewa}, {Gillessen}, {Pfuhl}, {Eisenhauer},
  {Genzel}, {Burkert}, {Dexter}, {Habibi}, {George}, {Ott}, {Waisberg}, \& {von
  Fellenberg}}]{Plewa2017}
{Plewa}, P.~M., {Gillessen}, S., {Pfuhl}, O., {et~al.} 2017, \apj, 840, 50

\bibitem[{{Pych}(2004)}]{Pych2004}
{Pych}, W. 2004, \pasp, 116, 148

\bibitem[{Ressler {et~al.}(2015)Ressler, Sukhatme, Franklin, Mahoney, Thelen,
  Bouchet, Colbert, Cracraft, Dicken, Gastaud, Goodson, Eccleston, Moreau,
  Rieke, \& Schneider}]{Ressler_2015}
Ressler, M.~E., Sukhatme, K.~G., Franklin, B.~R., {et~al.} 2015, Publications
  of the Astronomical Society of the Pacific, 127, 675

\bibitem[{Rieke {et~al.}(2015)Rieke, Wright, Böker, Bouwman, Colina, Glasse,
  Gordon, Greene, Güdel, Henning, Justtanont, Lagage, Meixner,
  N{\o}rgaard-Nielsen, Ray, Ressler, van Dishoeck, \& Waelkens}]{Rieke_2015}
Rieke, G.~H., Wright, G.~S., Böker, T., {et~al.} 2015, Publications of the
  Astronomical Society of the Pacific, 127, 584

\bibitem[{{Rousset} {et~al.}(2003){Rousset}, {Lacombe}, {Puget}, {Hubin},
  {Gendron}, {Fusco}, {Arsenault}, {Charton}, {Feautrier}, {Gigan}, {Kern},
  {Lagrange}, {Madec}, {Mouillet}, {Rabaud}, {Rabou}, {Stadler}, \&
  {Zins}}]{Rousset2003}
{Rousset}, G., {Lacombe}, F., {Puget}, P., {et~al.} 2003, in \procspie, Vol.
  4839, Adaptive Optical System Technologies II, ed. P.~L. {Wizinowich} \&
  D.~{Bonaccini}, 140--149

\bibitem[{{Sabha} {et~al.}(2012){Sabha}, {Eckart}, {Merritt}, {Zamaninasab},
  {Witzel}, {Garc{\'{\i}}a-Mar{\'{\i}}n}, {Jalali}, {Valencia-S.}, {Yazici},
  {Buchholz}, {Shahzamanian}, {Rauch}, {Horrobin}, \&
  {Straubmeier}}]{Sabha2012}
{Sabha}, N., {Eckart}, A., {Merritt}, D., {et~al.} 2012, \aap, 545, A70

\bibitem[{{Schartmann} {et~al.}(2012){Schartmann}, {Burkert}, {Alig},
  {Gillessen}, {Genzel}, {Eisenhauer}, \& {Fritz}}]{Schartmann2012}
{Schartmann}, M., {Burkert}, A., {Alig}, C., {et~al.} 2012, \apj, 755, 155

\bibitem[{{Sch{\"o}del} {et~al.}(2010){Sch{\"o}del}, {Najarro}, {Muzic}, \&
  {Eckart}}]{Schoedel2010}
{Sch{\"o}del}, R., {Najarro}, F., {Muzic}, K., \& {Eckart}, A. 2010, \aap, 511,
  A18

\bibitem[{{Sch{\"o}del} {et~al.}(2002){Sch{\"o}del}, {Ott}, {Genzel},
  {Hofmann}, {Lehnert}, {Eckart}, {Mouawad}, {Alexander}, {Reid}, {Lenzen},
  {Hartung}, {Lacombe}, {Rouan}, {Gendron}, {Rousset}, {Lagrange}, {Brandner},
  {Ageorges}, {Lidman}, {Moorwood}, {Spyromilio}, {Hubin}, \&
  {Menten}}]{Schoedel2002}
{Sch{\"o}del}, R., {Ott}, T., {Genzel}, R., {et~al.} 2002, \nat, 419, 694

\bibitem[{{Sciortino} {et~al.}(2019){Sciortino}, {Flaccomio}, {Pillitteri}, \&
  {Reale}}]{Sciortino2019}
{Sciortino}, S., {Flaccomio}, E., {Pillitteri}, I., \& {Reale}, F. 2019,
  Astronomische Nachrichten, 340, 334

\bibitem[{{Scoville} \& {Burkert}(2013)}]{Scoville2013}
{Scoville}, N. \& {Burkert}, A. 2013, apj, 768, 108

\bibitem[{{Shahzamanian} {et~al.}(2015){Shahzamanian}, {Eckart}, {Valencia-S.},
  {Witzel}, {Zamaninasab}, {Zaja{\v{c}}ek}, {Sabha}, {Garc{\'\i}a-Mar{\'\i}n},
  {Karas}, {Peissker}, {Karssen}, {Parsa}, {Grosso}, {Mossoux}, {Porquet},
  {Jalali}, {Horrobin}, {Buchholz}, {Dov{\v{c}}iak}, {Kunneriath}, {Bursa},
  {Zensus}, {Sch{\"o}del}, {Moultaka}, \& {Straubmeier}}]{Shahzamanian2015}
{Shahzamanian}, B., {Eckart}, A., {Valencia-S.}, M., {et~al.} 2015, The
  Messenger, 159, 41

\bibitem[{{Shahzamanian} {et~al.}(2016){Shahzamanian}, {Eckart}, {Zaja{\v
  c}ek}, {Valencia-S.}, {Sabha}, {Moser}, {Parsa}, {Peissker}, \&
  {Straubmeier}}]{Shahzamanian2016}
{Shahzamanian}, B., {Eckart}, A., {Zaja{\v c}ek}, M., {et~al.} 2016, \aap, 593,
  A131

\bibitem[{{Shimonishi} {et~al.}(2016){Shimonishi}, {Onaka}, {Kawamura}, \&
  {Aikawa}}]{Shimonishi2016}
{Shimonishi}, T., {Onaka}, T., {Kawamura}, A., \& {Aikawa}, Y. 2016, \apj, 827,
  72

\bibitem[{{Siess} {et~al.}(2000){Siess}, {Dufour}, \& {Forestini}}]{Siess2000}
{Siess}, L., {Dufour}, E., \& {Forestini}, M. 2000, \aap, 358, 593

\bibitem[{{Tielens} {et~al.}(1994){Tielens}, {McKee}, {Seab}, \&
  {Hollenbach}}]{Tielens1994}
{Tielens}, A.~G.~G.~M., {McKee}, C.~F., {Seab}, C.~G., \& {Hollenbach}, D.~J.
  1994, \apj, 431, 321

\bibitem[{{Tokunaga} \& {Vacca}(2007)}]{Tokunaga2007}
{Tokunaga}, A.~T. \& {Vacca}, W.~D. 2007, in Astronomical Society of the
  Pacific Conference Series, Vol. 364, The Future of Photometric,
  Spectrophotometric and Polarimetric Standardization, ed. C.~{Sterken}, 409

\bibitem[{{Trani} {et~al.}(2016){Trani}, {Mapelli}, {Bressan}, {Pelupessy},
  {van Elteren}, \& {Portegies Zwart}}]{Trani2016}
{Trani}, A.~A., {Mapelli}, M., {Bressan}, A., {et~al.} 2016, \apj, 818, 29

\bibitem[{Tsuboi {et~al.}(2016)Tsuboi, Kitamura, Uehara, Miyawaki, \&
  Miyazaki}]{tsuboi2016}
Tsuboi, M., Kitamura, Y., Uehara, K., Miyawaki, R., \& Miyazaki, A. 2016,
  Proceedings of the International Astronomical Union, 11, 115–118

\bibitem[{{Valencia-S.} {et~al.}(2015){Valencia-S.}, {Eckart}, {Zaja{\v c}ek},
  {Peissker}, {Parsa}, {Grosso}, {Mossoux}, {Porquet}, {Jalali}, {Karas},
  {Yazici}, {Shahzamanian}, {Sabha}, {Saalfeld}, {Smajic}, {Grellmann},
  {Moser}, {Horrobin}, {Borkar}, {Garc{\'{\i}}a-Mar{\'{\i}}n}, {Dov{\v c}iak},
  {Kunneriath}, {Karssen}, {Bursa}, {Straubmeier}, \&
  {Bushouse}}]{Valencia-S.2015}
{Valencia-S.}, M., {Eckart}, A., {Zaja{\v c}ek}, M., {et~al.} 2015, \apj, 800,
  125

\bibitem[{{Witzel} {et~al.}(2012){Witzel}, {Eckart}, {Bremer}, {Zamaninasab},
  {Shahzamanian}, {Valencia-S.}, {Sch{\"o}del}, {Karas}, {Lenzen}, {Marchili},
  {Sabha}, {Garcia-Marin}, {Buchholz}, {Kunneriath}, \&
  {Straubmeier}}]{Witzel2012}
{Witzel}, G., {Eckart}, A., {Bremer}, M., {et~al.} 2012, \apjs, 203, 18

\bibitem[{{Witzel} {et~al.}(2014){Witzel}, {Ghez}, {Morris}, {Sitarski},
  {Boehle}, {Naoz}, {Campbell}, {Becklin}, {Canalizo}, {Chappell}, {Do}, {Lu},
  {Matthews}, {Meyer}, {Stockton}, {Wizinowich}, \& {Yelda}}]{Witzel2014}
{Witzel}, G., {Ghez}, A.~M., {Morris}, M.~R., {et~al.} 2014, \apjl, 796, L8

\bibitem[{{Witzel} {et~al.}(2017){Witzel}, {Sitarski}, {Ghez}, {Morris},
  {Hees}, {Do}, {Lu}, {Naoz}, {Boehle}, {Martinez}, {Chappell}, {Sch{\"o}del},
  {Meyer}, {Yelda}, {Becklin}, \& {Matthews}}]{Witzel2017}
{Witzel}, G., {Sitarski}, B.~N., {Ghez}, A.~M., {et~al.} 2017, \apj, 847, 80

\bibitem[{{Wollman} {et~al.}(1977){Wollman}, {Geballe}, {Lacy}, {Townes}, \&
  {Rank}}]{Wollman1977}
{Wollman}, E.~R., {Geballe}, T.~R., {Lacy}, J.~H., {Townes}, C.~H., \& {Rank},
  D.~M. 1977, \apjl, 218, L103

\bibitem[{Yoshikawa {et~al.}(2013)Yoshikawa, Nishiyama, Tamura, Ishii, \&
  Nagata}]{Yoshikawa2013}
Yoshikawa, T., Nishiyama, S., Tamura, M., Ishii, M., \& Nagata, T. 2013, The
  Astrophysical Journal, 778, 92

\bibitem[{{Yusef-Zadeh} {et~al.}(2012){Yusef-Zadeh}, {Arendt}, {Bushouse},
  {Cotton}, {Haggard}, {Pound}, {Roberts}, {Royster}, \&
  {Wardle}}]{Yusef-Zadeh2012}
{Yusef-Zadeh}, F., {Arendt}, R., {Bushouse}, H., {et~al.} 2012, \apjl, 758, L11

\bibitem[{{Yusef-Zadeh} {et~al.}(2017{\natexlab{a}}){Yusef-Zadeh},
  {Sch{\"o}del}, {Wardle}, {Bushouse}, {Cotton}, {Royster}, {Kunneriath},
  {Roberts}, \& {Gallego-Cano}}]{Yusef-Zadeh2017-ALMAVLA}
{Yusef-Zadeh}, F., {Sch{\"o}del}, R., {Wardle}, M., {et~al.}
  2017{\natexlab{a}}, \mnras, 470, 4209

\bibitem[{{Yusef-Zadeh} {et~al.}(2017{\natexlab{b}}){Yusef-Zadeh}, {Wardle},
  {Kunneriath}, {Royster}, {Wootten}, \& {Roberts}}]{Yusef-Zadeh2017}
{Yusef-Zadeh}, F., {Wardle}, M., {Kunneriath}, D., {et~al.} 2017{\natexlab{b}},
  \apjl, 850, L30

\bibitem[{{Zaja{\v c}ek} {et~al.}(2017){Zaja{\v c}ek}, {Britzen}, {Eckart},
  {Shahzamanian}, {Busch}, {Karas}, {Parsa}, {Peissker}, {Dov{\v c}iak},
  {Subroweit}, {Dinnbier}, \& {Zensus}}]{Zajacek2017}
{Zaja{\v c}ek}, M., {Britzen}, S., {Eckart}, A., {et~al.} 2017, \aap, 602, A121

\bibitem[{{Zaja{\v{c}}ek} {et~al.}(2014){Zaja{\v{c}}ek}, {Karas}, \&
  {Eckart}}]{Zajacek2014}
{Zaja{\v{c}}ek}, M., {Karas}, V., \& {Eckart}, A. 2014, \aap, 565, A17

\bibitem[{{Zhukovska} {et~al.}(2018){Zhukovska}, {Henning}, \&
  {Dobbs}}]{Zhukovska2018}
{Zhukovska}, S., {Henning}, T., \& {Dobbs}, C. 2018, \apj, 857, 94

\end{thebibliography}

\begin{appendix}
\section{Appendix}
\subsection{Orbit plot of D23 and D31}
\begin{figure*}
\begin{subfigure}{.9\textwidth}
\centering
\includegraphics[width=.9\textwidth]{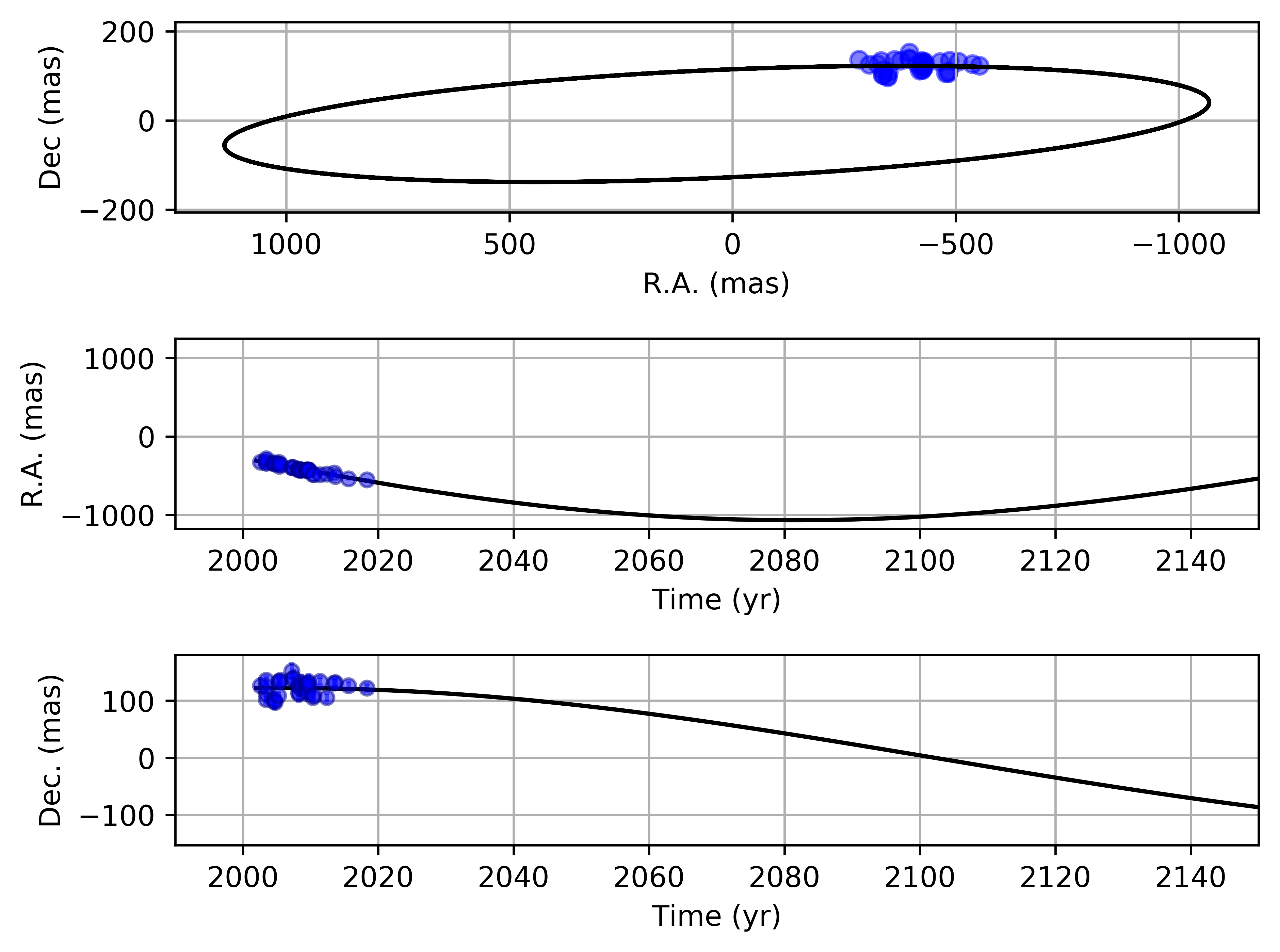}
%    \label{D23_orbit}
\end{subfigure}
\begin{subfigure}{.9\textwidth}
    \centering
\includegraphics[width=.91\textwidth]{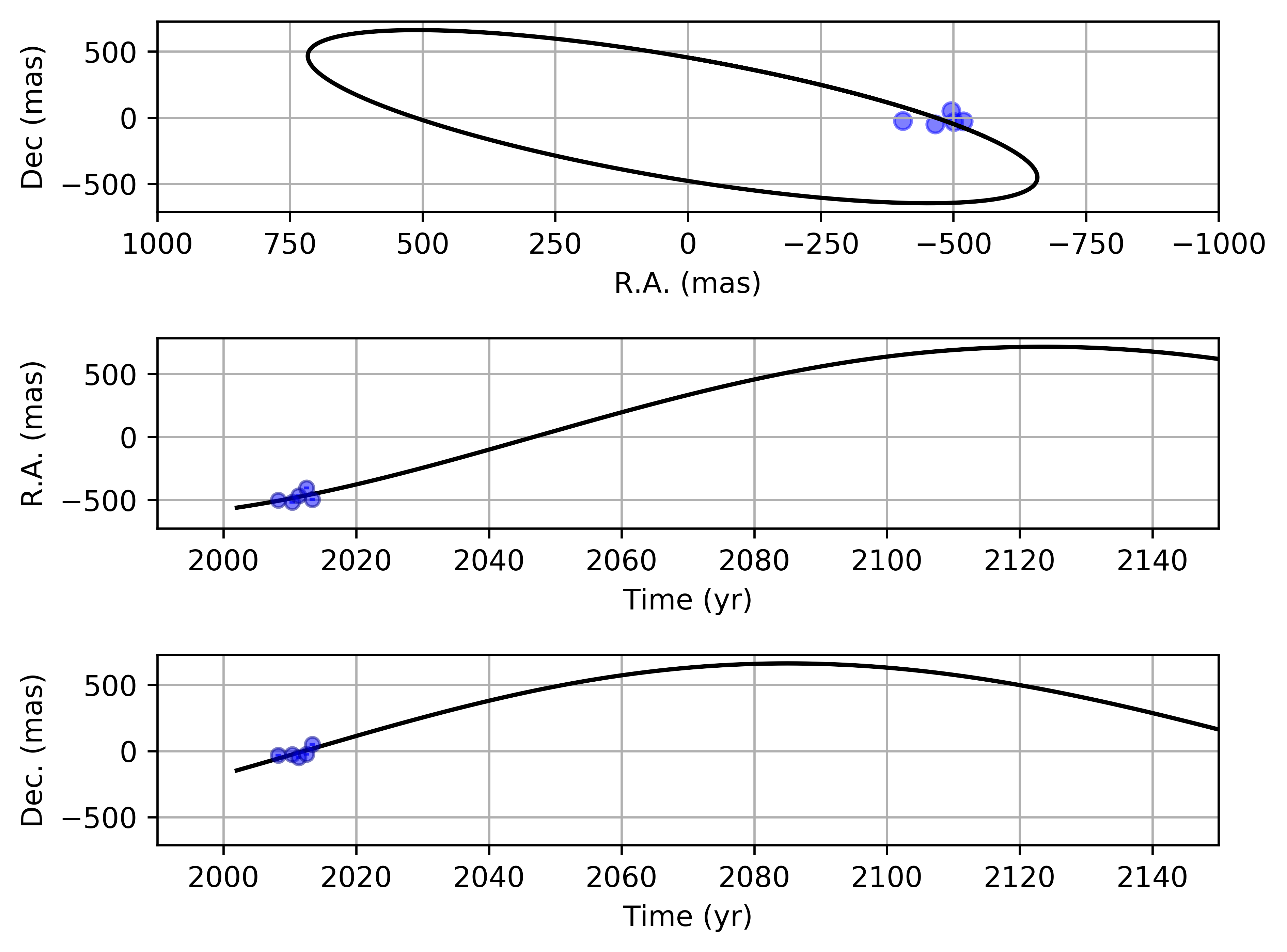}
%    \label{D31_orbit}
\end{subfigure}
\caption{Orbit overview of D23 and D3.1. The upper plot shows the D23-, the lower one the D3.1-orbit. The data is based on the Keplerian fit presented in Tab. \ref{tab:k_band_pos} and Tab. \ref{tab:L-band-pos}.}
\label{D23D31_orbit}
\end{figure*}
\subsection{Velocity plots of D2 and D3}
\begin{figure*}[htbp!]
\centering
\includegraphics[width=.55\textwidth]{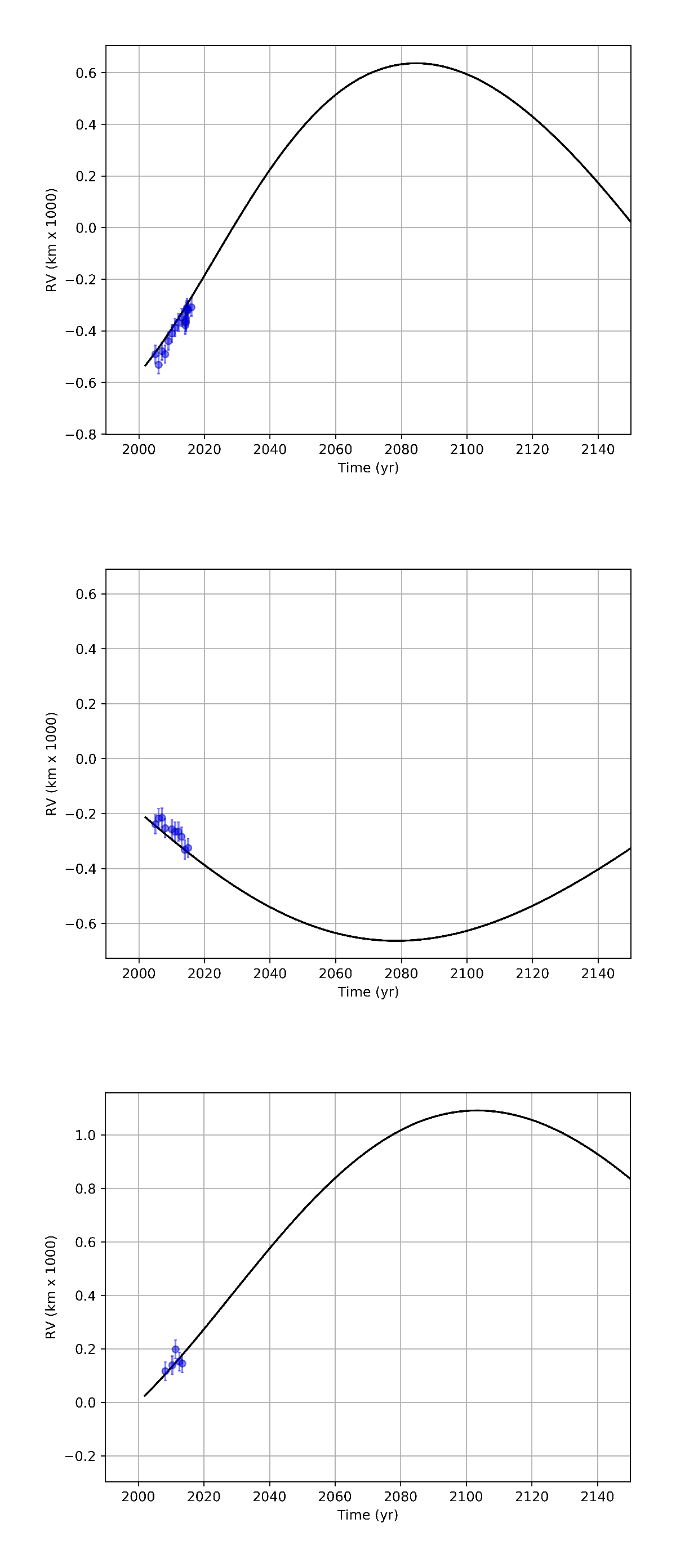}
    \label{D2_velocityorbit}
    \caption{Line of sight velocity fit of D2, D23, and D3.1. The upper plot shows the LOS fit of D2, the middle one is realted to D23, and the lower one the D3.1 fit. The data is extracted from the SINFONI data-cubes. Because the low amount of data for D3, the fit shows a poor response.}
\end{figure*}

\subsection{AIP ratio of the dusty sources}
\begin{figure*}[htbp!]
\centering
\includegraphics[width=1.\textwidth]{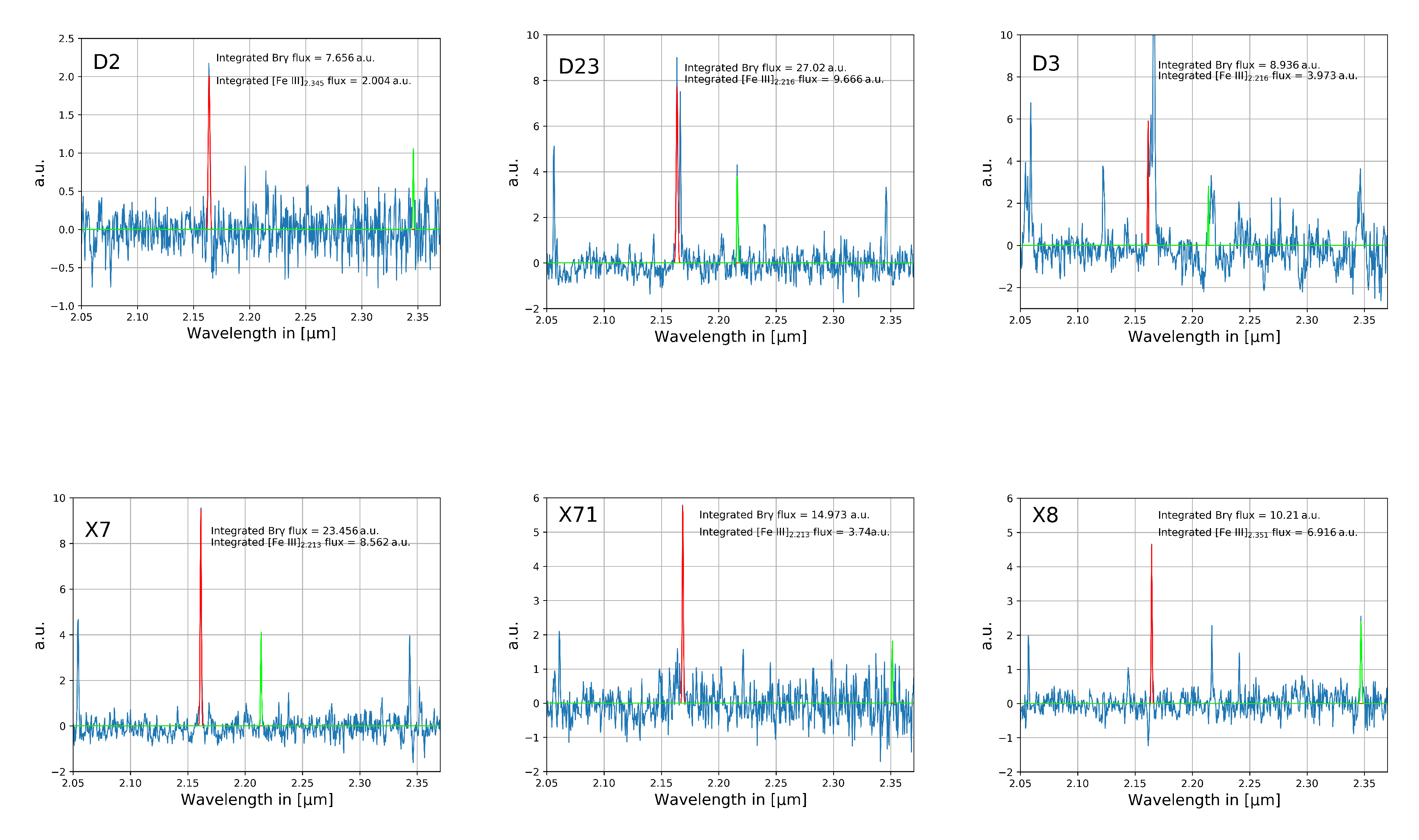}
\caption{Gaussian fit of the Doppler shifted Br$\gamma$ and [Fe III] line of D2, D23, D3, X7, X7.1, and X8. The normalized spectrum can be used to determine the [Fe III]/Br$_{\gamma}$ ratio.}
\label{fig:metallicity}
\end{figure*}

\section{Appendix}

\subsection{[FeIII] distribution of the D-sources}

\begin{figure*}[htbp!]
\centering
\includegraphics[width=.7\textwidth]{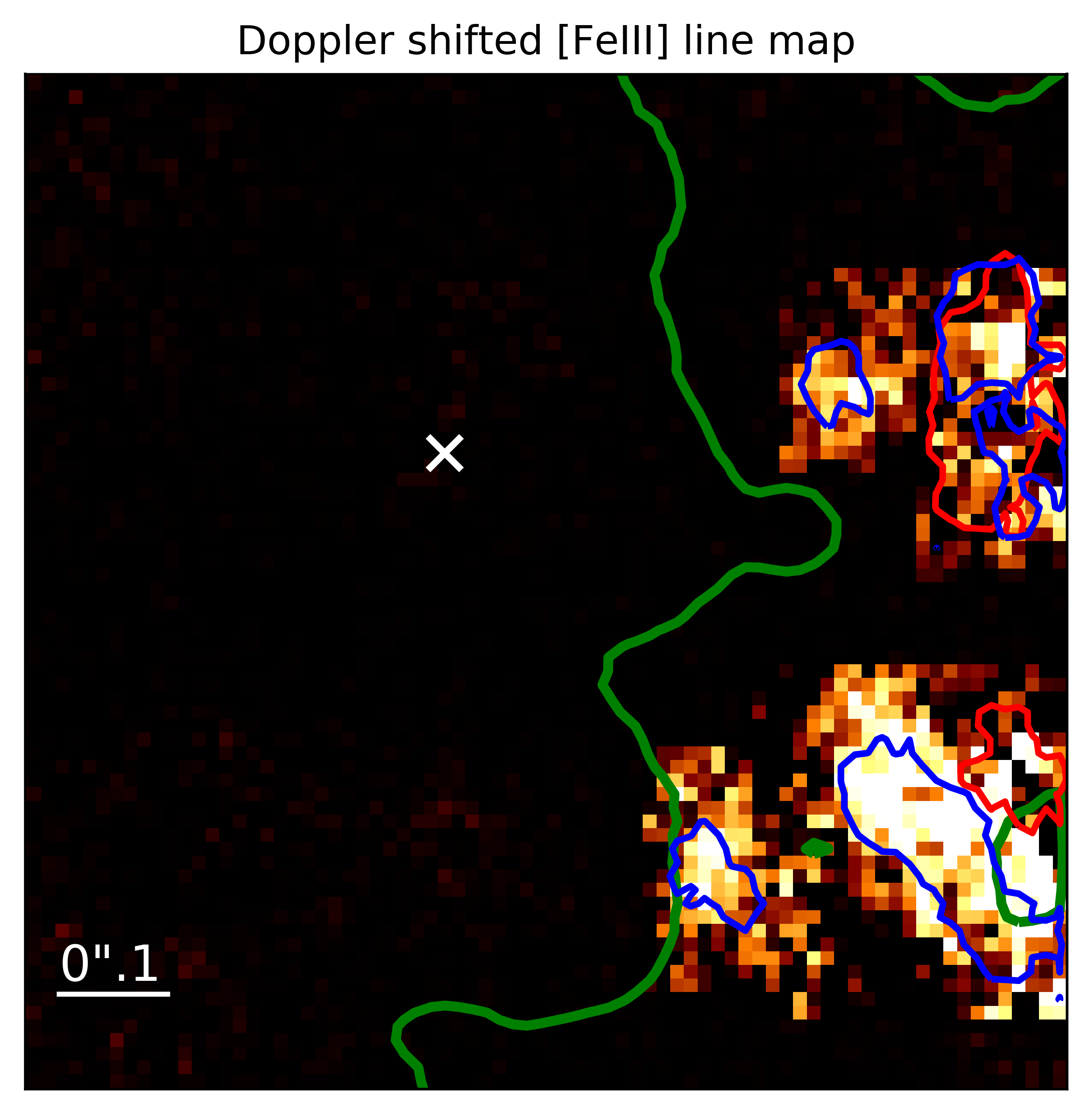}
\caption{[FeIII] emission related to Fig. \ref{fig:fe_line}. The FOV is consistent with the mentioned figure. The investigated sources west of SgrA* show a Doppler-shifted iron line emission that coincides with the Br$\gamma$ position of the D-sources.}
\label{fig:feIIIdistriGC}
\end{figure*}

\section{Appendix}

\subsection{NACO data}

\begin{table*}[hbt!]
\centering
\begin{tabular}{ccccc}
\hline
\hline
\multicolumn{5}{c}{NACO}\\
\hline
Date & Observation ID & \multicolumn{1}{p{1.5cm}}{\centering number \\ of exposures }   & \multicolumn{1}{p{1.5cm}}{\centering Total \\ exposure time(s) } & $\lambda$ \\
\hline
2002.07.31 & 60.A-9026(A) & 61 & 915    & K  \\
2002-08-19 &	60.A-9026(A)  &	76     &2280&  L$'$ \\
2002-08-30 &	60.A-9026(A)  &	80     &2400&  L$'$ \\
2003-05-10 &	71.B-0077(A)  &	56     & 2104& L$'$  \\
2004-04-25 &	073.B-0840(A) &	166    & 1743& L$'$  \\
2004-04-26 &	073.B-0840(A) &	263    & 3024& L$'$ \\
2005-05-14 &	073.B-0085(D) &	192    &5760&  L$'$  \\
2006-05-29 &	077.B-0552(A) &	522   &15660&  L$'$  \\
2007-04-01 &	179.B-0261(A) &	120    &1800&  L$'$  \\
2007-05-22 &	079.B-0084(A) &	30     &450	&  L$'$  \\
2008-05-26 &	081.B-0648(A) &	205    &3075&  L$'$  \\
2011.05.27 &   087.B-0017(A)  & 305    &4575& K    \\
2012-05-16 &	089.B-0145(A) &	30     &450	& L$'$   \\
2012.05.17 &    089.B-0145(A) & 169    & 2525 & K  \\
2013-05-09 &	091.C-0159(A) &	30     &450	& L$'$   \\
2013.06.28 & 091.B-0183(A) & 112 & 1680 & K          \\      
2013.09.01 & 091.B-0183(B) & 359 & 3590 & K          \\      
2015.08.01 & 095.B-0003(A) & 172 & 1720 & K          \\      
2016-03-23 &096.B-0174(A) &	60 & 900 & L$'$          \\
2018.04.22 & 0101.B-0052(B) & 120 & 1200& K          \\      
2018.04.24 & 0101.B-0052(B) & 120 & 1200& K          \\      
2018-04-22 &0101.B-0052(B)&	60 & 1800 & L$'$          \\
2018-04-24 &0101.B-0052(B)&	50 & 750 & L$'$           \\
\hline  
\end{tabular}
\caption{List of the used NACO data in the K- and L$'$-band. For every epoch, the number of exposures used for the final mosaics, the total exposure time,and the Project ID is listed. Note that NACO was decommissioned between 2013 and 2015.}
\label{data}
\end{table*}

\subsection{SINFONI data}

\begin{table*}[htbp!]
 	\centering
 	\begin{tabular}{cccccccc}
 	\hline\hline
 	\\	Date & Observation ID & Start time & End time & \multicolumn{3}{c}{Amount of on source exposures} & Exp. Time \\  \cline{5-7} & & & & Total & Medium & High &  \\
 	(YYYY:MM:DD) &  & (UT) & (UT) &  &  &  & (s)  \\ \hline %table heading
 	
 	\\	2006.03.17 & 076.B-0259(B) & 07:51:23 & 11:52:27 & 5  &  0   &  3  &    600  \\
 	2006.03.20 & 076.B-0259(B) & 07:33:06 & 07:42:27 &  1 &   1  &   0 &    600  \\
 	2006.03.21 & 076.B-0259(B) & 07:56:05 & 09:28:53 & 2  &   2  &  0  &    600  \\
 	2006.04.22 & 077.B-0503(B) & 08:47:59 & 08:57:24 & 1  &   0  &  0  &    600  \\
 	2006.08.17 & 077.B-0503(C) & 23:10:28 & 23:21:45 & 1  &   0  &  1  &    600  \\
 	2006.08.18 & 077.B-0503(C) & 23:19:54 & 03:46:01 & 5  &   0  &  5  &    600  \\
 	2006.09.15 & 077.B-0503(C) & 22:54:00 & 23:22:00 & 3  &   0  &  3  &    600  \\
 	2007.03.26 & 078.B-0520(A) & 06:46:26 & 06:57:32 &  1 &   1  &   0 &    600  \\
 	2007.07.24 & 179.B-0261(F) & 05:41:28 & 06:38:04 & 3 &   1  &  2  &    600  \\
 	2007.09.03 & 179.B-0261(K) & 23:42:50 & 01:11:12 & 4 &   1  &  3  &    600  \\
 	2007.09.04 & 179.B-0261(K) & 23:22:57 & 23:33:50 & 1 &   0  &  1  &    600  \\	
 	2008.04.06 & 081.B-0568(A) & 05:25:26 & 08:50:00 & 16 &   0  &  15   &    600  \\
 	2008.04.07 & 081.B-0568(A) & 08:33:58 & 09:41:05 &  4 &   0  &   4 &    600  \\
 	2009.05.21 & 183.B-0100(B) & 06:10:02 & 08:50:36 & 7 &   0  &  7   &    600  \\
 	2009.05.22 & 183.B-0100(B) & 04:51:22 & 05:29:13 &  4 &   0  &   4 &    400  \\
 	2009.05.23 & 183.B-0100(B) & 09:41:41 & 09:57:17 & 2 &   0  &  2  &    400  \\
 	2009.05.24 & 183.B-0100(B) & 06:08:03 & 07:38:59 & 3 &   0  &  3  &    600  \\
 	
 	\hline	\\
 	\end{tabular}	
 	\caption{SINFONI data between 2006 and 2009. We list the total amount of data. To ensure the best S/N ratio for our combined final data-cubes, we are just using single data-cubes with high quality.}
 	\label{tab:data_sinfo1}
 	\end{table*}
\begin{table*}[htbp!]

 	\centering
 	\begin{tabular}{cccccccc}
 	\hline\hline
 	\\	Date & Observation ID & Start time & End time & \multicolumn{3}{c}{Amount of on source exposures} & Exp. Time \\  \cline{5-7} & & & & Total & Medium & High & \\
 	(YYYY:MM:DD) & & (UT) & (UT) &  &  &  & (s) \\ \hline %table heading
 	
 	\\ 
 	2010.05.10 & 183.B-0100(O) & 06:03:00 & 09:35:20 & 3 &   0  &  3   &    600  \\
 	2010.05.11 & 183.B-0100(O) & 03:58:08 & 07:35:12 &  5 &   0  &   5 &    600  \\
 	2010.05.12 & 183.B-0100(O) & 09:41:41 & 09:57:17 & 13 &   0  &  13  &    600  \\
 	2011.04.11 & 087.B-0117(I) & 08:17:27 & 09:15:16 &  3 &   0  &   3 &    600  \\
 	2011.04.27 & 087.B-0117(I) & 05:10:41 & 09:23:09 & 10 &   1  &  9  &    600  \\
 	2011.05.02 & 087.B-0117(I) & 07:53:34 & 09:22:19 & 6 &   0  &  6  &    600  \\
 	2011.05.14 & 087.B-0117(I) & 08:47:28 & 09:25:07 & 2 &   0  &  2  &    600  \\
 	2011.07.27 & 087.B-0117(J)/087.A-0081(B) & 02:40:51 & 03:30:58 & 2 &   1  &  1  &    600  \\	
 	2012.03.18 & 288.B-5040(A) & 08:55:49 & 09:17:01 &  2 &   0  &   2 &    600  \\
 	2012.05.05 & 087.B-0117(J) & 08:09:14 & 08:41:33 & 3 &   0  &  3  &    600  \\
 	2012.05.20 & 087.B-0117(J) & 08:13:44 & 08:23:44 & 1 &   0  &  1  &    600  \\
 	2012.06.30 & 288.B-5040(A) & 01:40:19 & 06:54:41 & 12 &   0  &  10  &    600  \\
 	2012.07.01 & 288.B-5040(A) & 03:11:53 & 05:13:45 & 4 &   0  &  4  &    600  \\
 	2012.07.08 & 288.B-5040(A)/089.B-0162(I) & 00:47:39 & 05:38:16 & 13 &   3  &  8  &    600  \\
 	2012.09.08 & 087.B-0117(J) & 00:01:36 & 00:23:33 & 2 &   1  &  1  &    600  \\
 	2012.09.14 & 087.B-0117(J) & 01:21:30 & 01:43:27 & 2 &   0  &  2  &    600  \\	
 	2013.04.05 & 091.B-0088(A) & 08:55:49 & 09:43:38 &  2 &   0  &   2 &    600  \\
 	2013.04.06 & 091.B-0088(A) & 07:35:58 & 09:23:37 & 8 &   0  &  8  &    600  \\
 	2013.04.07 & 091.B-0088(A) & 07:27:49 & 09:40:35 & 3 &   0  &  3  &    600  \\
 	2013.04.08 & 091.B-0088(A) & 06:42:53 & 09:12:52 & 9 &   0  &  6  &    600  \\
 	2013.04.09 & 091.B-0088(A) & 07:08:17 & 10:01:46 & 8 &   1  &  7  &    600  \\
 	2013.04.10 & 091.B-0088(A) & 06:10:45 & 07:47:07 & 3 &   0  &  3  &    600  \\
 	2013.08.28 & 091.B-0088(B) & 00:40:17 & 03:11:43 & 10 &  1  &  6  &    600 \\
 	2013.08.29 & 091.B-0088(B) & 00:46:40 & 03:10:00 & 7 &  2   &  4  &    600 \\
 	2013.08.30 & 091.B-0088(B) & 02:31:22 & 03:24:46 & 4 &  2   &  0  &    600 \\
 	2013.08.31 & 091.B-0088(B) & 00:57:58 & 02:28:22 & 6 &  0   &  4  &    600 \\
 	2013.09.23 & 091.B-0086(A) & 23:40:31 & 01:49:33 & 6 &   0  &  0  &    600  \\
 	2013.09.25 & 091.B-0086(A) & 00:39:51 & 01:01:39 & 2 &   1  &  0  &    600  \\
 	2013.09.26 & 091.B-0086(A) & 23:55:53 & 00:39:13 & 3 &   1  &  1  &    600  \\	
 	
 	\hline	\\
 	\end{tabular}
 	
 	\caption{SINFONI data between 2010 and 2013.}
 	\label{tab:data_sinfo2}
 	\end{table*}

\begin{table*}[htbp!]

 	\centering
 	\begin{tabular}{cccccccc}
 	\hline\hline
 	\\	Date & Observation ID & Start time & End time & \multicolumn{3}{c}{Amount of on source exposures} & Exp. Time \\  \cline{5-7} & & & & Total & Medium & High & \\
 	(YYYY:MM:DD) & & (UT) & (UT) &  &  &  & (s) \\ \hline 
 	
 	\\ 
 	2014.02.27 & 092.B-0920(A) & 08:40:42  & 09:41:36  &  4 &   1  &  3  &    600   \\
 	2014.02.28 & 091.B-0183(H) & 08:34:58  & 09:54:37  &  7 &   3  &  1  &    400   \\
 	2014.03.01 & 091.B-0183(H) & 08:00:14  & 10:17:59  & 11 &   2  &  4  &    400  \\
 	2014.03.02 & 091.B-0183(H) & 07:49:05  & 08:18:54  &  3 &   0  &  0  &    400   \\
 	2014.03.11 & 092.B-0920(A) & 08:03:55  & 10:03:28  & 11 &   2  &  9  &    400   \\
 	2014.03.12 & 092.B-0920(A) & 07:44:34  & 10:07:45  & 13 &   8  &  5  &    400   \\
 	2014.03.26 & 092.B-0009(C) & 06:43:04  & 09:58:12  & 9  &   3  &  5  &    400   \\
 	2014.03.27 & 092.B-0009(C) & 06:32:49  & 10:04:12  & 18 &   7  &  5  &    400   \\
 	2014.04.02 & 093.B-0932(A) & 06:31:38  & 09:53:52  & 18 &   6  &  1  &    400    \\
 	2014.04.03 & 093.B-0932(A) & 06:20:45  & 09:45:02  & 18 &   1  &  17 &    400    \\
 	2014.04.04 & 093.B-0932(B) & 05:58:18  & 09:47:58  & 21 &   1  &  20 &    400       \\
 	2014.04.06 & 093.B-0092(A) & 07:51:42  & 08:43:15  &  5 &   2  &  3  &    400      \\
 	2014.04.08 & 093.B-0218(A) & 07:04:37  & 09:39:47  & 5  &   1  &  0  &    600   \\
 	2014.04.09 & 093.B-0218(A) & 07:43:44  & 09:31:25  &  6 &   0  &  6  &    600   \\
 	2014.04.10 & 093.B-0218(A) & 05:46:47  & 09:45:03  & 14 &   4  &  10  &    600   \\
 	    2014.05.08 & 093.B-0217(F) & 05:54:04  & 10:13:10  & 14 &   0  &  14  &    600   \\
 	    2014.05.09 & 093.B-0218(D) & 04:48:50  & 10:26:25  & 18 &   3  &  13  &    600   \\
 	2014.06.09 & 093.B-0092(E) & 05:22:34  & 08:59:59  &  14 &   3  &  0  &    400   \\
 	2014.06.10 & 092.B-0398(A)/093.B-0092(E) & 05:20:38  & 06:35:24  & 5 &   4  &  0   & 400/600 \\
 	2014.07.08 & 092.B-0398(A) & 02:22:28  & 04:47:18  & 6 &   1  &  3   &    600 \\
 	2014.07.13 & 092.B-0398(A) & 00:33:27  & 01:41:15  & 4 &   0  &  2   &    600 \\
 	2014.07.18 & 092.B-0398(A)/093.B-0218(D) & 02:53:11  & 03:03:18  & 1 &   0  &  0   &    600 \\
 	2014.08.18 & 093.B-0218(D) & 02:26:09  & 02:47:10  & 2 &   0  &  1   &    600 \\ 
 	2014.08.26 & 093.B-0092(G) & 00:23:04  & 01:00:16  &  4 &   3  &   0 &    400   \\
 	2014.08.31 & 093.B-0218(B) & 23:30:25  & 01:20:25  & 6 &   3   &   1 &    600 \\
 	2014.09.07 & 093.B-0092(F) & 01:28:11  & 01:42:45  & 2 &   0  &  0  &    400   \\
 	2015.04.12 & 095.B-0036(A) & 05:52:37  & 10:14:30  & 18 &  2 & 0 & 400 \\
 	2015.04.13 & 095.B-0036(A) & 05:44:12  & 10:06:37  & 13 &  7 & 0 & 400 \\
 	2015.04.14 & 095.B-0036(A) & 07:07:28  & 08:29:18  & 5  &  1 & 0 & 400 \\
 	2015.04.15 & 095.B-0036(A) & 05:52:26  & 10:16:55  & 23 &  13  & 10 & 400 \\
 	2015.08.01 & 095.B-0036(C) & 23:22:12  & 04:41:09  & 23 &   7  & 8  & 400 \\
 	2015.09.05 & 095.B-0036(D) & 23:18:11  & 02:23:03  & 17 &  11  & 4  & 400 \\

 	\hline	\\
 	\end{tabular}
 	
 	\caption{SINFONI data between 2014 and 2015.}
 	\label{tab:data_sinfo3}
 	\end{table*}
 	
\section{Appendix}

\subsection{K- and L-band positions of the D-sources}

\begin{table*}[htbp!]
        \centering
        \small
        \tabcolsep=0.15cm
        \begin{tabular}{ccccccccc}
            \hline
            \hline
            & \multicolumn{2}{c}{D2 positions} & \multicolumn{2}{c}{D23 positions}& \multicolumn{2}{c}{D3 positions} &\multicolumn{2}{c}{D3.1 positions}\\
            date & $\Delta$R.A. [mas]  & $\Delta$Dec. [mas] & $\Delta$R.A. [mas]  & $\Delta$Dec. [mas] & $\Delta$R.A. [mas]  & $\Delta$Dec. [mas] & $\Delta$R.A. [mas]  & $\Delta$Dec. [mas]\\
            \hline
            2002.578 & - & - & -324.76 $\pm$ 13.0 & 126.33 $\pm$ 13.0                                   & - & - & - & - \\
            2003.446 & - & - & -306.36 $\pm$ 6.5 & 125.18 $\pm$ 6.5                                     & - & - & - & - \\  
            2004.661 & - & - & -347.65 $\pm$ 13.0 & 99.29 $\pm$ 13.0                                    & - & - & - & - \\
            2004.669 & - & - & -336.28 $\pm$ 13.0 & 101.88 $\pm$ 13.0                                   & - & - & - & - \\
            2004.726 & - & - & -347.47 $\pm$ 6.5 & 97.07 $\pm$ 6.5                                      & - & - & - & - \\
            2005.268 & - & - & -350.00 $\pm$ 6.5 & 108.94 $\pm$ 6.5                                     & - & - & - & - \\
            2005.369 & - & - & -376.69 $\pm$ 6.5 & 134.30 $\pm$ 6.5                                     & - & - & - & - \\
            2005.372 & - & - & -332.92 $\pm$ 6.5 & 133.15 $\pm$ 6.5                                     & - & - & - & - \\
            2005.567 & - & - & -362.95 $\pm$ 6.5 & 135.83 $\pm$ 6.5                                     & - & - & - & - \\
            2007.213 & - & - & -396.50 $\pm$ 13.0 & 152.04 $\pm$ 13.0                                   & - & - & - & - \\ 
            2007.252 & - & - & -397.15 $\pm$ 13.0 & 137.21 $\pm$ 13.0                                   & - & - & - & - \\ 
            2007.454 & - & - & -396.20 $\pm$ 13.0 & 139.59 $\pm$ 13.0                                   & - & - & - & - \\ 
            2008.144 & -464.92 $\pm$ 13.0 & -93.8 $\pm$ 13.0  & -414.26 $\pm$ 6.5 & 125.75 $\pm$ 6.5    & -566.92 $\pm$ 10.44 & -23.80 $\pm$ 16.53 & -  & -  \\  
            2008.196 & -471.79 $\pm$ 13.0 & -82.21 $\pm$ 13.0 & -425.79 $\pm$ 6.5 & 114.30 $\pm$ 6.5    & -573.79 $\pm$ 15.08 & -12.21 $\pm$ 19.31 & -  & -  \\ 
            2008.267 & -470.03 $\pm$ 13.0 & -98.59 $\pm$ 13.0 & -419.38 $\pm$ 13.0 & 112.26 $\pm$ 13.0  & -572.03 $\pm$ 14.14 & -28.59 $\pm$ 17.84 & -501.0 $\pm$ 15.0 & 33.0 $\pm$ 15.0  \\
            2008.456 & -469.78 $\pm$ 13.0 & -85.77 $\pm$ 13.0 & -428.69 $\pm$ 6.5 & 122.64 $\pm$ 6.5    & -571.78 $\pm$ 11.70 & -15.77 $\pm$ 21.62 & - & - \\  
            2008.598 & -466.08 $\pm$ 13.0 & -86.89 $\pm$ 13.0 & -424.97 $\pm$ 13.0 & 132.00 2$\pm$ 13.0 & -568.08 $\pm$ 15.63 & -16.89 $\pm$ 23.87 & - & - \\
            2008.707 & -471.37 $\pm$ 13.0 & -85.80 $\pm$ 13.0 & -428.17 $\pm$ 6.5 & 126.80 $\pm$ 6.5    & -573.37 $\pm$ 14.28 & -15.80 $\pm$ 21.09 & - & - \\
            2009.298 & -459.5 $\pm$ 6.5 & -45.33 $\pm$ 6.5    & -424.62 $\pm$ 13.0 & 126.70 $\pm$ 13.0  & -575.50 $\pm$ 17.16 &  10.67 $\pm$ 17.78 & - & - \\
            2009.301 & -457.08 $\pm$ 6.5 & -48.45 $\pm$ 6.5   & -424.89 $\pm$ 13.0 & 123.67 $\pm$ 13.0  & -573.08 $\pm$ 13.05 &   7.55 $\pm$ 21.82 & - & - \\ 
            2009.334 & 449.71 $\pm$ 6.5 & -40.40 $\pm$ 6.5    & -430.34 $\pm$ 6.5 & 127.57 $\pm$ 6.5    & -565.71 $\pm$ 16.91 &  15.60 $\pm$ 18.24 & - & - \\
            2009.369 & - & - & -461.24 $\pm$ 13.0 & -62.54 $\pm$ 13.0                                   & - & - &   &   \\
            2009.501 & -457.89 $\pm$ 13.0 & -40.45 $\pm$ 13.0 & -428.58 $\pm$ 13.0 & 117.76 $\pm$ 13.0  & -573.89 $\pm$ 15.60 &  15.55 $\pm$ 19.01 & - & -  \\ 
            2009.509 & - & - & -424.48 $\pm$ 6.5 & 112.95 $\pm$ 6.5	                                   & - & - & -  & -  \\  
            2009.611 & -449.71 $\pm$ 6.5 & -40.40 $\pm$ 6.5 & -430.93 $\pm$ 6.5 & 130.90 $\pm$ 6.5      & -565.71 $\pm$ 16.12 &  15.60 $\pm$ 23.99 & - & - \\
            2009.715 & -439.88 $\pm$ 13.0 & -33.22 $\pm$ 13.0 & -426.42 $\pm$ 13.0 & 127.02 $\pm$ 13.0  & -555.88 $\pm$ 17.43 &  22.78 $\pm$ 26.15 & - & - \\
            2009.717 & -457.22 $\pm$ 13.0 & -48.24 $\pm$ 13.0 & -424.87 $\pm$ 13.0 & 133.18 $\pm$ 13.0  & -573.22 $\pm$ 19.22 &   7.76 $\pm$ 23.45 & - & - \\
            2010.350 & -454.89 $\pm$ 6.5 & -45.43 $\pm$ 6.5   &  -483.54 $\pm$ 6.5 & 105.78 $\pm$ 6.5   & -570.89 $\pm$ 15.25 &  10.57 $\pm$ 24.73 & -519.0 $\pm$ 15.0 & 33 $\pm$ 15.0\\
            2010.454 & -457.76 $\pm$ 6.5 & -35.38 $\pm$ 6.5   &  -484.46 $\pm$ 6.5 & 109.42 $\pm$ 6.5   & -573.76 $\pm$ 17.43 &  20.62 $\pm$ 25.42 & - & - \\
            2011.400 & -455.53 $\pm$ 6.5 & -25.50 $\pm$ 6.5   &  -486.99 $\pm$ 6.5 & 134.24 $\pm$ 6.5   & -599.53 $\pm$ 20.58 &  30.50 $\pm$ 26.21 & -466 $\pm$ 15.0  & 51 $\pm$ 15.0 \\
            2012.374 & -401.66 $\pm$ 6.5 & -8.35 $\pm$ 6.5    & -477.94 $\pm$ 6.5 & 105.63 $\pm$ 6.5    & -573.66 $\pm$ 17.22 &  47.65 $\pm$ 21.18 & - & - \\
            2013.487 & -412.51 $\pm$ 13.0 & 11.67 $\pm$ 13.0  &   -465.00 $\pm$ 6.5 & 131.23 $\pm$ 6.5  & -556.51 $\pm$ 20.25 &  39.67 $\pm$ 23.61 & - & - \\
            2013.670 & - & - & -506.79 $\pm$ 6.5 & 131.93 $\pm$ 6.5                                     & - & - & -  & -  \\
            2015.580 & -390.97 $\pm$ 13.0 & 61.57 $\pm$ 13.0  & -537.93 $\pm$ 6.5 & 126.65 $\pm$ 6.5    & -576.97 $\pm$ 19.99 &  61.57 $\pm$ 27.15 & - & - \\ 
            2018.306 & -307.31 $\pm$ 13.0 & 120.87 $\pm$ 13.0 & -553.72 $\pm$ 6.5 & 122.33 $\pm$ 6.5    & -549.31 $\pm$ 22.36 & 106.87 $\pm$ 24.55 & - & - \\ 
            \hline
        \end{tabular}
                \caption{K-band positions of D2, D23, D3, and D3.1 extracted from the NACO data between 2002 and 2018.}
        \label{tab:k_band_pos}
    \end{table*}

\begin{table*}[htbp!]
    \centering
    \small
    \tabcolsep=0.11cm
    \begin{tabular}{ccccccccc}
    \hline
    \hline
    & \multicolumn{2}{c}{D2 positions} &\multicolumn{2}{c}{D23 positions} & \multicolumn{2}{c}{D3 positions} &\multicolumn{2}{c}{D3.1 positions}\\
    date & $\Delta$R.A. [mas]  & $\Delta$Dec. [mas] & $\Delta$R.A. [mas]  & $\Delta$Dec. [mas] & $\Delta$R.A. [mas]  & $\Delta$Dec. [mas] & $\Delta$R.A. [mas]  & $\Delta$Dec. [mas]\\
    \hline
    2002.66 & -500.58 $\pm$ 5.64  & -176.85  $\pm$ 5.51 & -                 & -                 & -612.58 $\pm$ 10.44 & -36.85 $\pm$ 16.53 & - & - \\
    2002.72 & -497.88 $\pm$	4.77 & -180.9   $\pm$ 5.21 & -                 & -                  & -609.88 $\pm$ 15.08 & -40.90 $\pm$ 19.31 & - & - \\
    2002.75 & -487.08 $\pm$	4.42 & -182.79  $\pm$ 4.92 & -338.85 $\pm$13.6 &  119.47 $\pm$ 13.6 & -599.08 $\pm$ 14.14 & -42.79 $\pm$ 17.84 & - & - \\
    2002.83 & -497.88 $\pm$	5.04 & -183.6   $\pm$ 5.62 & -                 &               -    & -609.88 $\pm$ 11.70 & -43.60 $\pm$ 21.62 & - & - \\
    2003.44 & -491.13 $\pm$	3.99 & -176.58  $\pm$ 4.66 & -385.95 $\pm$13.6 & 113.85$\pm$13.6    & -603.13 $\pm$ 15.63 & -64.58 $\pm$ 23.87 & - & - \\
    2004.32 & -485.73 $\pm$	2.49 & -151.47  $\pm$ 8.54 & -                 &              -     & -611.73 $\pm$ 14.28 & -39.47 $\pm$ 21.09 & -550.73 $\pm$ 15.0 & -8.47 $\pm$ 15.0 \\
    2004.40 & -482.76 $\pm$ 4.04  & -155.79  $\pm$ 4.76 & -363.19 $\pm$13.6 & 119.24$\pm$13.6   & -608.76 $\pm$ 17.16 & -43.79 $\pm$ 17.78 & - & - \\
    2004.41 & -481.41 $\pm$ 4.52  & -159.57  $\pm$ 6.70 & -346.50 $\pm$13.6 & 118.68$\pm$13.6   & -607.41 $\pm$ 13.05 & -47.57 $\pm$ 21.82 & - & - \\
    2005.45 & -485.19 $\pm$ 4.50  & -133.92  $\pm$ 9.99 & -374.66$\pm$13.6  & 126.32$\pm$13.6   & -611.19 $\pm$ 16.91 & -21.92 $\pm$ 18.24 & -602.19 $\pm$ 15.0 & -16.92 $\pm$ 15.0\\
    2005.54 & -482.49 $\pm$	5.46 & -124.47  $\pm$ 5.92 & -                 &              -     & -608.49 $\pm$ 15.60 & -12.47 $\pm$ 19.01 & - & - \\
    2006.50 & -473.85 $\pm$ 4.86  & -110.43  $\pm$ 4.65 & -394.80$\pm$13.6  & 126.32$\pm$13.6   & -599.85 $\pm$ 16.12 & -12.43 $\pm$ 23.99 & -590.85 $\pm$ 15.0 & -6.43 $\pm$ 15.0 \\
    2007.34 & -460.62 $\pm$ 4.59  & -99.36   $\pm$ 3.65 & -416.99$\pm$13.6  & 126.27$\pm$13.6   & -572.62 $\pm$ 17.43 & -15.36 $\pm$ 26.15 & -603.62 $\pm$ 15.0 & 34.36 $\pm$ 15.0 \\
    2007.48 & -461.43 $\pm$ 3.09  & -96.12   $\pm$ 2.75 & -403.1$\pm$13.6   & 134.58$\pm$13.6   & -573.43 $\pm$ 19.22 & -12.12 $\pm$ 23.45 & - & - \\
    2008.49 & -453.33 $\pm$ 5.92  & -79.38   $\pm$ 6.93 & -432.47$\pm$13.6  & 129.57$\pm$13.6   & -579.33 $\pm$ 15.25 &  18.62 $\pm$ 24.73 & - & - \\
    2011.40 & -433.35 $\pm$ 10.09 & -41.04   $\pm$ 11.66& -                 &               -   & -545.35 $\pm$ 17.43 &  14.96 $\pm$ 25.42 & - & - \\
    2011.49 & -416.07 $\pm$ 11.86 & -34.02   $\pm$ 12.94& -479.54 $\pm$13.6 & 139.92$\pm$13.6   & -528.07 $\pm$ 20.58 &  21.98 $\pm$ 26.21 & - & - \\
    2012.46 & -423.63 $\pm$ 9.37  & -8.1     $\pm$ 12.31& -468.76$\pm$13.6  & 141.17$\pm$13.6   & -535.63 $\pm$ 17.22 &  61.90 $\pm$ 21.18 & -405.0 $\pm$ 15.0 & 25.0 $\pm$ 15.0 \\
    2013.44 & -414.99 $\pm$ 10.04 & 23.76    $\pm$ 13.49& -484.57$\pm$13.6  & 135.39$\pm$13.6   & -540.99 $\pm$ 20.25 &  79.76 $\pm$ 23.61 & - & - \\
    2016.31 & -370.44 $\pm$ 7.04  & 86.67    $\pm$ 9.93 & -519.09$\pm$13.6  & 114.32$\pm$13.6   & -510.44 $\pm$ 19.99 & 114.67 $\pm$ 27.15 & - & - \\
    2018.39 & -343.98 $\pm$ 7.24  & 123.66   $\pm$ 10.27& -510.75$\pm$13.6  & 121.32$\pm$13.6   & -539.98 $\pm$ 22.36 & 137.66 $\pm$ 24.55 & - & - \\
    2018.40 & -343.17 $\pm$ 6.39  & 125.55   $\pm$ 10.62& -514.39$\pm$13.6  & 136.87$\pm$13.6   & -539.17 $\pm$ 18.45 & 139.55 $\pm$ 29.51 & - & - \\
    \hline
    \end{tabular}
    \caption{Positions of D2, D23, D3, and D3.1 from the L-band NACO data set.}
    \label{tab:L-band-pos}
\end{table*}

\section{Appendix}

\subsection{Bubble around SgrA* - a NIR/Radio connection}

\begin{figure*}[htbp!]
\centering
\includegraphics[width=1.\textwidth]{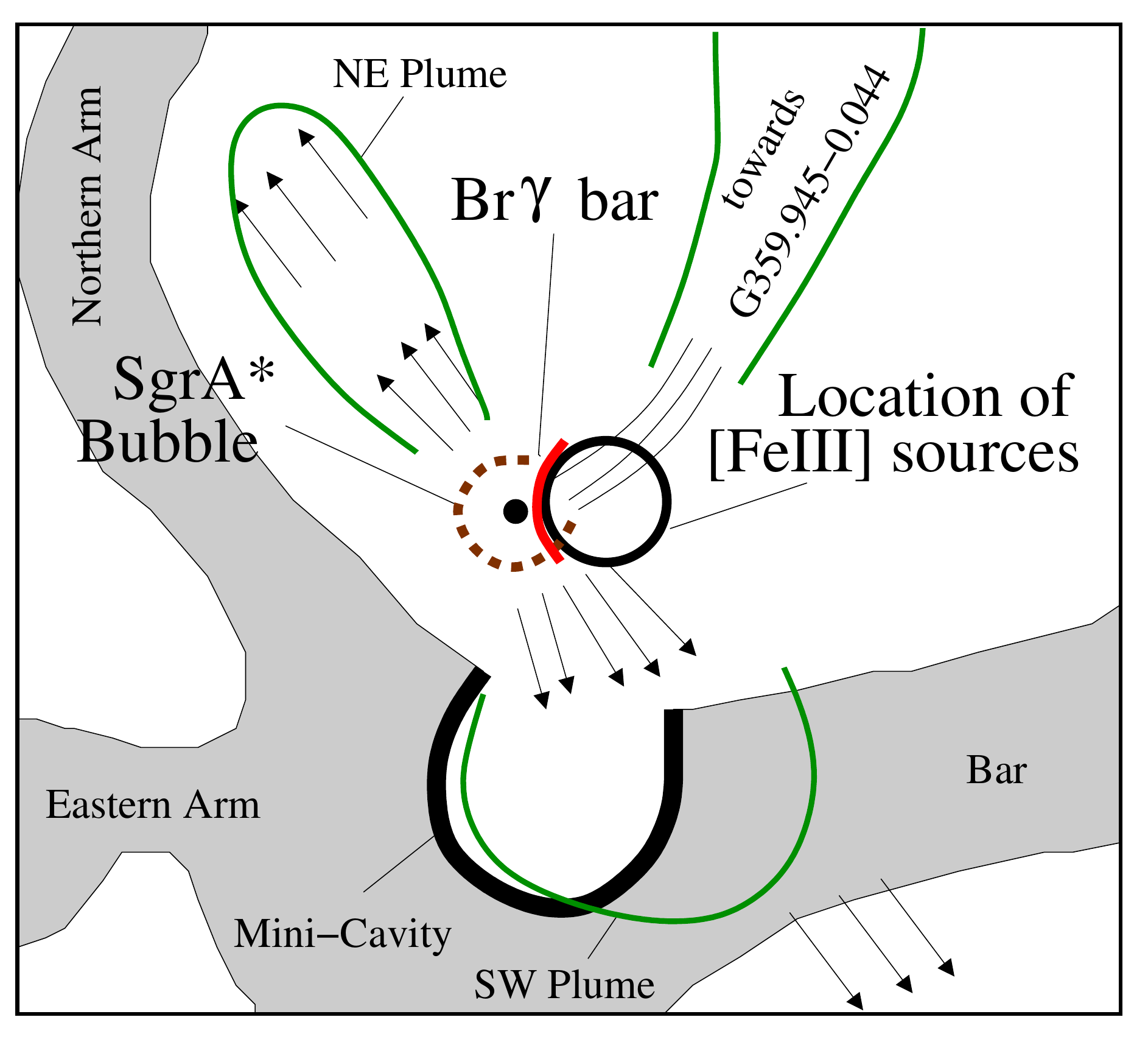}
\caption{This image is inspired by the results presented in \cite{Yusef-Zadeh2012}. It shows the radio emission and the detected features. We indicate the position of the bright Br$\gamma$-emission with a red bar at the position of the open SgrA* bubble. The black circle marks the position of the dusty sources, that show Doppler-shifted [FeIII] emission lines.}
\label{fig:RadioNIRbubble}
\end{figure*}

\section{Appendix}

\subsection{H- and K-band detection of the DSO/G2 object}

\begin{figure*}[htbp!]
\centering
\includegraphics[width=0.8\textwidth]{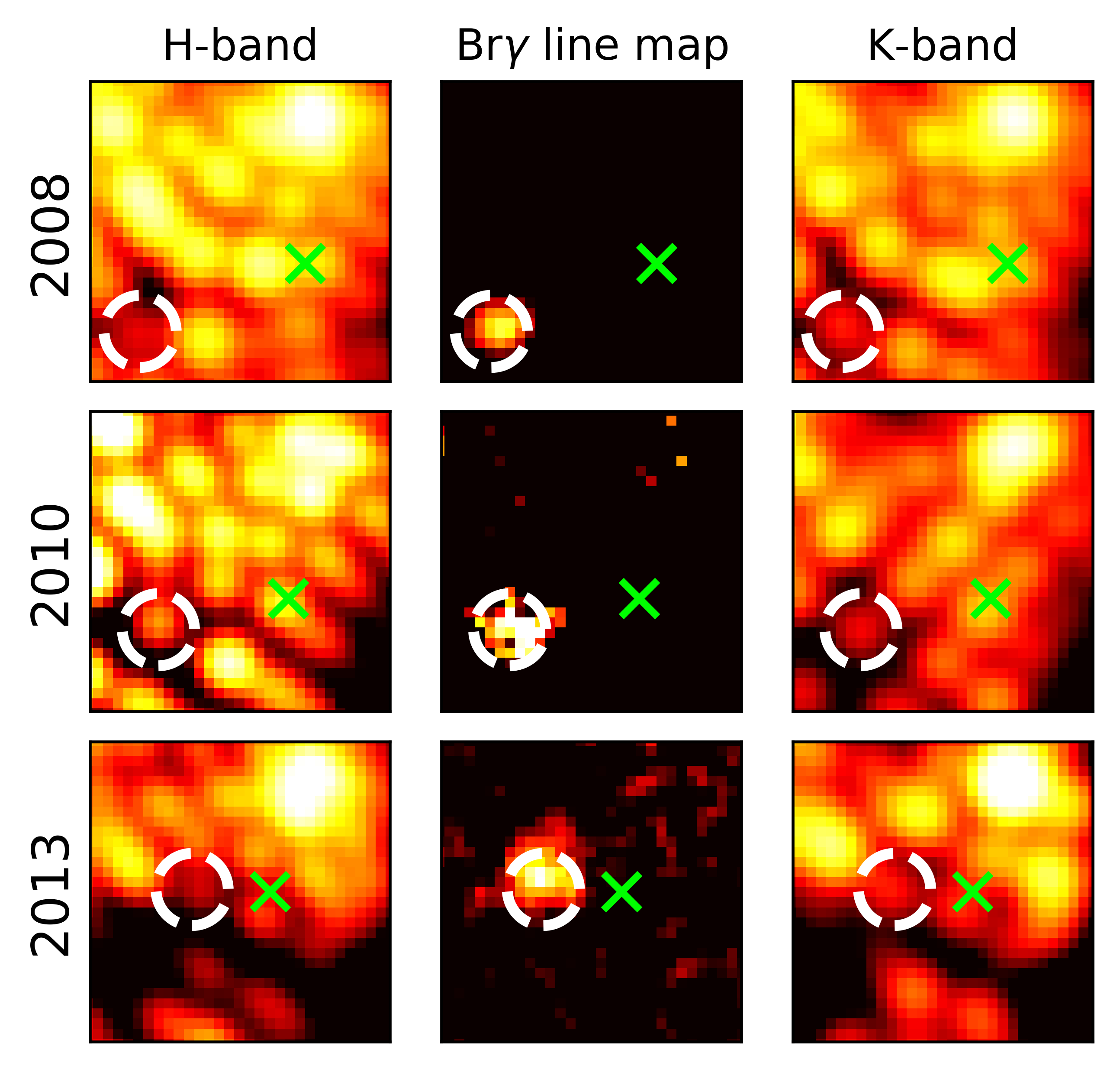}
\caption{H-, K-band detection of the DSO. For a positional comparison, we include the Doppler-shifted line-map detection. The data is based on the SINFONI H+K data-cubes of 2008, 2010, and 2013. In all images, the North is up and the East is to the left. Sgr A* is marked with a lime-colored {\it x}. The FOV in every image is 0".375$\times$0".375.}
\label{fig:hk_dso}
\end{figure*}

\end{appendix}
\end{document}